\documentclass[printer]{aa}
\usepackage{graphicx}
\usepackage{natbib,txfonts}
\usepackage[colorlinks=true, citecolor=blue]{hyperref}

\bibpunct{(}{)}{;}{a}{}{,} 

\def\teff{\textit{T}_{\text{eff}}}
\def\logg{\text{log}(\textit{g})}

\def\feh{[\text{Fe}/\text{H}]}

\def\vrad{V_{\text{rad}}}
\def\cms{cm~s$^{-2}$}

\begin{document}

\title{The Gaia-ESO Survey: Preparing the ground for 4MOST \& WEAVE galactic surveys\ \\ - Chemical evolution of lithium with machine learning \thanks{Table \ref{table:catalog} is only available in electronic form
at the CDS via anonymous ftp to \href{cdsarc.cds.unistra.fr}{cdsarc.cds.unistra.fr} (130.79.128.5)
or via \href{https://cdsarc.cds.unistra.fr/cgi-bin/qcat?J/A+A/}{https://cdsarc.cds.unistra.fr/cgi-bin/qcat?J/A+A/}}%
\fnmsep\thanks{\href{https://github.com/SamirNepal/Li_CNN_2022}{https://github.com/SamirNepal/Li\_CNN\_2022}}%
\fnmsep\thanks{Based on observations collected with ESO telescopes at the La Silla Paranal Observatory in Chile, for the Gaia-ESO Large Public Spectroscopic Survey (188.B-3002, 193.B-0936, 197.B-1074).}
}

\author{S. Nepal\inst{1, 2}\and G. Guiglion\inst{3, 1}\and R. S. de Jong\inst{1}\and M. Valentini\inst{1}\and C. Chiappini\inst{1}\and M. Steinmetz\inst{1}
\and M. Ambrosch\inst{4}
\and E. Pancino\inst{5}
\and R. D. Jeffries\inst{6}
\and T. Bensby\inst{7}
\and D. Romano\inst{8}
\and R. Smiljanic\inst{9}
\and M.L.L. Dantas\inst{9}
\and G. Gilmore\inst{10}
\and S. Randich\inst{5}
\and A. Bayo\inst{11}
\and M. Bergemann\inst{12,3}
\and E. Franciosini\inst{5}
\and F. Jim{\'e}nez-Esteban\inst{13}
\and P. Jofr{\'e}\inst{14}
\and L. Morbidelli\inst{5}
\and G.G. Sacco\inst{5}
\and G. Tautvai{\v s}ien{\. e}\inst{4}
\and S. Zaggia\inst{15}
}

\institute{Leibniz-Institut f\"ur Astrophysik Potsdam (AIP), An der Sternwarte 16, 14482 Potsdam, Germany \\
e-mail: \texttt{snepal@aip.de, guiglion@mpia.de}
\and 
Institut f\"ur Physik und Astronomie, Universit\"at Potsdam, Karl-Liebknecht-Str. 24/25, 14476 Potsdam, Germany
\and
Max Planck Institute for Astronomy, K\"onnigstuhl 17, 69117, Heidelberg, Germany
\and
Institute of Theoretical Physics and Astronomy, Vilnius University, Sauletekio av. 3, 10257 Vilnius, Lithuania
\and
INAF, Osservatorio Astrofisico di Arcetri, Largo Enrico Fermi 5, 50125 Firenze, Italy
\and
Astrophysics Group, Keele University, Keele, Staffordshire ST5 5BG, United Kingdom
\and
Lund Observatory, Department of Astronomy and Theoretical Physics, Box 43, SE-22100 Lund, Sweden
\and
INAF, Osservatorio di Astrofisica e Scienza dello Spazio, Via Gobetti 93/3, 40129 Bologna, Italy
\and
Nicolaus Copernicus Astronomical Center, Polish Academy of Sciences, ul. Bartycka 18, 00-716, Warsaw, Poland
\and
Institute of Astronomy, University of Cambridge, Madingley Road, Cambridge CB3 0HA, United Kingdom
\and
European Southern Observatory, Karl Schwarzschild-Stra\ss e 2, D-85748 Garching bei M\"{u}nchen, Germany
\and
Niels Bohr International Academy, Niels Bohr Institute, University of Copenhagen Blegdamsvej 17, DK-2100 Copenhagen, Denmark
\and
Spanish Virtual Observatory, Centro de Astrobiolog{\'i}a (INTA-CSIC), 28691 Villanueva de la Ca{\~n}ada, Madrid, Spain
\and
N{\'u}cleo de Astronom{\'i}a, Facultad de Ingenier{\'i}a y Ciencias, Universidad Diego Portales (UDP), Santiago de Chile
\and
INAF, Osservatorio Astronomico di Padova, vicolo dell'Osservatorio, 5 - 35122 PADOVA, Italy}

\date{Received 18 August 2022 / Accepted 5 December 2022}

\abstract
{With its origin coming from several sources (Big Bang, stars, cosmic rays) and given its strong depletion during its stellar lifetime, the lithium element is of great interest as its chemical evolution in the Milky Way is not well understood at present. To help constrain stellar and galactic chemical evolution models, numerous and precise lithium abundances are necessary for a large range of evolutionary stages, metallicities, and Galactic volume.}
{In the age of stellar parametrization on industrial scales, spectroscopic surveys such as APOGEE, GALAH, RAVE, and LAMOST have used data-driven methods to rapidly and precisely infer stellar labels (atmospheric parameters and abundances). To prepare the ground for future spectroscopic surveys such as 4MOST and WEAVE, we aim to apply machine learning techniques to lithium measurements and analyses.}
{We trained a convolution neural network (CNN), coupling Gaia-ESO Survey iDR6 stellar labels ($\mathrm{\teff}$, log(\textit{g}), [Fe/H], and A(Li)) and GIRAFFE HR15N spectra, to infer the atmospheric parameters and lithium abundances for $\sim$40\,000 stars. The CNN architecture and accompanying notebooks are available online via GitHub.}
{We show that the CNN properly learns the physics of the stellar labels, from relevant spectral features through a broad range of evolutionary stages and stellar parameters. The lithium feature at 6707.8\,\AA\, is successfully singled out by our CNN, among the thousands of lines in the GIRAFFE HR15N setup. Rare objects such as lithium-rich giants are found in our sample. This level of performance is achieved thanks to a meticulously built, high-quality, and homogeneous training sample.}{The CNN approach is very well adapted for the next generations of spectroscopic surveys aimed at studying (among other elements) lithium, such as the 4MIDABLE-LR/HR (4MOST Milky Way disk and bulge low- and high-resolution) surveys. In this context, the caveats of machine--learning applications should be appropriately investigated, along with the realistic label uncertainties and upper limits for abundances.}  


\titlerunning{Lithium with Machine--Learning}
\authorrunning{Nepal et al.}
\maketitle

\section{Introduction} \label{sec:intro}

The element lithium\footnote{Unless differently indicated, by lithium (Li) we refer to the main isotope of lithium, $^{7}$Li} (Li) is of particular interest in astrophysics given its complex origin and evolution. Lithium was produced during the big bang (BB), and its primordial abundance can be used to constrain the standard model of cosmology. The standard BB nucleosynthesis (SBBN) model predicts the primordial lithium abundance to be A(Li)\footnote{A(Li) = log (N$_\text{Li}$/N$_\text{H}$) + 12} $\sim 2.75$\,dex \citep{pitrou2018precision}. Attempts to obtain an astrophysical measurement of this primordial Li using old, warm ($\mathrm{\teff}$\,$>$\,5\,600\,K), metal-poor ([Fe/H]<-1.5\,dex) halo dwarf stars has resulted in observation of a thin spread of lithium abundance that is independent of metallicity and effective temperature -- referred to as the ``Spite plateau,'' with A(Li) $\sim 2.2$\,dex \citep{spite1982abundance, Bonifacio_1997MNRAS}. This difference of a factor of three between the theoretical prediction and observation brings on the famous cosmological lithium problem (e.g., \citealt{fields2011}).

At later times, Li is produced at two distinct sources; in the interstellar medium (ISM) via a spallative interaction of galactic cosmic rays and the ISM through the ~p\,+\,C,N,O or $\alpha$+C,N,O reaction channels \citep{Reeves_1970Natur} as well as in stellar sources such as asymptotic giant branch (AGB) stars \citep{McKellar_1940PASP}, and red giants \citep{Sackmann_1999ApJ}, as well as core-collapse supernovae and novae \citep{DAntona_1991A&A, Izzo_2015ApJ}. However, the stellar yields for the different sources are not well constrained and present large uncertainties \citep{matteucci1995, Romano_1999A&A, Romano_2001, prantzos2017, Randich_2021}.

One production channel for Li in the stars is known as the Cameron-Fowler mechanism \citep{Cameron1971} whereby $^{7}$Be is first formed in temperatures hotter than $4 \times 10^{7}$\,K  via the reaction $\mathrm{^{3}He + \alpha \rightarrow \,^{7}Be + \gamma }$. The fresh $^{7}$Be must then be quickly moved to cooler layers by convection, where it decays to $^{7}$Li and is conserved and eventually released to the ISM. This mechanism explains the existence of Li-rich giants \citep{Brown1989ApJS, Charbonnel_2000A&A, Hong-liang2022ChA&A}.
Lithium could also be produced via the $\nu-$process taking place in the external shells of collapsing massive stars \citep{WoosleyWeaver1995, Kusakabe2019}.

Additionally, Li can already be easily destroyed in stars by the proton capture reaction $\mathrm{^{7}Li(p,\alpha) ^{4}He}$ at temperatures as low as $2.5\times 10^{6}$\,K  as early as the pre-main sequence (PMS) and in later stages, whenever that temperature is reached \citep{Pinsonneault_1997ARA&A}. For example, the meteoritic A(Li) is $\sim$3.26\,dex \citep{Lodders_2009M&PSA}, which represents the initial ISM Li for the Sun; whereas the Solar photospheric abundance of only A(Li) $\sim$ 1.05\,dex \citep{Grevesse_2007SSRv} suggests an internal destruction by a factor $>150$.

In order to investigate the stellar and galactic evolution of lithium, we need a statistically robust and homogeneous sample, such that a large metallicity domain and different evolutionary stages are covered. In recent years, due to the availability of larger samples of stars (typically several hundred), it has become possible to study lithium abundance in the context of chemical evolution of the thick and thin disks, internal destruction in stars, galactic chemical evolution, and exoplanet connection \citep{Lambert_2004, Ramirez_2012, Delgado_2015, Bensby_2018}. For example, \cite{guiglion2016} used high-resolution spectra from ESO to homogeneously build a Li catalog composed of 7\,300 stars, while studying the lithium evolution in the Milky Way. Most recently, the number of stars with available Li abundances has rapidly increased thanks to large-scale Milky Way spectroscopic surveys such as Gaia-ESO \citep{Fu_2018, GES_Randich_2020, magrini_2021A&A, Romano_2021A&A}, LAMOST \citep{Gao_LAMOST_2019}, and GALAH \citep{Gao_GALAH_2020},  contributing significantly to our understanding of the evolution of Li.

\begin{figure*}[h!]
        \centering
        \includegraphics[width=\linewidth]{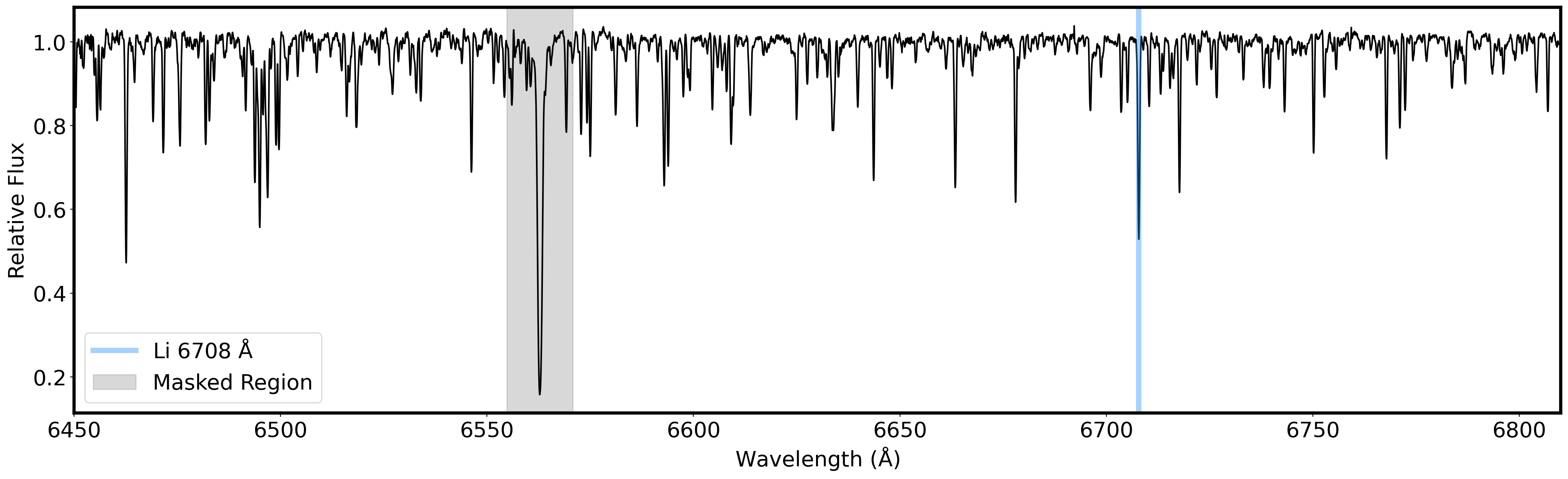}
        \caption[An example of GIRAFFE spectra]{Example GIRAFFE HR15N spectrum. This spectrum is of a star with labels: $\mathrm{\teff}=$ 4897\,K, log(\textit{g})= 2.55\,dex, [Fe/H]=-0.11\,dex, and A(Li)=2.63\,dex. Lithium spectral feature is shaded with blue, while the gray shaded region centred at H$_{\alpha}$ is masked and not used in the spectral analysis using CNN.}
        \label{fig:giraffe_spectra}
\end{figure*}

One way to precisely measure atmospheric parameters and chemical abundances in stellar atmosphere is to use stellar spectroscopy. Lithium abundance is usually derived from the Li doublet at 6\,707.8 \AA, shown in Fig. \ref{fig:giraffe_spectra}, which is the strongest Li feature in the optical wavelength regime. Other neutral Li lines at 6\,103 \AA\, and 8\,126 \AA\, have also been used for Li abundance analysis \citep{Gratton_1989A&A}, but these lines are very weak and  they are only detectable and measurable in high-resolution and/or at high-Li abundances. The 6\,707.8 \AA\, Li line strength has a strong dependence on the star's effective temperature and Li abundance. The Li doublet blends with the \ion{Fe}{I} line, thus making it challenging for classical spectroscopic pipelines to provide precise Li abundances at intermediate and low resolution or in the presence of noise. 

Over the last three decades, the community has generally measured Li abundances using classical spectroscopic pipelines\footnote{Classical pipelines refer to the tools that typically compare the observed spectrum to a model spectrum based on a line-list, a model atmosphere, and a prediction on the line shape as well as intensity (curve of growth) based on a model. These pipelines provide the stellar labels for training in the context of machine learning methods.} (SME, \citealt{SME_1996A&AS}; MOOG, \citealt{MOOG_sneden_2012}). In the era of future large spectroscopic surveys such as 4MOST \citep{deJong2019Msngr}, and WEAVE \citep{WEAVE_2016}, a number of $10^7$ spectra will be gathered and supplemented by the wealth of astrometric and photometric data provided by the Gaia satellite \citep{gaia2016,gaia2020,Lindegren2021}. The community will have to adapt their methods and machine learning is believed to be the way forward.

Machine learning (ML) tools are becoming popular for all research fields where it is necessary to quickly process large amount of data and/or automatically learn the complex correlations from high-dimensional data. One family of extremely versatile ML algorithms are neural networks (NN), which have become very popular and successfully applied in many other astronomy fields, such as gravitational lensing \citep{Petrillo_2017MNRAS}, the search for open clusters in Gaia data \citep{Castro-Ginard_2020A&A}, detecting outliers in astronomical imaging data sets \citep{Margalef-Bentabol_2020MNRAS} detecting gravitational waves \citep{Lin_2021PhRvD}, photometric redshift predictions \citep{Lima2022A&C}, and many more. Neural networks have actually been used in astrophysical applications for a long time, even though their architecture was relatively simple compared to the modern networks. For example: \citealt{Bailer-Jones_1997MNRAS} used NN to parametrize $\mathrm{\teff}$, log(\textit{g}), and [M/H] from stellar spectra and \citet{Bailer-Jones_1998MNRAS} used NN and principal component analysis (PCA) to classify spectral types.

Such machine learning approaches have also started to play an important role in the derivation of stellar labels. Such methods transfer the knowledge from a reference set of data, a so-called "training sample," to a larger set of data to derive the stellar labels. The reference set of data can be constructed from either empirical data or by employing spectral synthesis models. The Cannon \citep{Ness_Cannon_2015ApJ} is one of the pioneering data-driven spectroscopic analysis tools, while the Payne \citep{Ting_Payne_2019ApJ} has demonstrated that we can combine physical stellar models using neural networks as a function to generate spectra, instead of a quadratic polynomial function (as in the case of Cannon). It is important to note that the Payne uses noiseless synthetic spectra as the training set. A modification of the Payne tool, named data driven-Payne \citep{Xiang_2019ApJS}, has also been applied to the LAMOST low-resolution spectra.

A few recent studies used a class of neural networks called convolutional neural networks (CNN; \citealt{LeCun_backpropagation_1989, lecun-bengio-95a}) to derive atmospheric parameters and chemical abundances from both high- and low-resolution stellar spectra. Such CNNs are very efficient at feature extraction, hence, they can be used to learn about the spectral features in stellar spectra and relate it to the atmospheric parameters and chemical abundances. \citet{Fabbro_2018MNRAS} developed the StarNet pipeline based on a CNN and a synthetic training set. \citet{Bialek_2020MNRAS} applied StarNet to Gaia-ESO Survey UVES instrument spectra by training the CNN with various synthetic spectral grids while mitigating the ``synthetic gap." \citet{Leung_2019MNRAS} developed the astroNN tool (capable of handling missing labels) trained on observational data to derive 22 stellar parameters and chemical abundances based on APOGEE DR14 spectra and labels. \citet{Zhang_2019PASP} used StarNet to estimate the atmospheric parameters and chemical abundances of LAMOST low-resolution spectra, based on the high resolution APOGEE labels. \citet{guiglion2020} performed similar label transfer from APOGEE DR16 to the intermediate-resolution RAVE survey, in addition to combining astrometry and photometry as additional inputs. \citet{guiglion2020} showed that it is possible to improve the quality of predicted effective temperature and surface gravity by lifting the degeneracy in log(\textit{g}) using the absolute magnitudes. Very recently, novel methods such as auto-encoders and generative domain adaptation have also been implemented for stellar spectroscopy (e.g., in \citealt{Teaghan_2021ApJ, Klemen_2021MNRAS}). These research efforts and the developments in future spectroscopic surveys, computational power, and improved ML techniques are the motivation for preparing the ML ground for future spectroscopic surveys.

The main aim of this work is to provide reliable atmospheric parameters and Li abundances for a large sample of spectra and use it to study lithium evolution in the Milky Way. We adopted a CNN as a supervised ML method and our training labels are as follows: effective temperature, $\mathrm{\teff}$, surface gravity, log(\textit{g}), iron abundance, [Fe/H], and lithium abundance, A(Li). Any supervised ML method demands a very careful choice of training labels, as the trends and biases present in the training data are also learned and, hence, easily transferred to the predicted labels. This paper goes together with the work of \citet{Ambrosch_2022arXiv}, which focuses on the chemical evolution of Al and Mg abundances with CNN from GES GIRAFFE HR10 and HR21 spectra.

The paper is organized as follows. In Sect.~\ref{sec:obs_data}, we present the spectral data set adopted in this study. In Sect.~\ref{sec:CNN}, we detail the CNN procedure. The catalog of lithium abundances is presented in Sect.~\ref{sec:result}, while its validation is done in Sect.~\ref{sec:validation}. We present two scientific application of our catalog in Sect.~\ref{sec:gal_arcaeo} and we summarize our work and draw some future prospects in Sect.~\ref{sec:conclusion}.

\section{Observation and data} \label{sec:obs_data}

Our preliminary goal is to prepare the ground for 4MOST and WEAVE Li analyses. We  looked for public spectra similar to the red arm of these two surveys, with associated high-quality lithium and atmospheric parameters. We adopted the Gaia-ESO Survey (GES, \citealt{GES_2012Msngr, Randich_2013Msngr}) data.   Spectra was gathered by GES for all major Galactic components (halo, bulge, and thin and thick disks), including a large number of open and globular clusters, as well as calibration observations such as benchmark stars, radial velocity ($\mathrm{\vrad}$) standards, and asteroseismic CoRoT/K2 fields (see \citealt{Bragaglia_2022A&A, Pancino_2017A&A, Stonkute_2016MNRAS, valentini_corot_2016}). For this study, we use the spectra and parameters and abundances from the internal Data Release 6 (iDR6)\footnote{\href{http://ges.roe.ac.uk}{http://ges.roe.ac.uk}, \href{http://casu.ast.cam.ac.uk/gaiaeso/}{http://casu.ast.cam.ac.uk/gaiaeso}}.

The spectra were obtained using the GIRAFFE instrument of the Fibre Large Array Multi Element Spectrograph (FLAMES; \citealt{Pasquini_2002Msngr}) located at Very Large Telescope (VLT) Observatory at Cerro Paranal (ESO) in Chile. We used the H665.0/HR15N setup that includes the Li doublet at $6\,708\,$\AA. The HR15N setup is centred at ~6\,650\,\AA, and covers the domain ~[6\,470-6\,790]\,\AA\, with a resolving power R=19\,200, very similar to the WEAVE and 4MOST HR red arm. The GES-iDR6 also comprises Li abundances for $\sim6\,400$ UVES spectra, which, however, we do not use in this work.

The spectroscopic analysis within GES was performed by multiple data analysis nodes which use different spectroscopic tools, but adopting the same line list and model atmospheres (\citealt{Smiljanic_2014A&A, Lanzafame2015, heiter_2021, Gilmore2022, randich2022}; Worley et al., in prep.). The atmospheric parameters from each of the nodes are homogenized to provide a single measurement and associated uncertainty as the node-to-node dispersion. The different methods can  be  summarized  into three categories:  i)  equivalent width (EW) analysis where the atmospheric parameter determination is based on the excitation and ionization balance of the Fe lines; ii) spectral synthesis method that estimates atmospheric parameters from a $\chi ^{2}$\ fit to the  observed spectra; and iii) multilinear regression method that derives atmospheric parameters and abundances by projecting the observed spectrum into vector functions that are constructed as the best linear combination of synthetic spectra from a grid. Here, we adopted the GES-iDR6 atmospheric parameters, $\mathrm{\teff}$, and log(\textit{g}), as well as the [Fe/H] abundance ratio.

GES-iDR6 provides one-dimensional local thermodynamical equilibrium (1D LTE) abundances for $^{7}$Li, measured using the EW measurement of the spectral feature at 6707.8 \AA. The measured EWs are converted to lithium abundances using curves of growth (only one GES node contributed to Li determinations; see Sect. 2.1 of \citealt{Romano_2021A&A}, and \citealt{Franciosini_2022}). For the GIRAFFE spectra, the Li line is blended with a nearby FeI line at 6\,707.4 \AA, hence, a correction was applied. When the Li spectral line is very weak or not visible, an upper limit to the abundance is provided. GES also provides a flag for Li abundances ($\mathrm{UPPER\_COMBINED\_LI1}$, 0=detection, 1=upper limit); an upper limit is provided when the 6\,707.8\,\AA\ Li line is undetected, as a result of too low values for the signal-to-noise ratio (S/N) or too little lithium (see \citealt{Franciosini_2022} for details).

\subsection{Training and observed sample} \label{sec:t_sample}

\begin{figure*}[h]
        \centering
        \includegraphics[width=\linewidth]{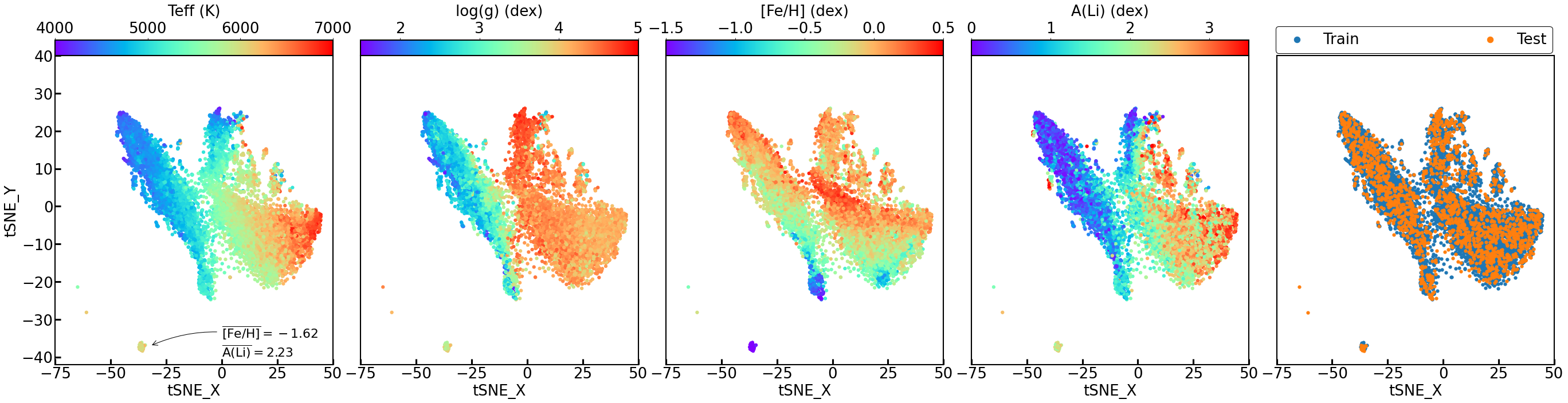}
        \caption{2D projection of t-SNE output for the 7\,031 spectra of the training sample, colored by the labels $\mathrm{T_{eff}}$, log(\textit{g}), [Fe/H] and A(Li) respectively. The right-most plot shows the t-SNE as the train and test sets to highlight their similar distribution across the label range. In the left subplot, we show the mean [Fe/H] and A(Li) for the highlighted island that consists of Spite plateau-like stars in the globular cluster NGC\,6752.}
        \label{fig:tsne_training_color}
\end{figure*}

To build the training sample\footnote{Throughout the paper, "training sample" refers to the whole data use for training and cross-validation purpose; "train set" and "test set" refer to 75\% and 25\% of the "training sample," respectively.}, we applied several selection criteria. Starting with the total of 41\,710 HR15N spectra, we selected objects with S/N\,$>$\,40/pix (see Sect. \ref{sec:noise_rotation} below) and applied the following cuts for labels: $\mathrm{4\,000<\teff<7\,000\,}$K, $\mathrm{1.0<log(\textit{g})<5.0\,}$dex, $\mathrm{-2.0<[Fe/H]<0.5\,}$dex and $\mathrm{0<A(Li)<4.0\,}$dex. We further cleaned the training sample by applying uncertainty cuts of  e$\mathrm{\teff}<100\,$K, e$\mathrm{log(\textit{g})<0.3\,}$dex, $\mathrm{e[Fe/H]<0.2\,}$dex, and $\mathrm{eA(Li)<0.5\,}$dex. We rejected stars with Li upper limits. We also applied an uncertainty cut on the radial velocity $\mathrm{E\_VRAD<}$ 0.5\,km\,s$^{-1}$ (see Sect. \ref{sec:rad_v}). Spectra with GES flags for data reduction and analysis problems (TECH) and for peculiarities affecting the spectra (PECULI) were also rejected (see \citealt{Gilmore2022} for more details). During the training, some variable and high proper motion stars were identified with significant variability in flux seen in their multiple observations. As GES provides the same homogenized labels for these multiple observations, these objects were subsequently removed from the training. The training sample is then composed of 7\,031 spectra and respective labels. The remaining 33\,119 spectra, not included in the training sample, comprise the observed sample. We do not provide labels for 1560 spectra due to missing $\mathrm{\vrad}$ or very high $\mathrm{\vrad}$ values, shifting the spectrum out of the desired wavelength range after correction. 

Next, we applied radial velocity correction to the GES continuum-normalized spectra and removed the random cosmic features. Any pixel value exceeding median of the continuum by over five sigma is replaced by a median of the continuum. Negative pixel values are replaced by a median of the continuum$+$lines. The spectra were then re-sampled to a common wavelength coverage $\lambda \in$ [6\,450 - 6\,810] \AA,\ while keeping the original pixel separation of 0.05 \AA.

The HR15N sample consists of many young objects that have strong H$_{\alpha}$ emission lines. Since dealing with this is out of the scope of the current work, we masked the region of 16\,\AA\, around H$_{\alpha}$.
The only requirement for the observed sample was that the radial velocity should be present in the recommended radial velocity catalog provided with the Gaia-ESO survey iDR6. Spectra with S/N values as low as 2 are present in the observed sample. The implication of such a low S/N on the CNN predictions are discussed later (see Sect.~\ref{sec:generalize}). As GES provides repeated observations, some stars have multiple spectra available with varying S/N values. These repeated spectra are present in both training and observed samples and provide a good test for the consistency of the CNN.

\subsection{Pre-processing training and observed sample}

We used Scikit-learn \citep{scikitlearn} for pre-processing. Using the train\_test\_split function, we adopted 25\% of the total training sample data as test set (leading to 1\,758 spectra and associated labels). The test set is not directly used for training of the CNN model, but it is only used to monitor the performance of the trained models at the end of each epoch (see Sect.~\ref{sec:hyper}). The train set is then composed of 5\,273 spectra (75\% of the training sample). Train and test sets are uniformly distributed across the label range, as homogeneity is crucial to help the CNN generalizing instead of over- or underfitting. We refer to Sect.~\ref{sec:tSNE} for a further discussion on homogeneity. 

We normalized the stellar labels to values between 0 and 1, using the MinMax normalization function. Normalizing all the stellar labels within same value range helps train the CNN with easier and faster convergence to the loss function global minimum.

\subsection{The t-SNE method for homogeneity check and outlier detections} \label{sec:tSNE}

To check the homogeneity of our train and test sets, we apply the t-distributed stochastic neighbour embedding (t-SNE; \citealt{tSNE2008}), an unsupervised ML method. It works by assigning similar objects in the high-dimensional space with a higher probability distribution and, hence, modeling them closer together in the lower dimensional map, while dissimilar objects are mapped further apart. Overall, t-SNE has been widely used in astrophysical applications \citep{Matijevic_2017A&A, tSNE_Anders}. For example, \cite{tSNE_Anders} successfully applied t-SNE to their study of the stellar abundance space and identifying substructures as well as chemically peculiar stars.

We plotted the t-SNE maps (perplexity = 50)\footnote{Perplexity is a parameter that sets the number of effective nearest neighbours; a higher value is usually recommended for larger samples.} for the whole training data set (7\,031 spectra with $\sim$7\,000 pixels each) in Fig. \ref{fig:tsne_training_color}. The axes value themselves have no physical meaning, while the nearby points represent similar spectra. The right-most plot shows how well the train and test sets follow each other in the t-SNE. This is only possible if they are homogeneously distributed across the range of labels. The figure shows a few outliers identified by the t-SNE; we checked these spectra and found them to have low S/N and we see they are affected by bad cosmic ray removal. The island at ~$\mathrm{tSNE\_X=-25}$ and ~$\mathrm{tSNE\_Y=-45}$, consists of Spite plateau-like stars ~($\mathrm{\overline{[Fe/H]}=-1.62\,}$dex, ~$\mathrm{\overline{A(Li)}=2.23\,}$dex) in the globular cluster NGC\,6752, which represents the most metal-poor group in the training sample. The figure also shows how spectra and atmospheric parameters are correlated. This reveals that they are intrinsically linked by a high-complexity mapping, which the CNN will have to learn during its training.


\section{Convolutional neural network for stellar parametrization} \label{sec:CNN}

\subsection{Architecture of the CNN} \label{sec:cnn_arch}

We built our CNN model with the open source deep learning library Keras \citep{chollet2015keras}, using the TENSORFLOW backend \citep{tensorflow2015}. Keras provides a Python interface in a compact and easy manner to develop high-level artificial neural networks. Then, TENSORFLOW developed by the Google Brain Team, is an open-source software library for ML. We trained the CNN with the gradient-based Adam optimizer \citep{Kingma_ADAM_2014arXiv}.

\begin{figure}[h]
        \centering
        \includegraphics[width=\linewidth]{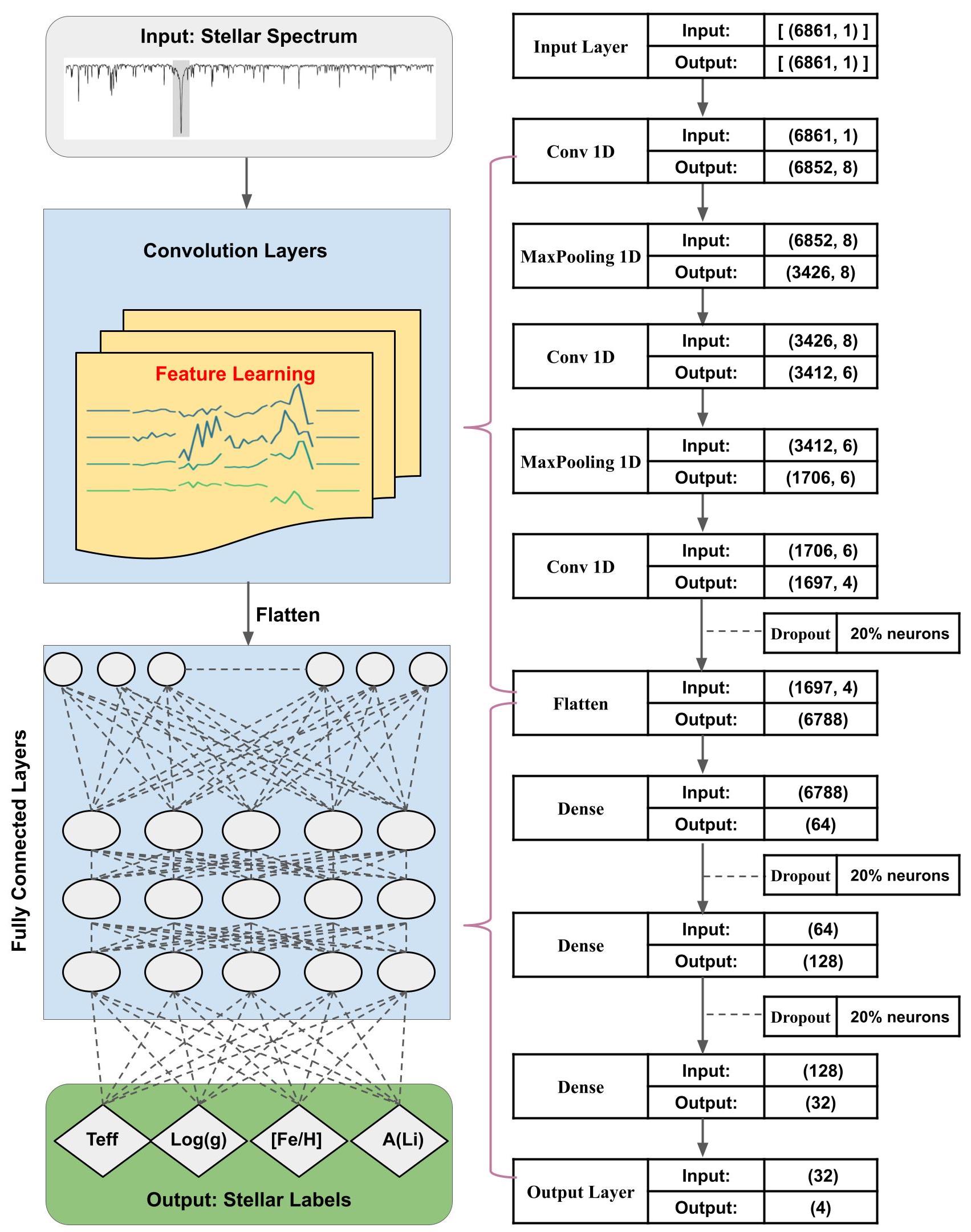}
        \caption{Architecture of the CNN adopted for this study is shown as a block diagram on the left and its detailed structure with layers is shown on the right panel. The model can be divided into four distinct sections: input layer, convolution layers, fully connected layers, and output layer, with a total of 448\,134 trainable parameters. The numbers, for example, (6\,861, 1) and (6\,852, 8), represent the shape of input and output of first Conv1D layer.}
        \label{fig:cnn_model}
\end{figure}

In deep learning methods, the final choice of the architecture is usually an outcome of a lot of experimentation with various setups and tuning of hyperparameters. The architecture of the CNN makes a significant impact on the training and prediction performances. The implementation of various architectures for stellar spectra parametrization can be found in the literature, we refer to the work referenced in Sect. \ref{sec:intro} for further details. For this project, we built on the work of \citet{guiglion2020} and optimized the architecture.

Figure \ref{fig:cnn_model} shows the architecture of our CNN. The pre-processed spectrum is provided as input and as output the CNN predicts $\mathrm{\teff}$, log(\textit{g}), [Fe/H] and A(Li). The model has three convolution layers and four (3 + 1) dense layers, including the output layer (discussed in Appendix~\ref{sec:cnn_technical}). Studies such as \citet{Leung_2019MNRAS, Fabbro_2018MNRAS} have also adopted a similar architecture as a good trade-off between desired precision and computation time.

Further details on the CNN architecture, the choice of hyperparameters, and model generalization (avoiding over- or underfitting) of the CNN can be found in Appendix~\ref{sec:cnn_technical}.

\subsection{Training the CNN} \label{sec:train}

Our CNN model architecture, as illustrated in Fig. \ref{fig:cnn_model}, has a total of 448\,134 trainable parameters. These parameters include all the weights and biases for the different layers present in the model. The training process optimizes the values for the parameters by minimizing the value of a loss function and judges the performance of the training by calculating a metric on the test data. We use the mean squared error (MSE) as the loss function as well as the metric. The EarlyStopping callback, defined in Sect. \ref{sec:hyper}, monitors the metric and the best model weights are saved. We trained an ensemble of 30 models\footnote{The training of the models required a time period of 16 to 26 minutes using only normal CPU on the \href{https://colab.aip.de}{COLAB cloud service at AIP} for compute and storage.}\footnote{We adopted 30 models for the Ensemble method as a good trade-off between the reliable statistics and computational load.}, where for each model, weights were randomly initialized. The training for the models stopped at different epochs due to the stochastic nature of the learning algorithm.

\begin{figure}[h!]
        \centering
        \includegraphics[width=\linewidth]{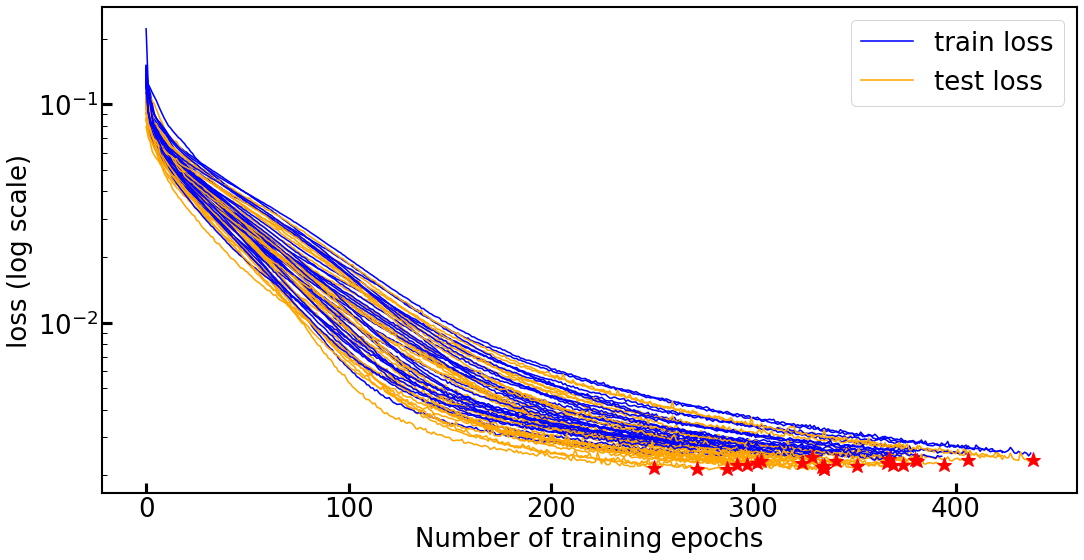}
        \caption{Value of the loss functions for the train (blue) and test (orange) sets for the 30 CNN runs as a function of the epoch. The red stars identify the selected 24 models.}
        \label{fig:loss_acc}
\end{figure}

In Fig. \ref{fig:loss_acc}, we show the progress of the training by plotting the evolution of the loss functions of the training (blue) and test (orange) sets for the 30 models. The loss curves show that the training was smooth and provides a good fit as the training and test loss decreases to a point of stability, with a small gap between the two final loss values.

The models with higher test loss than the 80\textsuperscript{th} percentile value are discarded, and the predictions from the selected 24 models are averaged as the final result. The dispersion is provided as the label uncertainties. (See Sect. \ref{sec:e_trend} for more on uncertainties.)

\begin{figure*}[h!]
        \centering
        \includegraphics[width=\linewidth]{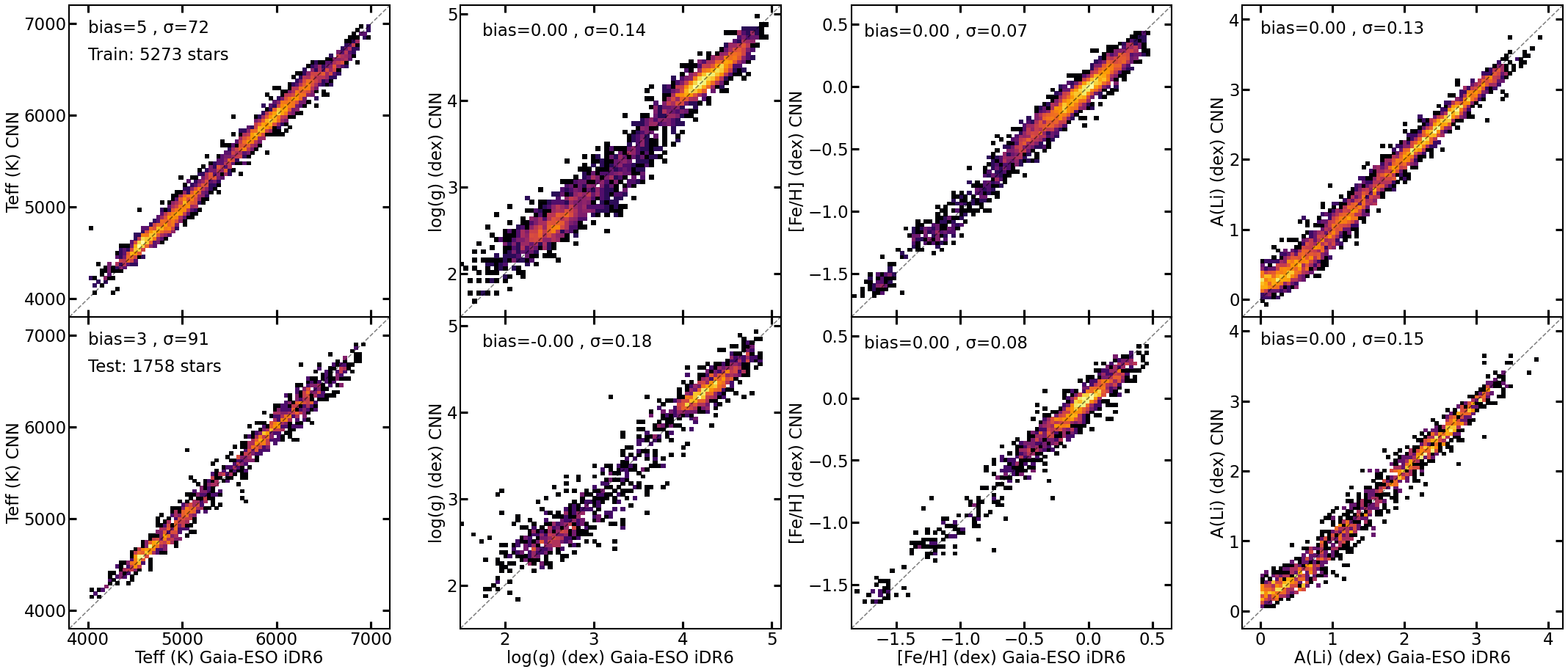}
        \caption{2D histograms showing 1-to-1 comparison between the GES-iDR6 labels (CNN input, x-axis) and CNN predictions (y-axis) for the train (top row) and test (bottom row) sets. The bias=mean(CNN-iDR6) and $\sigma$=std(CNN-iDR6) are also calculated.}
        \label{fig:train_test_onetoone}
\end{figure*}

\subsubsection{Result of the training}

In Fig. \ref{fig:train_test_onetoone}, we show a comparison of the input GES-iDR6 labels to the CNN prediction for the train and test sets. The figure shows a well-behaved 1-to-1 relation with no apparent systematic trends. The bias and scatter values represent the mean and the standard deviation of the residuals. The results show no bias (negligible for $\mathrm{\teff}$). The scatter is comparable for the train and test sets, with slightly higher scatter for scarcely populated label regions such as log(\textit{g}) $<$ 2.0\,dex and [Fe/H] $<$ -0.5\,dex. Overall, the test set follows the train set, showing that the trained models do not over-fit. Even though the wavelength range in the GIRAFFE HR15N setup is not optimal for determination of atmospheric parameters \citep{Lanzafame2015}, and despite masking the H$_{\alpha}$ line, which is an important spectral feature for the estimation of $\mathrm{\teff}$ and log(\textit{g}), the CNN shows very good performances. This indicates that the trained CNN models have learned significantly from the available spectral features.

\begin{figure}[h]
        \centering
        \includegraphics[width=\linewidth]{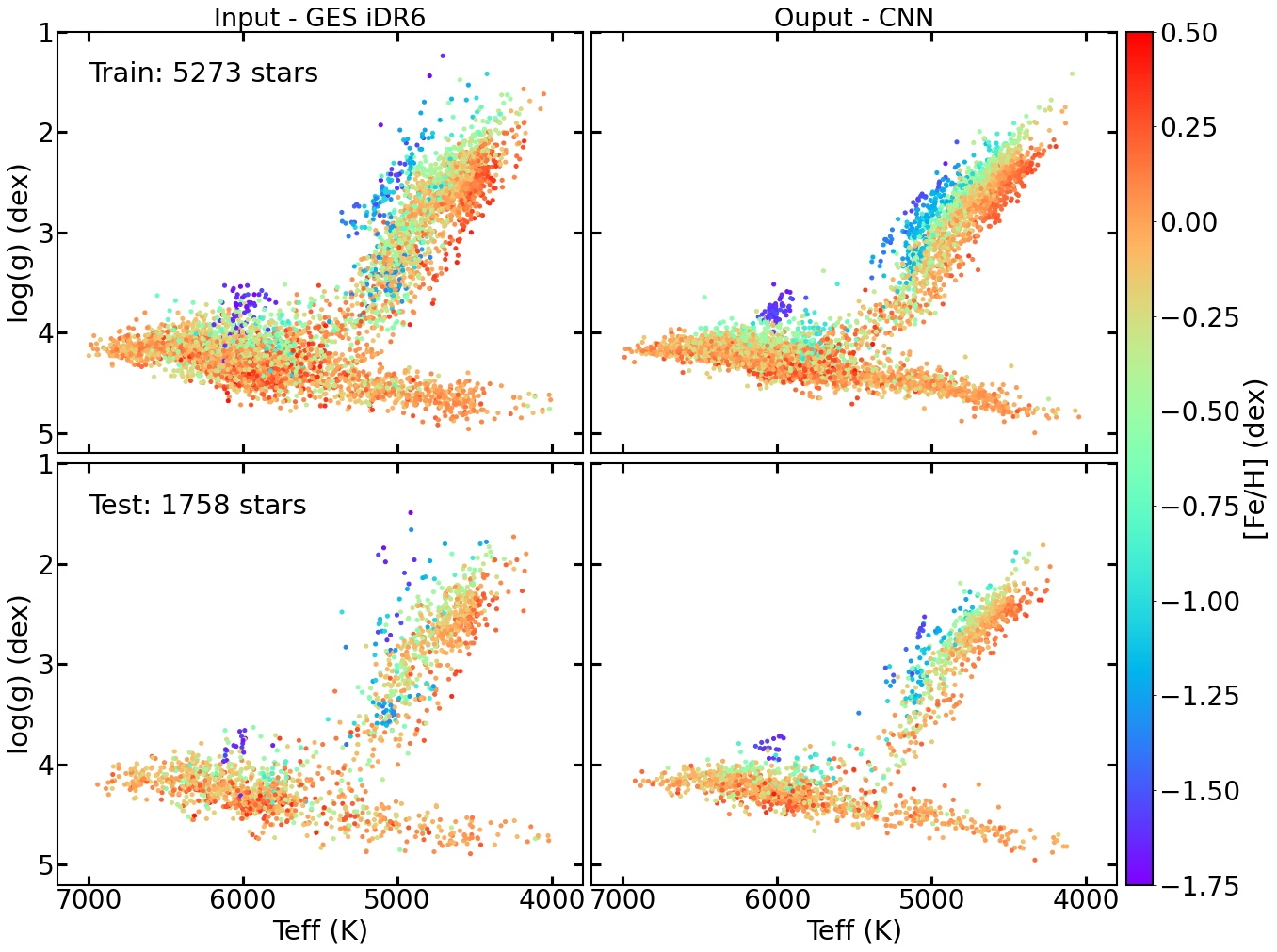}
        \caption{Kiel Diagrams for the input and CNN output colored by [Fe/H]: top two panels show the train set stars using iDR6 input labels on the left and CNN output on the right. Bottom two panels show the same for the test set.}
        \label{fig:hr_in_out}
\end{figure}

In Fig. \ref{fig:hr_in_out}, we present Kiel diagrams ($\mathrm{\teff}$ v.s. log(\textit{g})) for the train (top panels) and test (bottom panels) sets. The left columns show the input iDR6 labels and the right columns show the labels as predicted by the CNN. We see that the main features of the Kiel diagram are well recovered. The dwarfs and giants are clearly separated with a smooth transition from main-sequence turn-off to the subgiants and the metallicity gradient in the giant branch is very well described for both the train and test sets. The dwarfs, which span a large $\mathrm{\teff}$ range from 7000\,K to 4000\,K, are adequately parametrized even for the very hot and the very cool regime. The metal-poor giants, around 5\,000\,K, show much less scatter for the CNN output compared to the GES-iDR6. Two distinct issues can explain this difference: \textbf{1.} This region is very sparsely populated in the training data, so the one way to improve CNN prediction would be to add more training data in this region. \textbf{2.} No benchmark stars are present in this region, namely, there are no metal-poor giants (see Sect. \ref{sec:gbs} for details). Similar lower scatter, at the metal-poor end for giants when predicted by the ML methods have been reported by \citet{Ness_Cannon_2015ApJ} (see Fig 12 and \citet{Ting_Payne_2019ApJ}, see Fig 7); both studies compared their results with isochrones to find their ML results at this region in better agreement with stellar isochrones compared to the surveys, suggesting discrepancies due to calibration issues.

\begin{figure}[h]
        \centering
        \includegraphics[width=\linewidth]{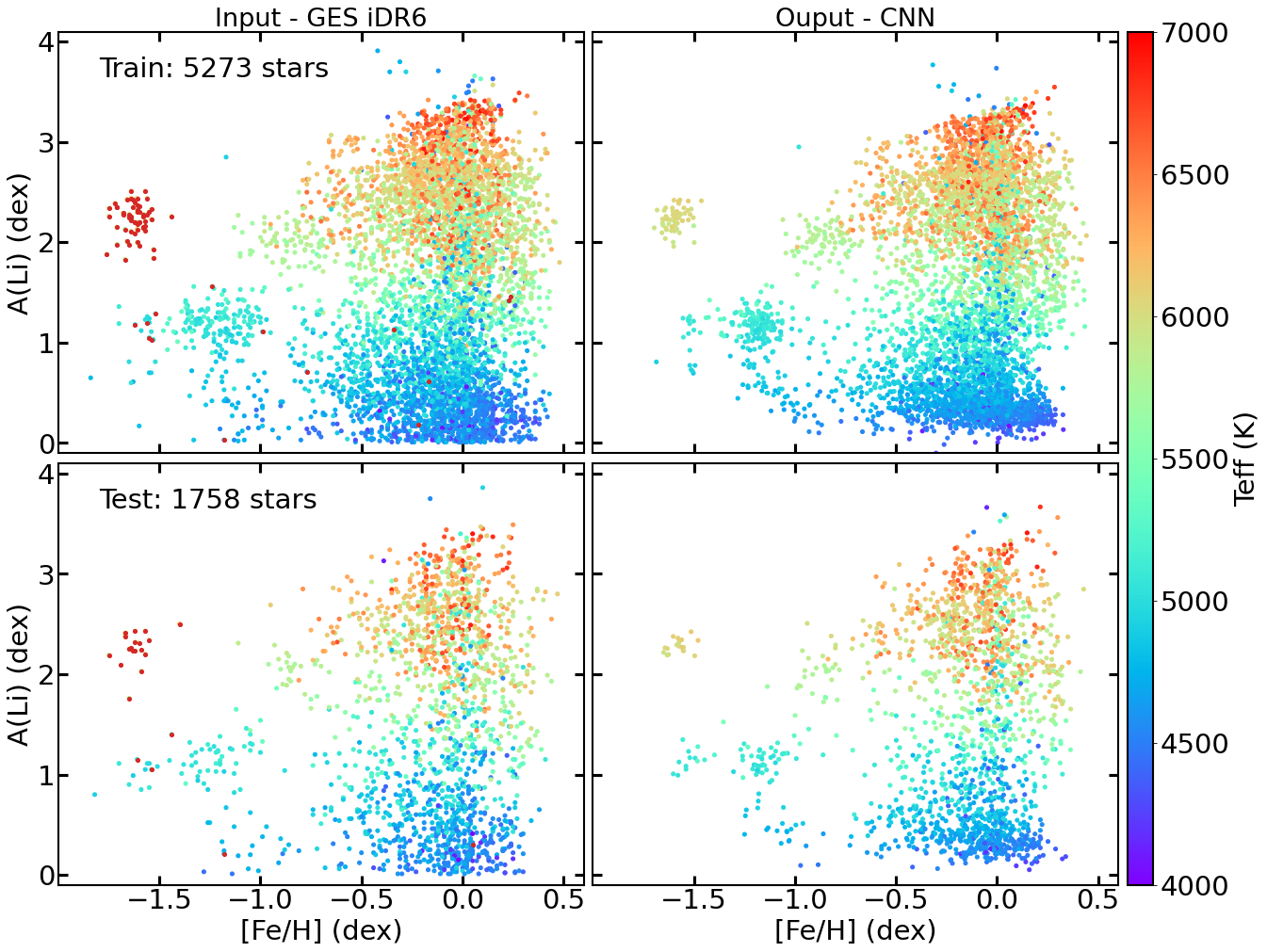}
        \caption{[Fe/H] vs A(Li) for the iDR6 input and CNN output colored by $\mathrm{\teff}$: top two panels show the train set stars using iDR6 input labels on the left and CNN output on the right. Bottom two panels show the same for the test set.}
        \label{fig:feh_li_in_out}
\end{figure}

In Fig. \ref{fig:feh_li_in_out}, we present the lithium abundance trends, colored by $\mathrm{\teff}$, for both train and test sets. The main features are also very well recovered. The most metal-poor globular cluster NGC\,6752 with [Fe/H]$<$-1.5\,dex and A(Li)$\sim$2.2\,dex is well located for both train and test sets. We also find good agreements for globular clusters such as NGC\,1281 and NGC\,2808, seen around -1.5$<$[Fe/H]$<$-1.0\,dex and A(Li)$\sim$1.2\,dex. The $\mathrm{\teff}$ dependence for Li, with higher Li abundance for hotter stars and lower Li abundance for cooler stars, is also seen. The highest Li abundances, at the metal-rich regime, seen for the hottest stars and the coolest PMS stars, are also recovered for both train and test sets. It is consistent, for instance, with \citet{Romano_2021A&A}, who use GES iDR6 to infer the highest, undepleted Li abundances for both field (hot stars) and cluster (hot MS and cool PMS) stars.

\begin{figure*}[ht!]
        \centering
        \includegraphics[width=\linewidth]{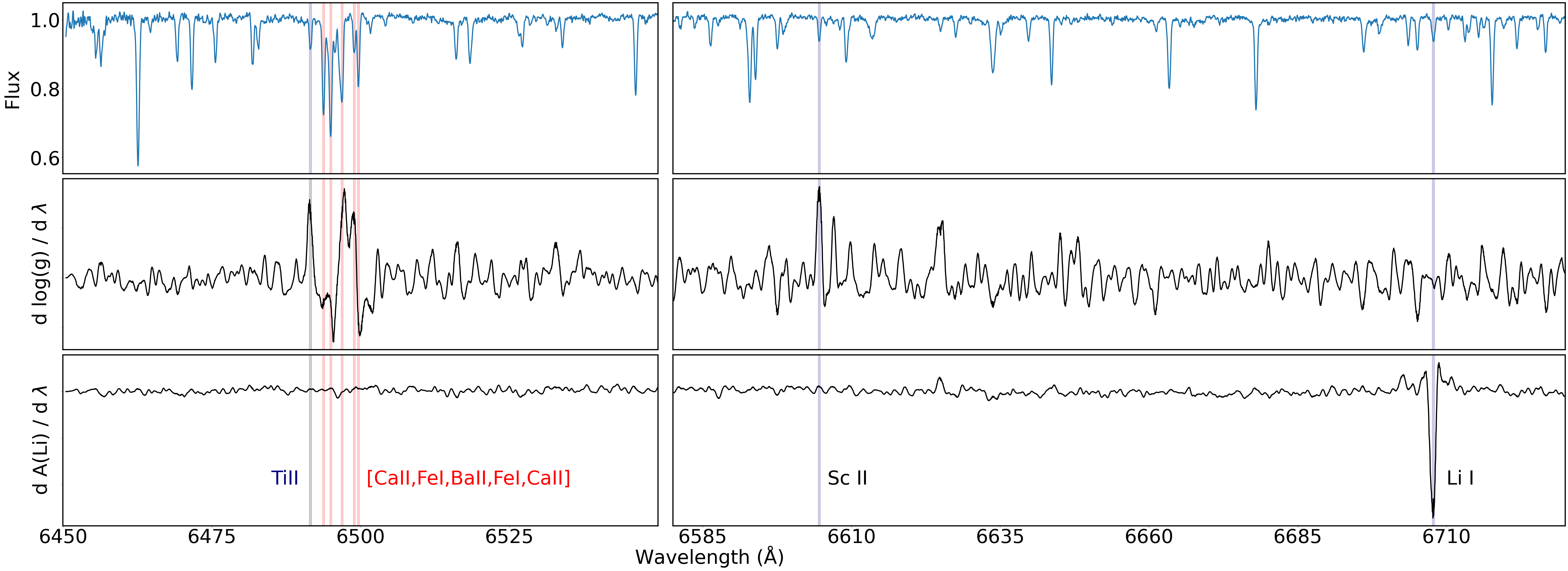}
        \caption{Gradients of the output labels with respect to input pixels for the solar twins in the training sample. Selected as $\mathrm{\teff\ = 5\,777\pm25\,}$K, 
        log(\textit{g}) = 4.44$\pm$0.10\,dex and [Fe/H] = 0.0$\pm$0.05\,dex, there are 13 stars. The top row shows the mean input spectrum and the second and third row represent the gradient/response for log(\textit{g}) and A(Li), respectively. Left column shows wavelength region [6\,450 - 6\,550]~\AA\, and right shows [6\,580 - 6\,730]~\AA\, as we mask the H$_{\alpha}$ region. Various spectral features that are discussed in the text are labeled.}
        \label{fig:grad}
\end{figure*}

\subsubsection{Examining if the CNN can learn from spectral features} \label{sec:grad}

Treating our neural network as a mathematical function that maps input spectra to output labels, it is desirable to check how each part of the input spectrum influences the output labels. In other words, if we can calculate the sensitivity of output labels to each of the input fluxes, we can understand whether the CNN is learning from the spectral features. Calculating gradients is one such method for generating a sensitivity map for a spectrum by performing partial derivatives of each of $\textit{T}_{\text{eff}}$, log(g), [Fe/H], and A(Li) with respect to every input neuron (or wavelength), namely, ${{\partial Label}/{\partial \lambda}}$. The gradient-based optimizing algorithm Adam \citep{Kingma_ADAM_2014arXiv} calculates a negative gradient of the weight matrix at each iteration to reduce the loss function, hence, calculating gradients are inherent to neural networks. The gradient of an output label is a kind of back-propagation of the model through the CNN and is obtained by using the simple chain rule of derivative moving backward from output to the hidden layers and finally to the input layer. This is achieved via a set of techniques called automatic differentiation\footnote{For further details on automatic differentiation and gradients, see \url{https://www.tensorflow.org/guide/autodiff}}, which makes it possible to evaluate the derivative of the function represented by the CNN. We used the GradientTape function from Tensorflow to calculate the gradients.

In Fig. \ref{fig:grad}, we show as an example, the gradients of log(\textit{g}) and A(Li) for the 13 solar twins in our training sample. We make following representative observations: First, the gradient of the lithium label with respect to $\lambda$ is only active at the lithium line and almost flat elsewhere. This shows the ability of our CNN to discard all other wavelengths and learn from this singular feature. The CNN then properly measures lithium abundances, instead of simply inferring them from correlations among the labels. Second, \citet{Damiani2014} showed that the quintet feature, between 6\,490-6\,500\,\AA\ consisting of blended FeI, CaI, BaII, and TiI lines, is highly sensitive to gravity. The TiII 6\,491.56\,\AA\, line, on the bluer side of the quintet, was also considered as an important line for their spectral indices. Here, the CNN gradients ${{\partial\, log(\textit{g})}/{\partial\, \lambda}}$ show that these wavelength regions are indeed very sensitive to log(\textit{g}). Finally, \citet{Jofre2015_alpha} listed the ionized Scandium, ScII, line at 6\,604.6\,\AA\ as a Golden Line for FGK dwarfs and giants but not for metal-poor stars and M giants. Our log(\textit{g}) gradients also show very high response at this wavelength region.

Such diagnostic checks confirmed that CNN properly learns from spectral features and these gradients could allow for the identification of new sensitive spectral features that are presently not used by standard classical pipelines. Then, the classical pipelines and the CNN could be used in a sort of feedback manner to improve their mutual output.

\subsubsection{Sensitivity to the radial velocity} \label{sec:rad_v}

\begin{figure}[h!]
        \centering
        \includegraphics[width=0.9\linewidth]{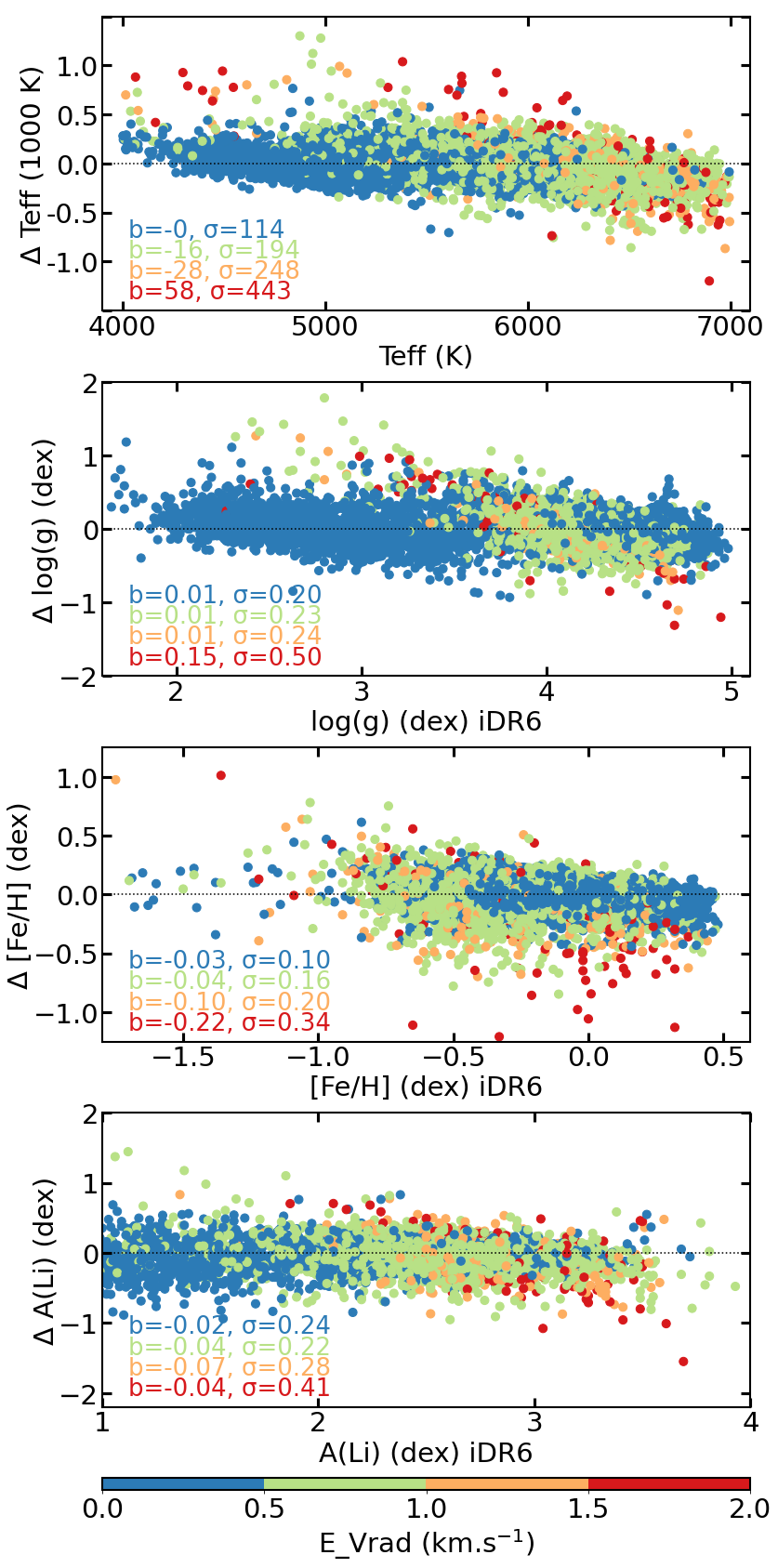}
        \caption{Residuals (CNN-iDR6) as a function of labels for the selected observed sample. The stars are color-coded into four bins based on their reported uncertainties in radial velocities. For each label, the bias=mean(CNN-iDR6) and $\sigma$=std(CNN-iDR6) in the four VRAD uncertainty bins are listed.}
        \label{fig:rv_dependence}
\end{figure}

Accurate and precise radial velocities are crucial for obtaining a reliable estimate of the atmospheric parameters and chemical abundances, as it matches the observed spectrum to the line-list which is the ground truth for any EW or spectral fitting methods. The radial velocities (and associated uncertainties) of the GIRAFFE HR15N spectra were estimated by GES, by spectral fitting of the observations to model spectra \citep{Gilmore2022}. The radial velocity is measured using the HR15N spectra, but an offset is applied to it during the homogenization process to bring radial velocities measured from different setups to the same scale. The offsets are measured considering HR10 (5\,340\,\AA\,-\,5\,620\,\AA) setup as a zero-point of the radial velocity scale; GES made sure that HR10 radial velocities are in good agreement with Gaia radial velocity standards. However, such a combination of different setups can be a source of small systematics. While GES reports the highest Vrad precision achieved to be on the order of 0.25\,km\,s$^{-1}$ (see \citealt{Gilmore2022}), over 80\% of the HR15N sample have Vrad errors larger than 0.25\,km\,s$^{-1}$ and with a third of the sample above 0.55\,km\,s$^{-1}$.

Figure \ref{fig:rv_dependence} shows the residual (CNN-iDR6) plots for the selected observed sample, colored in bins of GES radial velocity uncertainties. We clearly see that the dispersion increases with increasing $\mathrm{\vrad}$ uncertainties and a large bias is visible for stars with large E\_VRAD, for instance, as shown by the red dots. Due to such results, we apply a cut at E\_VRAD $<$0.5\,km\,s$^{-1}$ in our training sample. \citet{Jackson2015} report that $\mathrm{\vrad}$ precision for GIRAFFE spectra worsens for $\mathrm{\teff} > $ 5\,200\,K, as a result of paucity of strong narrow lines in hotter stars. We also observe that E\_VRAD $>$ 0.5\,km\,s$^{-1}$ are mostly for stars hotter than $5\,500$\,K in iDR6. The HR10 re-calibration is a function of $\mathrm{\teff}$, log(\textit{g}), and [Fe/H], and this could create tiny $\mathrm{\vrad}$ corrections that the CNN is able to detect. We avoid a deeper investigation as it is outside the scope of this paper.

However, we showed that ML pipelines can be very sensitive to small wavelength shifts in the input data. For upcoming surveys such as 4MOST and WEAVE, which will observe in multiple setups, a precise radial velocity estimation will be more important as ML techniques will be extensively used due to the larger volume of observations. Also, another source of $\mathrm{\vrad}$ errors for GES could be the fact that the different wavelength ranges were calibrated independently \citep{randich2022}.
The expected accuracy of 4MIDABLE-HR radial velocities is expected to be $<$1.0\,km\,s$^{-1}$\citep{deJong2019Msngr}. Further tests on real 4MOST spectra will be necessary in order to estimate the CNN sensitivity to Vrad.

\subsubsection{Inferring lithium abundances without a lithium line} \label{sec:no_li_line}

\begin{figure}[h!]
        \centering
        \includegraphics[width=.70\linewidth]{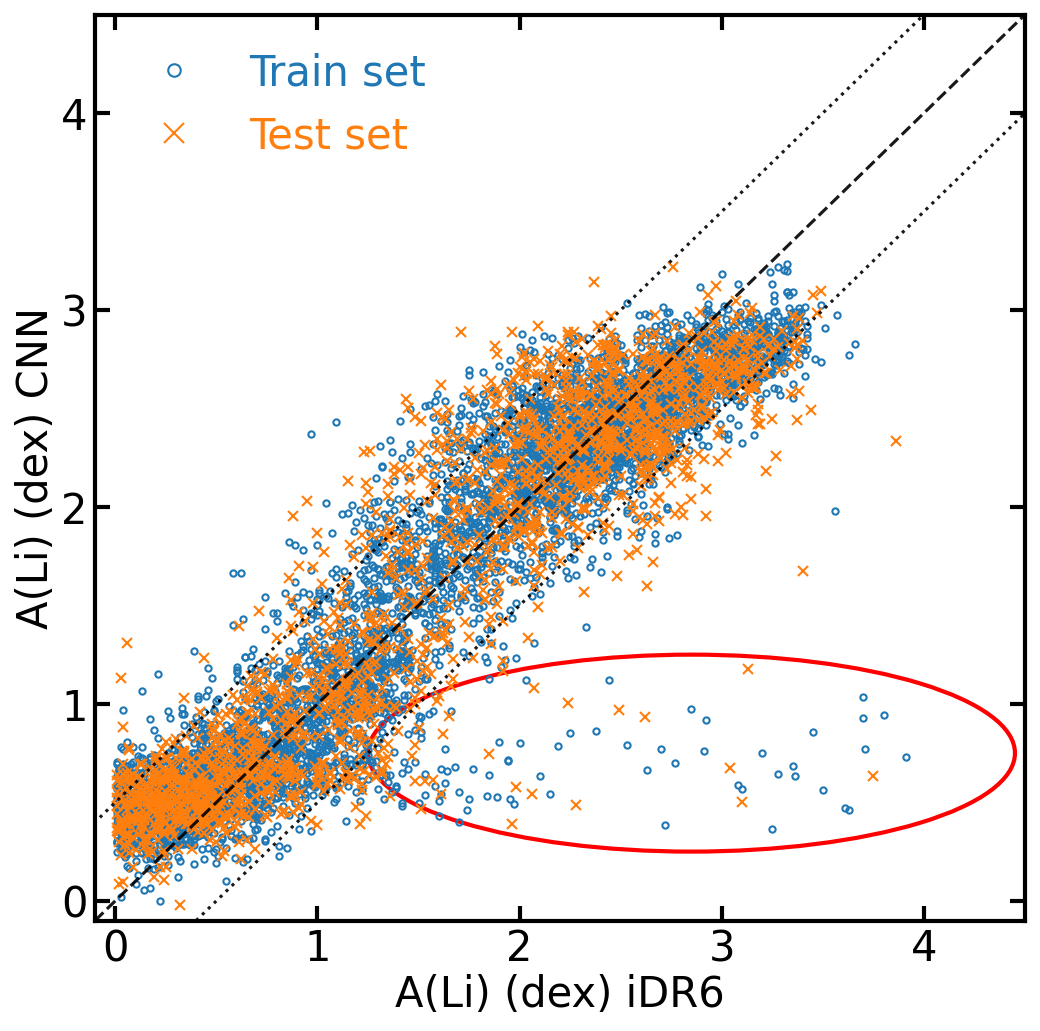}
        \caption{CNN vs GES-iDR6 A(Li) for the CNN trained using spectra masked at 6\,707.8 \AA\, Li line. Blue and orange represent train and test sets respectively. The dashed line is the 1-to-1 line, and two dotted lines are at $\pm$\,0.5\,dex. The red ellipse shows the incorrectly inferred Li-rich giants.}
        \label{fig:no_li_line}
\end{figure}

ML algorithms are efficient at learning astrophysical correlations, for example, inferring oxygen abundances from spectra with no oxygen feature \citep{Ting_2017ApJ_no_Oxygen, Ting_2018ApJ_no_Oxygen}. Lithium abundance is highly correlated to the $\mathrm{\teff}$, and depends a lot on the surface gravity (see e.g., Fig. \ref{fig:tsne_training_color}). To test whether it is possible to infer lithium based on pure astrophysical correlations, we trained a CNN with the same GIRAFFE training sample, but masking the 6\,707.8 \AA~lithium line. In Fig. \ref{fig:no_li_line}, we compare the CNN Li abundance with GES-iDR6 Li abundance, finding very poor performance compared to Fig. \ref{fig:train_test_onetoone}, with a large scatter of over 0.5 dex throughout the label range for both the train and the test sets. Here, we note that the A(Li) output by CNN comes purely from the correlations among labels and it is not a measurement from the spectral feature. Hence, we see an underprediction at higher values and an overprediction at lower values, also known as regression dilution. The Li-rich giants (see Sect. \ref{sec:li_g}) are completely missed when inferring lithium solely from astrophysical correlations. We visually inspected the Li sensitivity map, as we did in Sect. 3.2.2, and most of the HR15N features are used to infer Li. Then, Li must be then measured from Li spectral feature instead of being inferred based on correlations.

\section{Catalog of stellar parameters \& Li abundances} \label{sec:result}
\subsection{CNN parametrization of the GES GIRAFFE spectra}

We used CNN models to predict the atmospheric parameters and lithium abundances for the observed sample spectra. Prediction using a trained model is very fast and takes only $\sim$20 seconds for the four labels, $\mathrm{\teff}$, log(\textit{g}), [Fe/H], and A(Li), for all 33\,119 observed sample spectra. The prediction for the selected 24 models takes only about nine minutes. An average of the 24 predictions is computed as the final result and the dispersion as an uncertainty.

\begin{figure*}[h!]
        \centering
        \includegraphics[width=\linewidth]{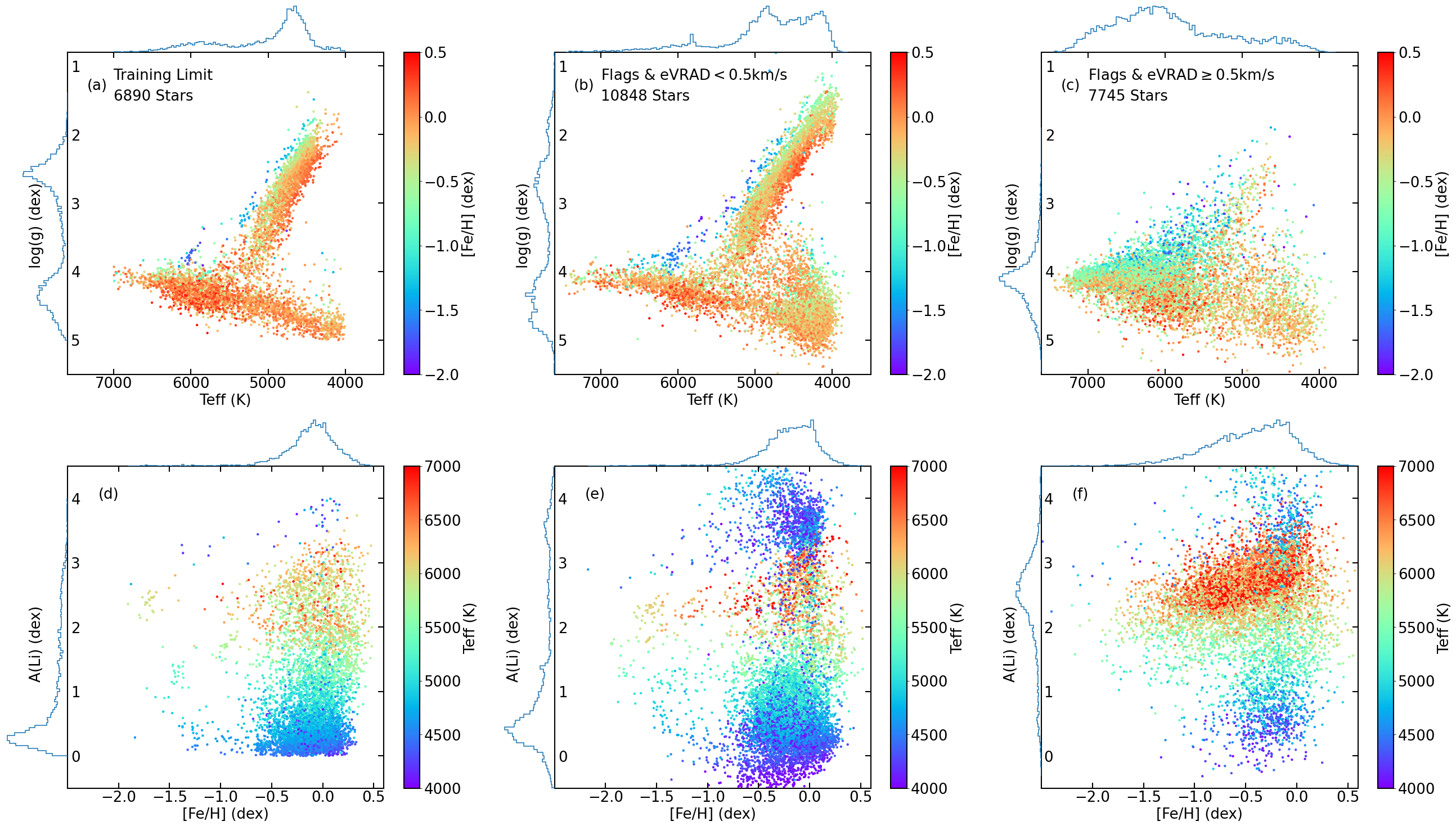}
        \caption{ Results for the observed sample. Top row: Kiel diagram for the observed sample stars with S/N$>$10/dex and labels within training limits color-coded with [Fe/H]. \textbf{(b)} Same plot as (a) but for stars with S/N$>$10/pix, GES-iDR6 flags and E\_VRAD $<$ 0.5\,km\,s$^{-1}$. \textbf{(c)} Same selection as (b) but for E\_VRAD $\geq$0.5\,km\,s$^{-1}$. Each subplot shows a histogram of the labels on the left and top axis. Bottom row: A(Li) vs [Fe/H] color-coded with $\mathrm{\teff}$ for the same stars as the Kiel diagram on top.}
        \label{fig:result_obs}
\end{figure*}

\begin{figure*}[h!]
        \centering
        \includegraphics[width=\linewidth]{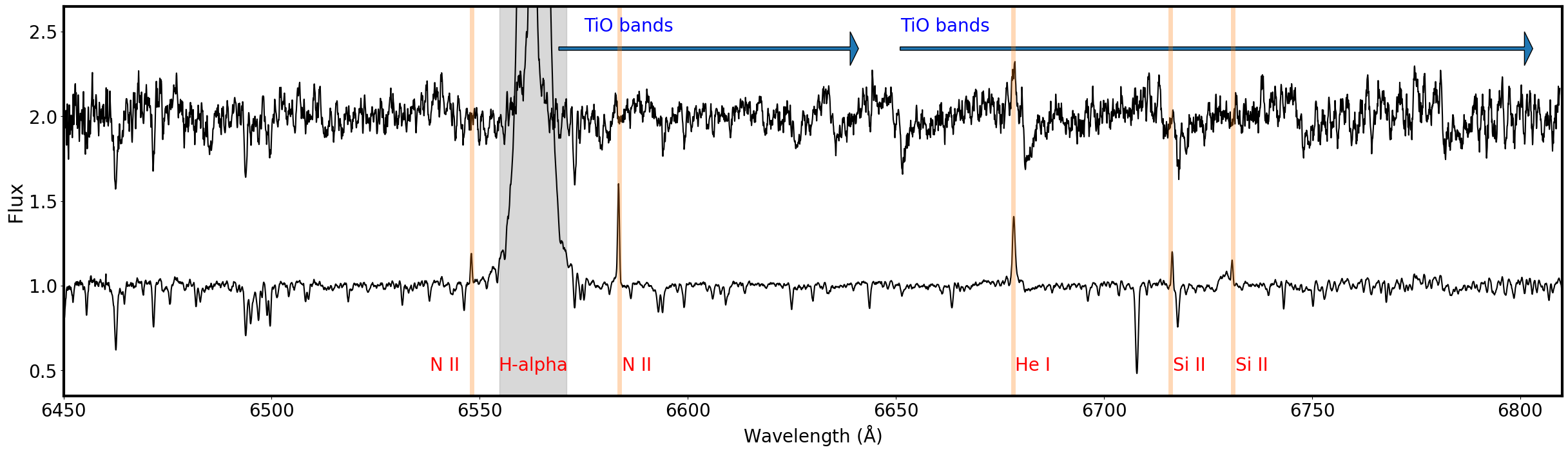}
        \caption{HR15N spectra with emission lines highlighted in yellow. From left to right: Lines: 6\,548~\AA\, NII, 6\,563~\AA\, H$_{\alpha}$, 6\,583~\AA\, NII, 6\,678~\AA\, HeI, 6\,716~\AA\, SiII, and 6\,731~\AA\, SiII. For the upper spectrum, the region for the strong molecular bands of TiO starting at 6\,569~\AA\, and 6\,651~\AA\, are seen. The relative flux values for top spectrum are increased by a unit for the ease of plotting.}
        \label{fig:emission_example}
\end{figure*}

For the stars within the training set limits, a typical Kiel diagram is seen, similar to Fig. \ref{fig:result_obs} (a), with clear distinction between the main sequence and the giants, along with the metallicity gradient for the giants as well as the turn-off stars. At the cool end, we see few stars with~$\logg\sim$\,4.0: we checked the spectra for these stars and found the presence of emission lines. An example of a HR15N spectrum with emission lines and molecular bands is shown in Fig. \ref{fig:emission_example}. For the second column Kiel diagram in Fig. \ref{fig:result_obs}, we see similar trends as in the case of training limits, except there is a cool dwarf clump. The group consists of very young clusters members, with emission lines and TiO molecular bands (M dwarfs). As there were no cool M dwarfs ($\mathrm{\teff} <$ 3\,500\,K) in the training set, some systematics may be present in the parametrization of these stars. However, GES is still refining the flags, thus further exploration of the particular flags is out of the scope of this project. In the third column of the Kiel diagram, the observed sample with radial velocity uncertainties $>$0.5\,km\,s$^{-1}$ are presented. Most of these stars lie in the warm dwarf region, as uncertainties in VRAD increase with $\mathrm{\teff}$ (as discussed in Sect. \ref{sec:rad_v}). The metallicity gradient is also seen for these warm dwarf stars.

In Fig. \ref{fig:result_obs} (d-f), we also present lithium abundance trends with respect to [Fe/H]. We see that most of the stars in the panels (d) and (e) are cool Li-poor stars, with a peak at solar [Fe/H]. For the observed sample stars in the training set limits, we see a clear trend with $\mathrm{\teff}$, with only a few cool stars with A(Li)$>$3.0\,dex. In plot (e), an increase of cool stars with high lithium is seen. These are young cluster members, for which the Li depletion has not been completed. In plot (f) we see the stars with GES flags and E\_VRAD$>$0.5\,k\,s$^{-1}$. Most of these stars are hotter stars with $\mathrm{\teff} >$ 5\,500\,K (see Sect. \ref{sec:rad_v}). Some of these warm, lithium-rich stars are likely to represent the warm group of stars on the left side of lithium dip.

\begin{figure*}[h!]
        \centering
        \includegraphics[width=\linewidth]{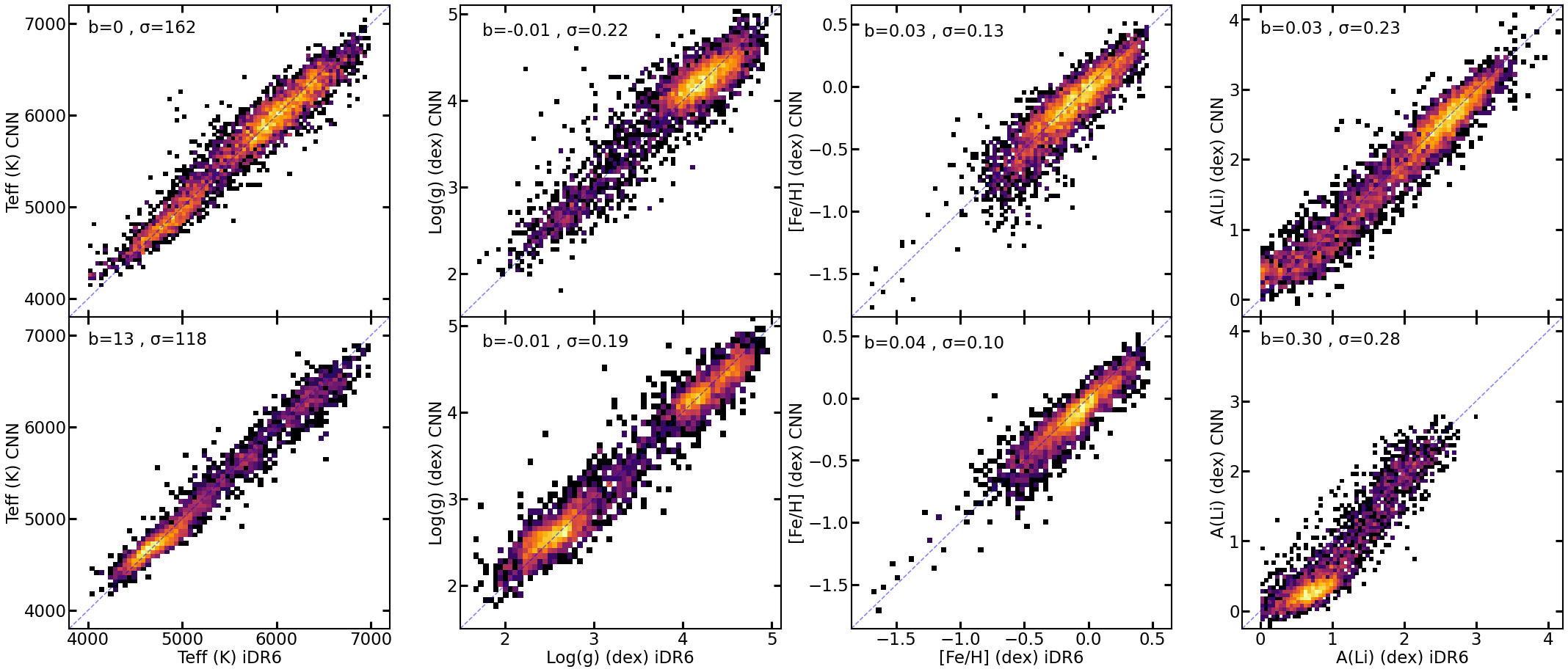}
        \caption{One-to-one comparison for observed sample stars with, S/N>20/pix, eVrad less than 1\,k\,s$^{-1}$, no PECULI or TECH flags and within the training label range. Here, bias=mean(CNN-iDR6) and $\sigma$=std(CNN-iDR6). Top row: Stars with Li measurements. Bottom row: Stars with Li upper limit. Most of the stars in the observed sample with Li measurement have low S/N spectra, hence, the higher scatter for $\mathrm{\teff}$, $\logg$), and $\feh$.}
        \label{fig:1_to_1_obs}
\end{figure*}

In \figurename~\ref{fig:1_to_1_obs}, we present the comparison of CNN predicted labels with iDR6 labels for a selection of the observed sample with S/N$>$20/pix, E\_VRAD$<$1.0\,km\,s$^{-1}$ and no TECH and PECULI flags. In the first row, we show $4\,481$ observed sample stars with iDR6 Li abundance with the flag $\mathrm{UPPER\_COMBINED\_LI1 = 0}$. The second row shows comparison for $3\,099$ stars, with Li upper limits given by $\mathrm{UPPER\_COMBINED\_LI1 = 1}$. There is an upper limit provided by GES  on the Li abundance when the 6\,707.8\,\AA\ Li line is undetected (too low S/N or too low lithium). For stars with GES Li measurement, we see a very good one to one match with no bias. There is a scatter of $162\,$K for $\mathrm{\teff}$, $0.22\,$dex for log(\textit{g}), $0.13\,$dex for [Fe/H] and $0.23\,$dex for A(Li). For the stars with GES Li upper limit, a very good one to one match with iDR6 measurement is seen with a small bias of 13\,K for $\mathrm{\teff}$ and no bias for log(\textit{g}) and [Fe/H]. A larger bias and scatter for A(Li) is observed, but this is expected as the iDR6 values are upper limits, and we provide lithium measurement for these stars. The scatter for $\mathrm{\teff}$, log(\textit{g}), and [Fe/H]  is higher for the Li measurement stars as most of these spectra ($\sim80\%$) have S/N$<$40/pix, while the most of the Li upper limits have higher S/N; this is because stars with higher S/N and Li measurements, that is, those without a limit, are included in the training set. Also, most of the stars with an upper limit for lithium are giants that have already evolved past their Li depletion phase (defined in Sect. \ref{sec:gal_evo}).

Our catalog of atmospheric parameters ($\mathrm{\teff}$, $\logg$), $\feh$, and lithium abundances for $\sim 40\,000$ stars is summarized in \tablename~\ref{table:catalog}. Of course, the apt use of this catalog will depend on the scientific application, but we encourage the reader to use lithium abundances within the training set limits (flag\_li = 1), and Li uncertainties below 0.15 dex (S/N>20). Similarly, atmospheric parameters are reliable only within the training set limits (flag\_x = 1). In addition, we make the CNN code, spectra and labels available to the community online via GitHub\footnote{\href{https://github.com/SamirNepal/Li_CNN_2022}{https://github.com/SamirNepal/Li\_CNN\_2022}}.

\begin{table}
\caption{Atmospheric parameters, Li abundances, 
and boundary flags of the publicly available online catalog for ~$\sim\,40\,000$ stars.}
\resizebox{0.48\textwidth}{!}{
\centering
\begin{tabular}[c]{l l l l l}
\hline
\hline
Col  & Format & Units & Label  &  Description \\
\hline
1   & char  & -   & cname      & GES ID            \\
2   & char  & -   & spectra\_name  & Name of Spectrum            \\
3   & float & K    & teff       & Effective temp. ($\mathrm{\teff}$)   \\
4   & float & K    & eteff      & Uncertainty of $\mathrm{\teff}$  \\
5   & int   & -    & flag\_teff & Boundary flag for $\mathrm{\teff}$    \\
6   & float & \cms & logg       & Surface gravity              \\
7   & float & \cms & elogg      & Uncertainty of $\logg$             \\
8   & int   & -    & flag\_logg & Boundary flag for $\logg$    \\
9  & float & dex  & feh        & $\feh$ ratio                 \\
10  & float & dex  & efeh       & Uncertainty of $\feh$              \\
11  & int   & -    & flag\_feh  & Boundary flag for $\feh$     \\
12  & float & dex  & li         & Li abundance            \\
13  & float & dex  & eli        & Uncertainty of Li          \\
14  & int   & -    & flag\_li   & Boundary flag for Li \\
15  & int   & pixel$^{-1}$ & snr & Signal-to-noise ratio \\
\hline 
\hline
\end{tabular}}
\label{table:catalog}
\end{table}

\subsection{Effects of noise and rotation on CNN predictions} \label{sec:noise_rotation}

The CNN was trained with spectra with S/N\,$>$\,40/pix, as this provides a balance in the training sample size and good quality. Noise is an unavoidable aspect of observational data (see Sect. \ref{sec:generalize}). In poor S/N spectra, the spectral features can be affected by the noise and can lead to a poor training performance as the CNN starts to learn the unwanted correlations due to noise. We find the mean difference between GES input and CNN output is uniform for different S/N ranges and do not see any significant increase with decreasing S/N (for both the training and observed samples). We conclude that CNN does not show any significant bias as a function of S/N (See Appendix \ref{sec: snr_trend} and \figurename~\ref{fig:bias_disp_snr} for further details).

Another important aspect concerns the stellar rotational velocity. As the projected rotational velocity (\textit{v}sin\textit{i}) increases, the spectral lines get wider and shallower and there is an increase in the line blending (with conserved EW). Classical spectroscopic pipelines must take into account rotational broadening during analysis of a spectrum.

\begin{figure}[h]
        \centering
        \includegraphics[width=\linewidth]{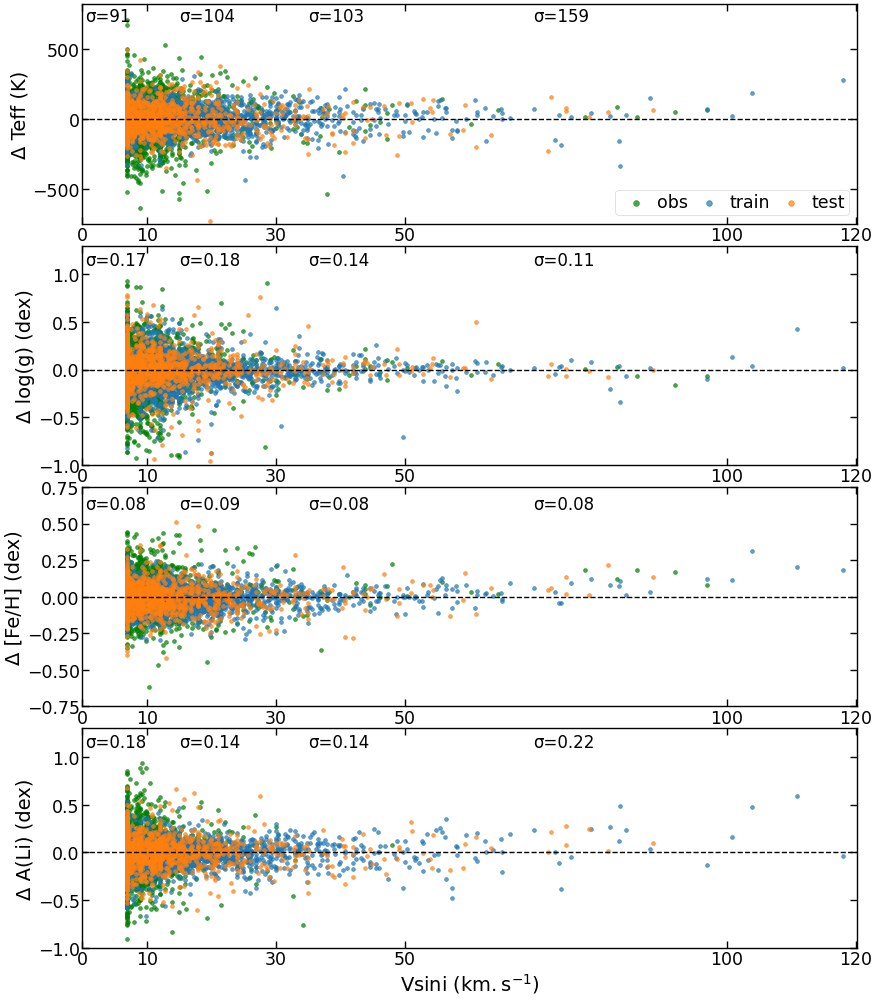}
        \caption{Residuals ($\Delta$label = GES - CNN) as a function of \textit{v}sin\textit{i} (km\,s$^{-1}$) for the train (blue), test (orange), and selected observed sample (green) stars. The observed sample is selected within training label limits, S/N>10/pix, ~E\_VRAD\,<\,0.5\,km\,s$^{-1}$, with no GES flags and with Li measurement. The mean scatter of the residuals ($\sigma$) in the \textit{v}sin\textit{i} bins ($\leq$10, (10,30], (30,50] \& $>$50) is also shown for each label.}
        \label{fig:vsini_residual}
\end{figure}

Our training sample of 7\,031 spectra has a distribution of rotational velocities (in k\,s$^{-1}$) as follows: ~[\textit{v}sin\textit{i}\,$\leq$\,10]=\,62\%, [10\,$<$\,\textit{v}sin\textit{i}\,$\leq$\,30]=\,34\%, [30\,$<$\,\textit{v}sin\textit{i}\,$\leq$\,50]= 3\%, and [\textit{v}sin\textit{i}\,$>$\,50]=\,1\%.
Assuming that stars with \textit{v}sin\textit{i} $>$ 10\,km\,s$^{-1}$ are fast-rotators, the training sample has a significant number of such spectra. In fact, the CNN can learn from spectral features about the rotational broadening effects, even if \textit{v}sin\textit{i} is not used as a stellar label. As shown in Fig. \ref{fig:vsini_residual}, for~\textit{v}sin\textit{i}\,$<$\,50\,km\,s$^{-1}$, there is no significant change in dispersion (between input and output labels) and we observe no visible trends with the increasing rotation, even for hot stars with $\mathrm{\teff}$ > 6\,000, indicating an excellent CNN performance. For very fast rotators at ~\textit{v}sin\textit{i}\,$>$\,50\,km\,s$^{-1}$, the line shapes are significantly altered; we see an increase in dispersion, to 159\,K and 0.22\,dex, for $\mathrm{\teff}$ and A(Li). Also for [Fe/H], for \textit{v}sin\textit{i}\,$>$\,70\,km\,s$^{-1}$, we see a trend of under-prediction by CNN. We conclude that CNN does not suffer from significant systematics due to rotational broadening, thus it allows us to accurately parametrize fast-rotating stars.

\begin{figure*}[h!]
        \centering
        \includegraphics[width=\linewidth]{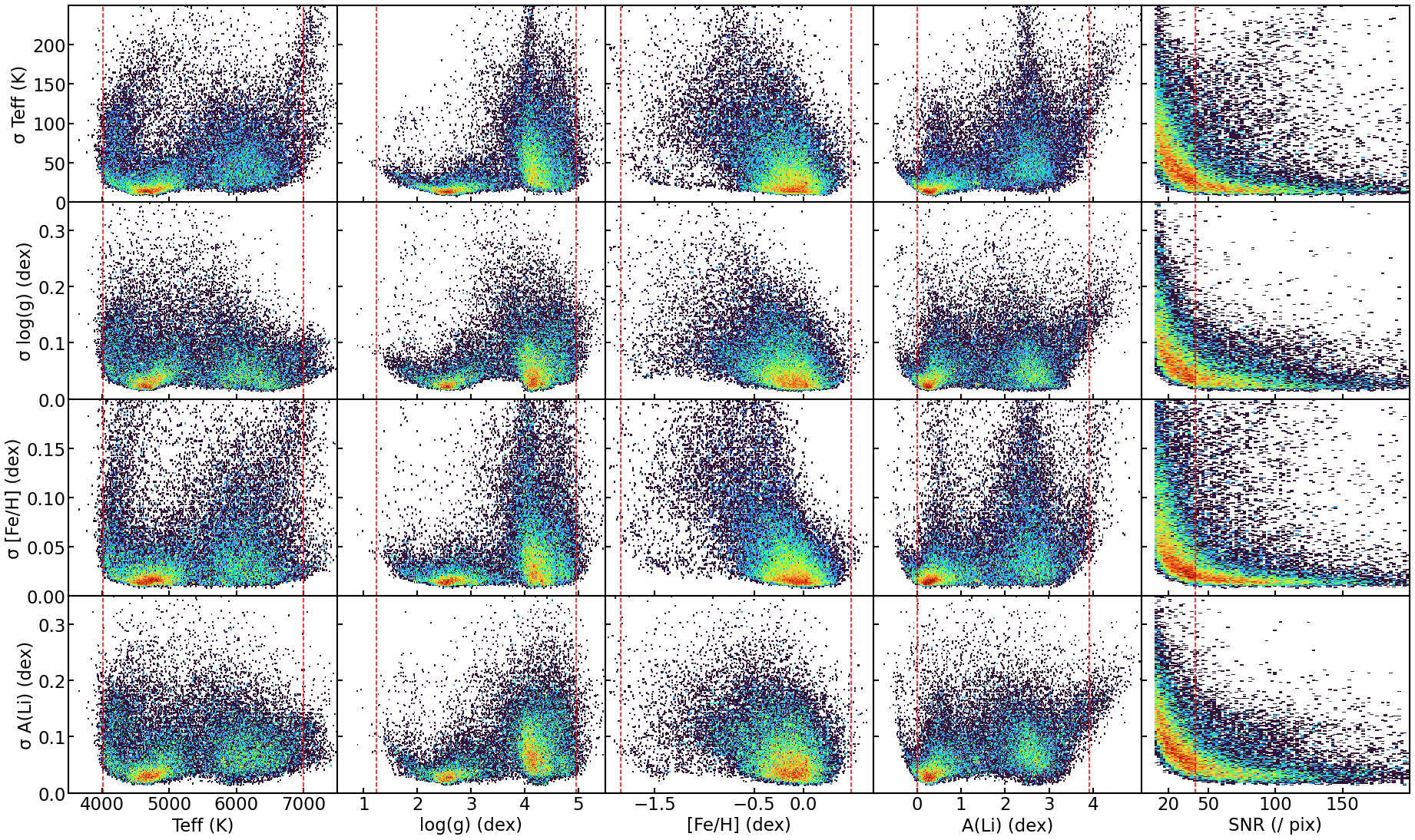}
        \caption[2D histograms showing CNN uncertainties (internal precision) as a function of the 4 labels and S/N for the observed sample with S/N$>$10/pix, i.e., 31\,272 spectra.]{2D histograms showing CNN uncertainties (internal precision) as a function of four labels ($\mathrm{\teff}$, log(\textit{g}), [Fe/H], A(Li)) and S/N for the observed sample with S/N$>$10/pix, i.e., 31\,272 spectra. The red dashed line shows the limits of the training labels. The x-axis represents the labels and the y-axis shows the uncertainty ($\sigma$).}
        \label{fig:error_dist_func}
\end{figure*}

\subsection{CNN internal uncertainty and estimation of precision and accuracy} \label{sec:e_trend}

The CNN internal uncertainties are calculated as the dispersion of the predictions from 24 selected models and is representative of the internal precision of the CNN. In Fig. \ref{fig:error_dist_func}, we present the uncertainty distributions for atmospheric parameters and Li abundance for the 31\,272 observed sample stars with S/N$>$10/pix. Overall, the uncertainties are low and similar to the training sample and reflect that our models provide stable results. We find larger uncertainties for lower S/N spectra and for stars with labels outside the training limits.

The train, test, and observed sets show similar uncertainties, if the observed sample is restrained to the training sample limits. The uncertainties are very low, with medians of about 19\,K for $\mathrm{\sigma \teff}$, 0.03\,dex for $\sigma$log(\textit{g}), 0.017\,dex for $\sigma$[Fe/H] and 0.035\,dex for $\sigma$A(Li) for the train, test, and observed sets (within the training sample limits). It comes from the fact that the training sample covers a higher S/N range and also includes spectra without any TECH or PECULI flags. The increased error for the whole observed sample is simply the irreducible uncertainty due to the sampling of the noise in the training set. We note that nearly 60\% of the observed sample have S/N below the training minimum of 40 per pixel. The train, test, and observed sets follow each other well, meaning that the CNN models are able to generalize properly.

\begin{figure*}[h!]
        \centering
        \includegraphics[width=\linewidth]{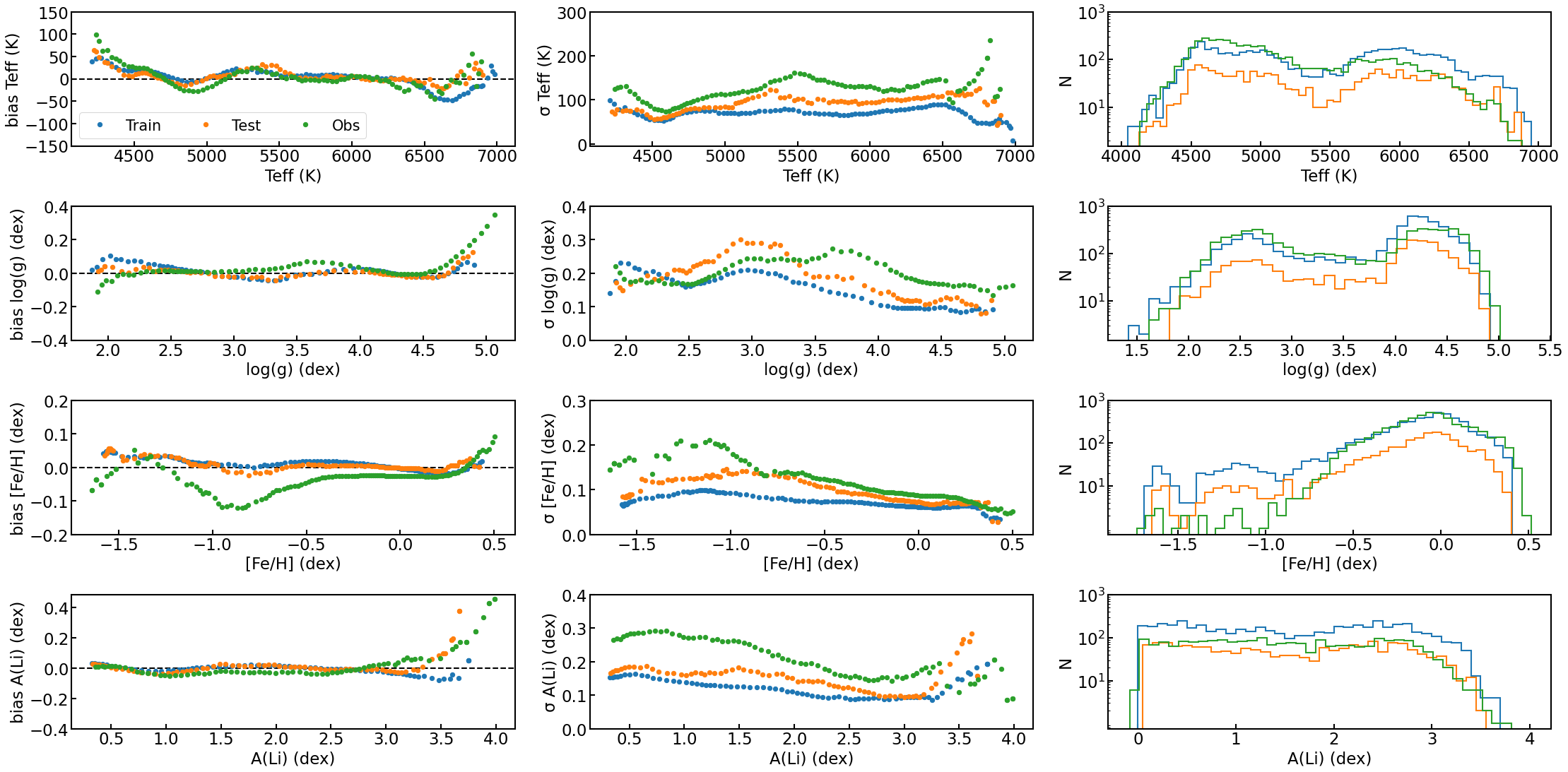}
        \caption{Running mean bias and mean dispersion as a function of labels for the train (blue), test (orange), and observed (green) sets calculated in bin sizes: 250\,K for $\mathrm{\teff}$, 0.3\,dex for log(\textit{g}), [Fe/H], and A(Li). The curves are representative of the real accuracy and precision of our CNN predictions. Bias = mean(CNN-iDR6) and $\sigma$=std(CNN-iDR6) for each bin. On the right column we present the distribution of the train, test and observed sets in logarithmic y-axis. The observed sample is selected within the training set, with S/N$>$20/pix and no GES flags; for A(Li), we selected only stars with Li measurements, instead of those with upper limit Li estimates.}
        \label{fig:running_error}
\end{figure*}

The CNN internal uncertainties may, however, be underestimated. To show a realistic approximation of the accuracy and precision of the method, in Fig. \ref{fig:running_error} we present the bias (running mean difference) and sigma (running mean dispersion) curves for our train, test, and observed sample predictions, compared to GES-iDR6 labels. The observed sample is selected within the training set limits, with S/N$>$20/pix and no GES flags, and GES lithium detection. The bias curves corresponds to the accuracy and the sigma curves correspond to the precision of CNN.

For $\mathrm{\teff}$, between $\mathrm{4\,400< \teff\ < 6\,600}$\,K, the accuracy is within 25\,K and increases only at the edges of the training set limits due to sparse training data. We report a good precision within 100 for the train and test sets and within 120 for the observed sample, affected by the lower S/N data. Similarly, for log(\textit{g}), an excellent accuracy is seen within 0.1\,dex across the label range except at the edges, due to the low statistics. A similar effect is seen in the precision curves within 0.2\,dex across the range except $\logg$\,<\,2.0\,dex and 3.0\,<\,$\logg$\,<\,4.0\,dex, which are less populated. For [Fe/H]$<$-1.0\,dex, with just 19 stars that have available GES-iDR6 values in the observed sample, the bias and $\sigma$ curves cannot be adequately interpreted. For [Fe/H]$>$-1.0\,dex, we achieve a very good accuracy within 0.05\,dex and precision within 0.1\,dex. For A(Li), the observed sample bias curve follows the train set, with an excellent accuracy within 0.05\,dex except at A(Li)$>$3.5\,dex, where we have very few stars. The precision of the train and test sets are within 0.2\,dex, while the observed sample is within 0.3\,dex as $\sim$90\% of the stars have S/N$<$40/pix. For future applications, such sigma and bias curves could be used to provide realistic precision and accuracy estimates.

\section{Validation of CNN predictions} \label{sec:validation}

\subsection{Validation with Gaia benchmark Stars} \label{sec:gbs}
\begin{figure*}[h]
        \centering
        \includegraphics[width=\linewidth]{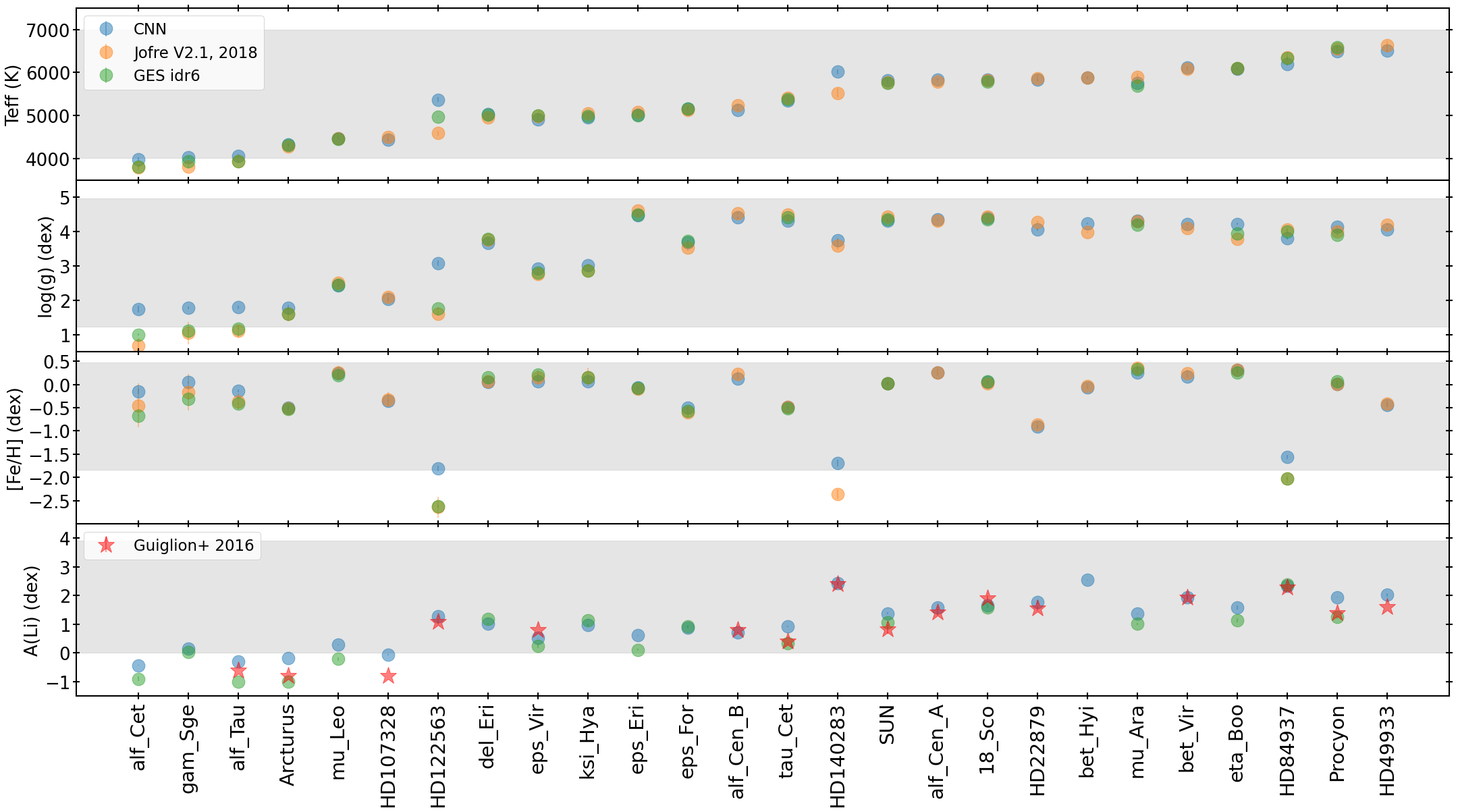}
        \caption{Comparison of CNN prediction for the Gaia Benchmarks Stars (GBS). The reference $\mathrm{\teff}$, log(\textit{g}), and [Fe/H]  come from \cite{gbs_v2_2018} and A(Li) from \cite{guiglion2016}. The GES-iDR6 values are also shown for comparison. On the x-axis, we present the GBS names sorted by increasing $\mathrm{\teff}$ and on the y-axis, we present the four labels. The shaded region for each label represents the training set limits. The CNN predictions and error bars are mean of the estimates for the multiple spectra. CNN error bars are too small to be seen.}
        \label{fig:gbs_horiz}
\end{figure*}

The Gaia benchmark star (GBS; \citealt{gbs_2015_heiter, gbs_2014_BC, gbs_2014_jofre}) sample provides precise stellar parameters and chemical abundances, derived from the best available spectra with very high-resolution and S/N along with the requirements of having accurate parallaxes, angular diameters from interferometry, bolometric flux, and stellar masses. The GBS are selected to represent typical Milky Way FGK stars covering different regions of the Hertzsprung–Russell diagram and a wide range of metallicities. Benchmark stars are commonly used as validators or calibrators by large spectroscopic surveys, such as GES \citep{Pancino_2017A&A}. In Fig. \ref{fig:gbs_horiz}, we compare CNN predictions with the GBS catalog Version 2.1 \citep{gbs_v2_2018} which contains 36 benchmark stars in total. The benchmarks stars were excluded from the training sample. There were 26 benchmark stars from the GBS in GES-iDR6, with high S/N, for which we compare the $\mathrm{\teff}$, log(\textit{g}), and [Fe/H] to the CNN predictions. As the GBS catalog does not provide lithium abundances, we used the AMBRE$\ $Li abundances from \cite{guiglion2016}, which has 15 stars in common between the GBS and GES-iDR6. The AMBRE$\ $Li catalog provides Li abundances derived from high-resolution (R = 40\,000) ESO spectra using an optimization pipeline GAUGUIN, based on a synthetic spectra grid and a Gauss-Newton algorithm.

The benchmark stars in Fig. \ref{fig:gbs_horiz}, are sorted by increasing $\mathrm{\teff}$, and most of the stars are within the training set limits. We find that for most of the GBS, the CNN results compare very well. The cool giants alf\_Cet, gam\_Sge and alf\_Tau have $\mathrm{\teff}$ and log(\textit{g}) outside the training limits, hence, we see a spread in log(\textit{g}) and [Fe/H]. The GBS catalog also reports higher uncertainty for these three stars and the CNN [Fe/H] measurements are within the uncertainty limits. There are three metal-poor stars, HD122563, HD140283, and HD84937, with [Fe/H] less than -2.0\,dex. HD122563 is the most metal-poor star with [Fe/H] = -2.62\,dex for which we see the highest differences in $\mathrm{\teff}$, log(\textit{g}) and [Fe/H], although CNN estimate for A(Li) agrees with the AMBRE value. For HD140283, with [Fe/H] = -2.36\,dex, we see a difference of $\sim$500 for $\mathrm{\teff}$ and 0.7\,dex in [Fe/H], while the estimates for log(\textit{g}) and A(Li) are in a good match. For HD84937, CNN predictions for $\mathrm{\teff}$, log(\textit{g}) and A(Li) are in a very good agreement with GBS and AMBRE measurements, but we note a difference of 0.5\,dex for [Fe/H]. In the case of lithium, for most of the GBS stars, CNN predictions compare well with AMBRE abundances within $1-\sigma$. For stars with A(Li) below the training set limit of 0.0\,dex, we see a difference of up to 0.8\,dex in CNN and AMBRE/iDR6 predictions; for stars that are within the training limit and have A(Li)$<$1.5\,dex, a small difference ($\sim$0.25\,dex) in CNN, iDR6, and AMBRE measurements are seen. Overall, the CNN performs very well across the training label range and differences are seen only for stars outside the training range. Future spectroscopic surveys should be careful to target more metal-poor stars and cool giants. Also, the benchmark stars should include more metal-poor stars and cool giants.

\begin{figure}[h!]
        \centering
        \includegraphics[width=\linewidth]{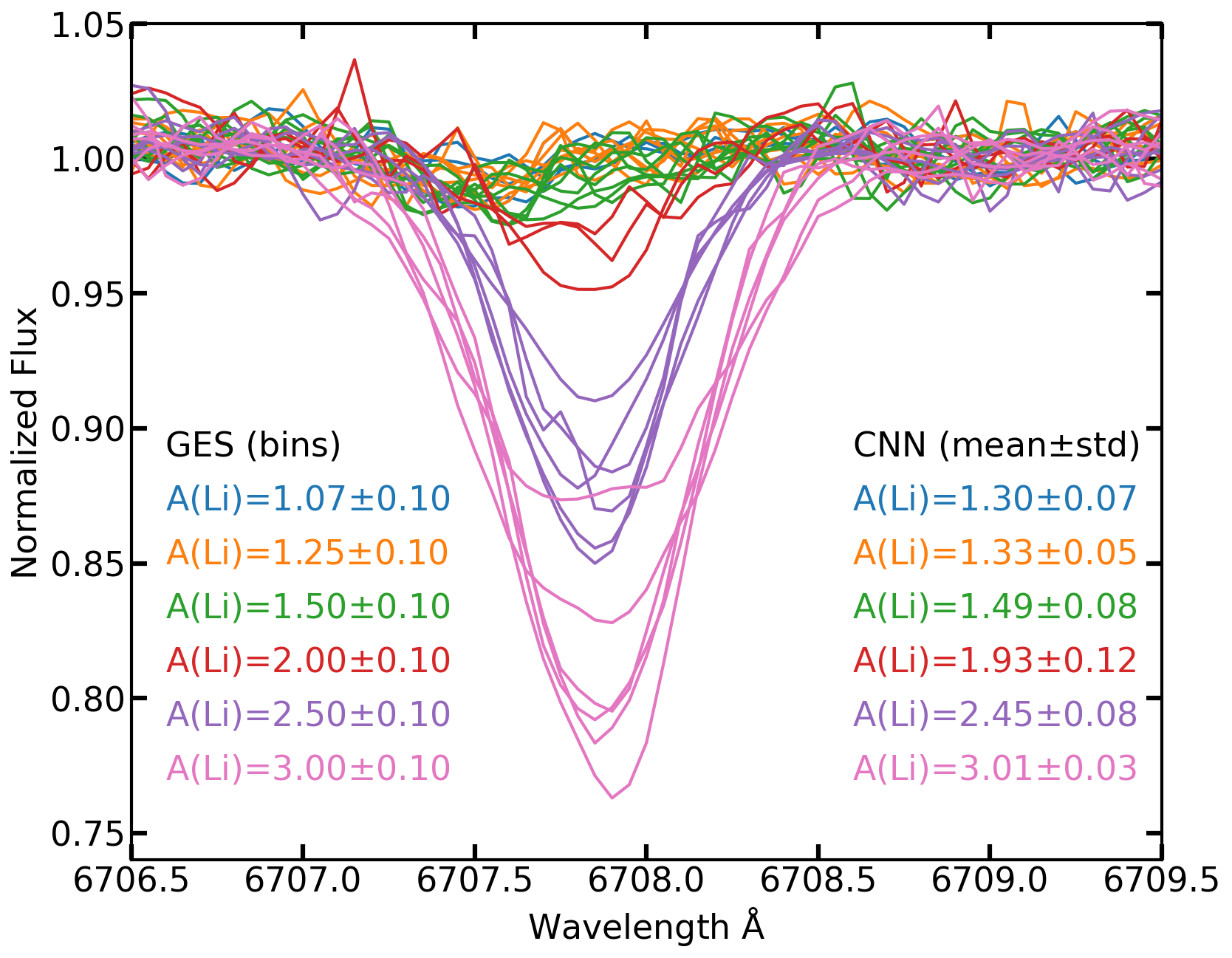}
        \caption[Li features for Solar Twins with varying lithium abundance.]{Li features for the "solar twins" with varying Li abundance. The solar twins in the training sample are selected with S/N$>$90/pix and with GES $\mathrm{\teff\ = 5\,777\pm150\,}$K, log(\textit{g}) = 4.44$\pm$0.15\,dex and [Fe/H] = 0.0$\pm$0.15\,dex. The colors represent the A(Li) bins, as listed on the left. On the right, we show the mean of CNN prediction for the shown spectra in each bin.}
        \label{fig:li_solar_twins}
\end{figure}

In Fig. \ref{fig:li_solar_twins}, we present the HR15N spectra around the 6707.8 \AA\ lithium line for some solar twins, in different A(Li) regimes. The solar twins are selected from the training sample with S/N$>$90/pix and with $\mathrm{\teff\ = 5\,777\pm150}\,$K, log(\textit{g}) = 4.44$\pm$0.15\,dex and [Fe/H] = 0.0$\pm$0.15\,dex. CNN provides robust measurements for A(Li)$\geq$1.25\,dex. Below this limit, CNN suffers from a positive bias, namely, the Solar abundance reported by GES is A(Li)=1.07, while CNN measures 1.3\,dex. For A(Li) of 1.07\,dex (blue) and 1.25\,dex (orange), the spectral features look almost identical within the noise. For these spectra, we see that the maximum flux absorption is $\mathrm{\sim1.5\%}$ and most of the signal comes from an Fe blend.

An accurate measurement for lithium below 1.25\,dex in Solar twins at resolution $\mathrm{R\sim20\,000}$ with CNN is then challenging and basically Li $<$ 1.25\,dex should be considered as limit in the dwarf regime. This could explain the difference in CNN, iDR6 and AMBRE measurements for the lithium measured in some of the benchmark stars. We carried out the same exercise for a typical RC star (around Solar [Fe/H]), and given the line is deeper, the CNN performs with no significant bias up to Li=0\,dex. It is representative of the well-known temperature dependence of the lithium line-shape. For 4MOST-LR/HR, it will be important to generalize this type of detection limit to the whole parameter space of the sample.

\subsection{Validation with GALAH-DR3}

\begin{figure*}[h!]
        \centering
        \includegraphics[width=\linewidth]{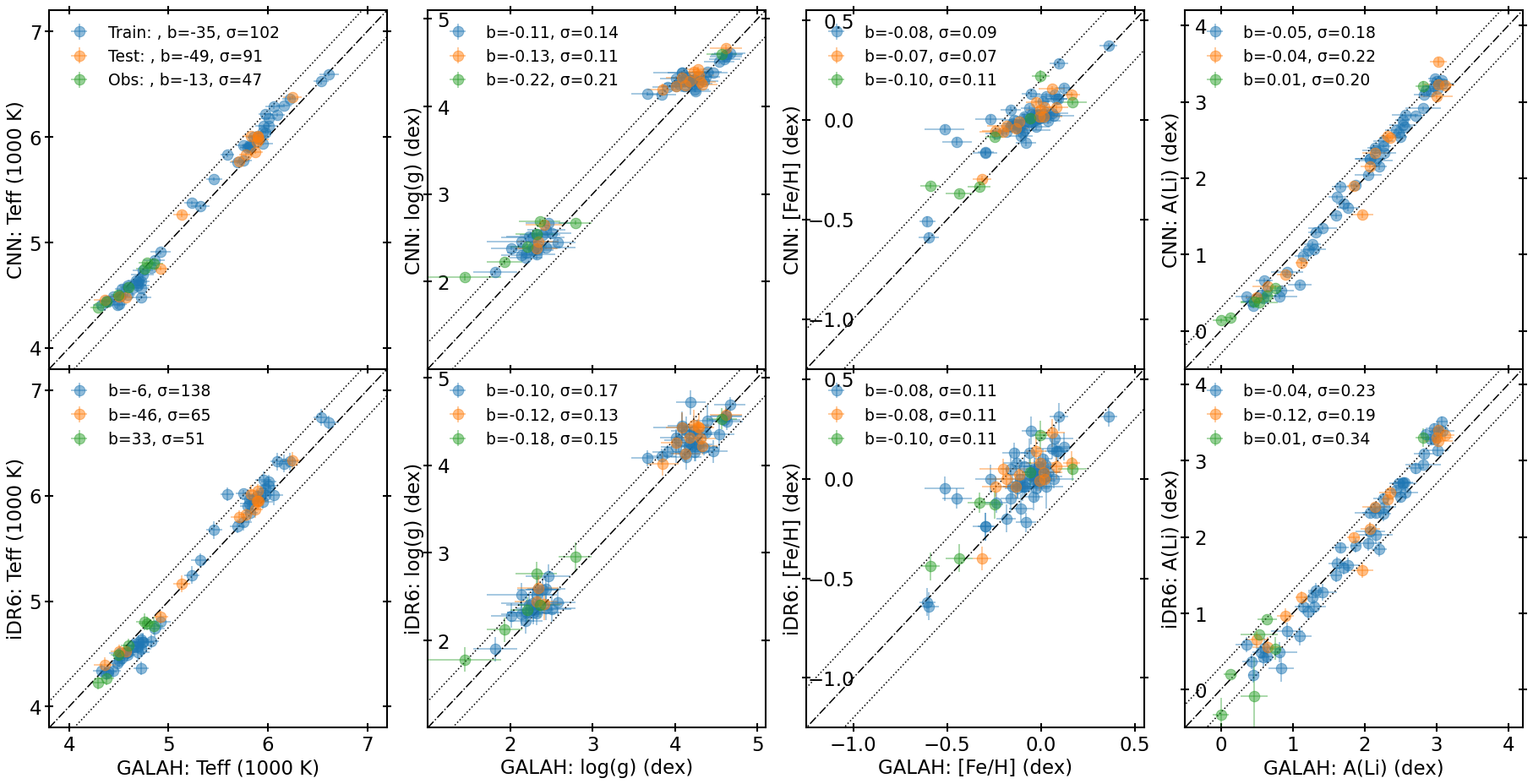}
        \caption{Comparison of CNN results for stars in common with GALAH-DR3 {\cite{galah_dr3_2021}}. GES-iDR6 sample has stars selected with S/N$>$30/pix,  within the training label limits, eVRAD$<0.5$km\,s$^{-1}$ and no GES flags and GALAH stars are selected with snr\_c3\_iraf $>$30/pix, flag\_sp = 0, flag\_fe\_h = 0, and flag\_Li\_fe = 0. The dash-dot line is the 1-to-1 line and two dotted lines are at $\pm$ 250 for $\mathrm{\teff}$, $\pm$ 0.3\,dex for log(\textit{g}), $\pm$0.2\,dex for [Fe/H], $\pm$0.3\,dex for A(Li). The error bars show the errors reported in GES-iDR6 and GALAH-DR3; CNN uncertainties are too small to be seen.}
        \label{fig:cnn_galah}
\end{figure*}

The Galactic Archaeology with HERMES (GALAH, \citealt{galah_dr3_2021}) survey provides stellar parameters and chemical abundances, including lithium, using the spectrum synthesis code Spectroscopy Made Easy (SME) and 1D MARCS model atmospheres, along with additional photometry and astrometry. GALAH spectra are obtained at a higher resolution of R$\sim$28\,000, compared to the GIRAFFE at R$\sim$20\,000, and in four non-contiguous spectral bands between 4\,700\,\AA\, and 7\,900\,\AA. In Fig. \ref{fig:cnn_galah}, we present a comparison of CNN results for GES-iDR6 HR15N stars in common with the third data release GALAH-DR3 \cite{galah_dr3_2021}. The selected GES/CNN sub-sample has 73 HR15N stars in common with GALAH with available $\mathrm{\teff}$, log(\textit{g}), [Fe/H], and A(Li). For GES/CNN we only consider the stars within the training set limits, S/N$>$30/pix, eVRAD $<$ 0.5\,km\,s$^{-1}$, and no GES flags. For GALAH stars, we followed the GALAH recommended S/N and flags, namely, snr\_c3\_iraf $>$ 30/pix, flag\_sp = 0, flag\_fe\_h = 0, and flag\_Li\_fe = 0 (the flags\,$=$\,0 represent no identified problems with determination of stellar parameters and iron and lithium abundances, respectively). The CNN atmospheric parameters and lithium predictions agree very well with GALAH, within 250 for $\mathrm{\teff}$, 0.3\,dex for log(\textit{g}), 0.2\,dex for [Fe/H], 0.3\,dex for A(Li). For the case of A(Li)\,$<$\,1.0\,dex, the spread in 1-to-1 relation is less for the case of CNN versus GALAH, indicating that CNN results are in better agreement with GALAH than the iDR6 measurements. Given the higher resolution for GALAH, it should be able to capture weaker lithium lines, hence providing more precise lithium values at A(Li)\,$<$\,1.0\,dex. We see that CNN works better at low lithium than standard pipelines in the cool regime (see also \figurename~\ref{fig:1_to_1_obs}). Also, CNN can also efficiently deal with the noise. We see systematic $\mathrm{\teff}$ offsets in GALAH vs iDR6 with lower iDR6 measurements for cooler stars, and higher for hotter stars. This is also seen in the GALAH vs CNN comparison. A similar systematic offset is seen for lithium, with lower CNN/iDR6 measurements for A(Li) $<$ 2.5\,dex and higher CNN/iDR6 measurements for A(Li) $>$ 2.5\,dex. Overall, GALAH, and CNN are in a good agreement and the offsets seen are systematic between GALAH and GES-iDR6.

\subsection{Validation with Asteroseismic gravities}

\begin{figure}[h!]
        \centering
        \includegraphics[width=0.7\linewidth]{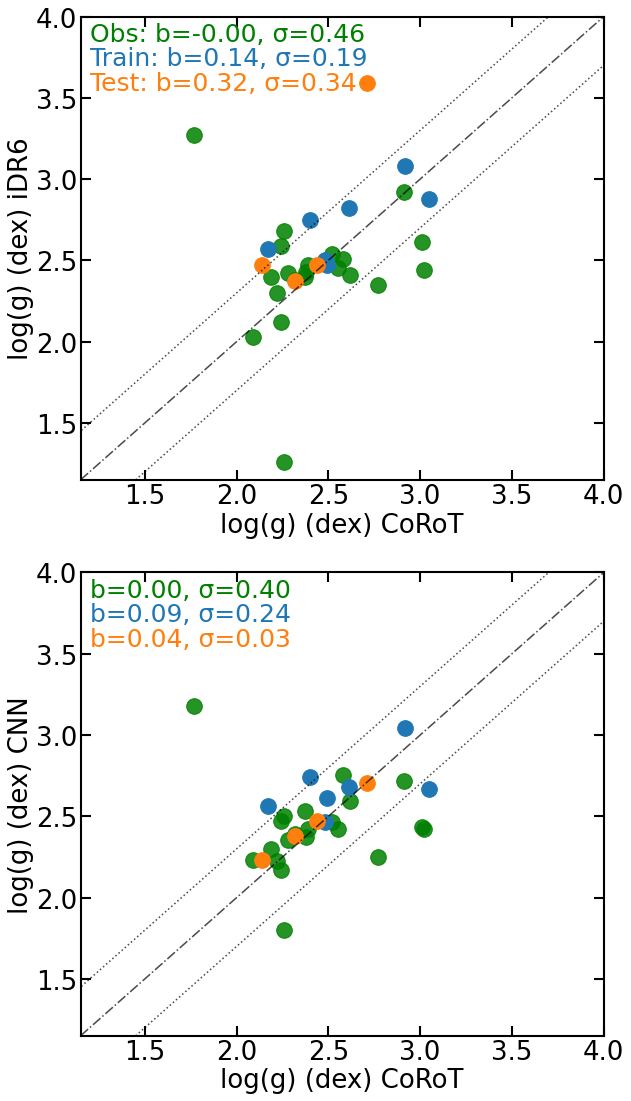}
        \caption{Comparison with asteroseismic results. Left: CoRoT-GES vs GES-iDR6 labels, Right: CoRoT-GES vs CNN predictions. Blue, orange, and green symbols represent the train, test, and observed sample selected within the training set limits (S/N$>$30/pix, eVRAD$<$0.5\,km\,s$^{-1}$, no GES flags) and with CoRoT-GES flag OFLAG\_GIR=0. The bias=mean(CNN-CoRoT) and $\sigma$=std(CNN-CoRoT) are provided. The dash-dotted line is the 1-to-1 line and the two dotted lines are at $\pm$\,0.3\,dex.}
        \label{fig:cnn_corot}
\end{figure}

Here, we are aiming to compare CNN surface gravities with precise asteroseismic gravities. In Fig. \ref{fig:cnn_corot}, we present a comparison of log(\textit{g}) for 32 stars present in the CoRoT-GES sample of \citet{valentini_corot_2016} with the CNN predictions. We selected only stars with good asteroseismic results given by flag OFLAG\_GIR=0 from \cite{valentini_corot_2016} and CNN/iDR6 stars are selected within the training label limits, S/N$>$30/pix, eVRAD$<$0.5\,km\,s$^{-1}$, and no GES flags. Fig. \ref{fig:cnn_corot} shows that there is an intrinsic bias between GES-iDR6 and CoRoT labels due to the different methods for deriving log(\textit{g}). The CNN results are consistent with the GES-iDR6 values, and they show a similar trend. The comparison shows presence of some outliers and we discuss two such outliers below.

For the star CNAME=19264480+0032497, with $\mathrm{\teff}$ = 4\,815\,K and log(\textit{g}) = 3.59\,dex in iDR6, the CNN results (4\,635\,K and 2.83\,dex) agree better with CoRoT-GES values (4\,550\,K and 2.71\,dex). The star has a high projected rotational velocity (\textit{v}sin\textit{i}) of 27.6\,km\,s$^{-1}$, which can be a cause behind this difference. About 35\% of our training sample have stars with \textit{v}sin\textit{i} $>$ 10\,km\,s$^{-1}$, hence, CNN are able to learn about the rotationally broadened spectral features.

For the star CNAME=19240528+0152010, the iDR6 predictions are $\mathrm{\teff}$ = 4\,663\,K, log(\textit{g}) = 3.27\,dex, and [Fe/H] = 0.01\,dex, which is in agreement with CNN output (4\,872\,K, 3.2\,dex, and 0.04\,dex), while there is a discrepancy with Corot predictions (4\,514\,K, 1.77\,dex, and -0.46\,dex). A significantly lower log(\textit{g}) and [Fe/H] is provided by CoRoT-GES. We compare the spectrum of this star with another star for which the atmospheric parameters are similar to our CNN result and for which the CNN, iDR,6 and CoRoT-GES results agree. Both spectra look similar (besides the slightly lower log(\textit{g}) of the second spectrum), showing that Corot atmospheric parameters for this star should be taken with caution.

Such a comparison between the CNN predictions and Corot tells us that CNN is able to properly parametrize giants, while considering the HR15N is not an optimal setup for precisely constraining log(\textit{g})s. We also show that CNN can correct inaccurate labels that are misclassified by standard pipelines; it is illustrative of the anomaly detection capability of CNNs.

\section{Constraining the chemical evolution of lithium in the Milky Way} \label{sec:gal_arcaeo}
\subsection{Galactic evolution of lithium} \label{sec:gal_evo}

\begin{figure*}[ht!]
        \centering
        \includegraphics[width=\linewidth]{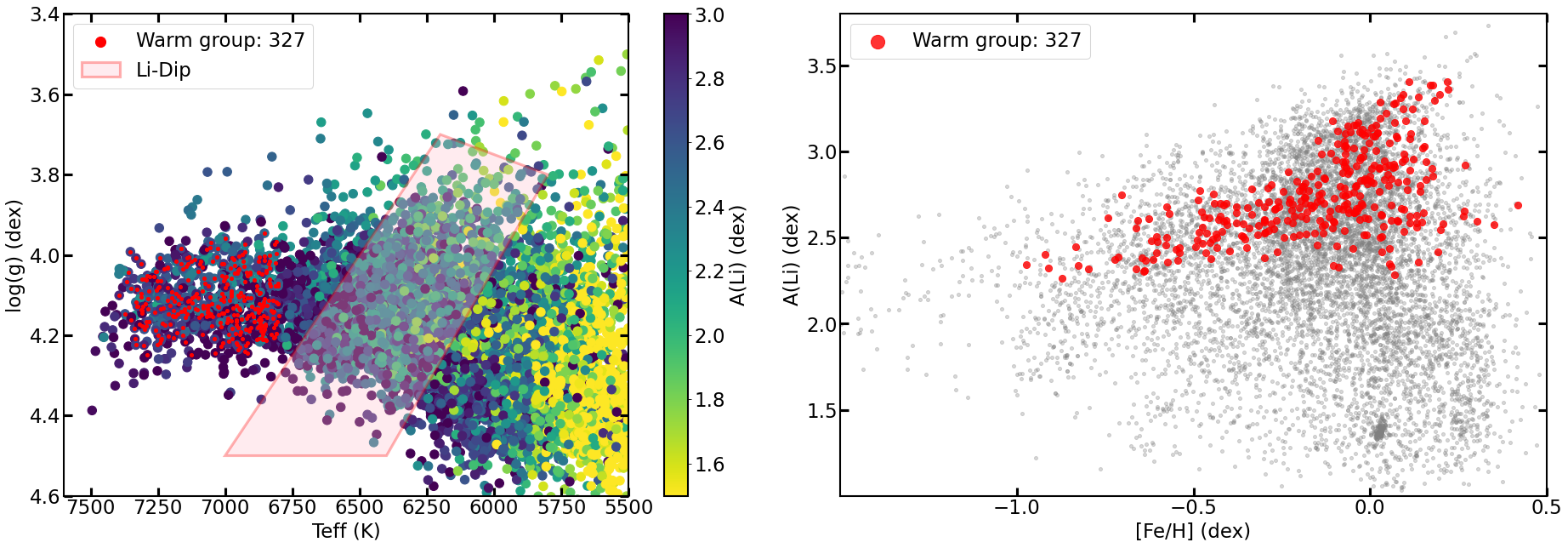}
        \caption{Effective temperature vs. surface gravity diagram, with the stars color-coded according to their Li abundance (left). The approximate location of the Li-dip region according to \cite{Gao_GALAH_2020} is highlighted in pink. The red points represent the warm stars, $\mathrm{\teff} >$6\,800 and S/N$>$75/pix. The\text{[Fe/H]} vs. Li abundance trend for the warm stars shown as red points. Gray dots represent the other stars shown in the left plot (right).}
        \label{fig:ISM_li_dip}
\end{figure*}

Recently, a number of studies have challenged the possibility to use main-sequence stars ($\teff>5\,500\,$K) to trace the lithium ISM abundance. \citet{guiglion2019} suggested that the upper boundary of lithium in the super-solar metallicity main-sequence stars do not reflect the original ISM content -- but, rather, lithium depletion due to an interplay between stellar evolution and radial-migration (see also \citealt{Miglio_2021} and references therein). \citet{GES_Randich_2020} investigated this Li decrease using GES stars both on the warm side of the lithium dip ($\teff>6\,800\,$K) in metal-rich open clusters together with PMS stars from very young clusters\footnote{An updated list of clusters comprising also the OCs released in iDR6 can be found in Table 2 of \citet{Romano_2021A&A}} (age < 100 Myr). They showed a lithium plateau of A(Li)$\sim$3.4\,dex at 0.1<[Fe/H]<0.3\,dex. Their conclusion supported the scenario of \citet{guiglion2019} which has recently been confirmed by \citet{Dantas_2022arXiv}.

Stars on the hot side of this dip have not undergone any Li depletion and they are the best candidates for the study of the galactic evolution of lithium with metallicities, ages, and galactocentric distances. However, atomic diffusion might
have changed the original Li abundances in the atmospheres of (some) solar-metallicity stars \citep{Romano_2021A&A, Charbonnel_2021}. Indeed, the lithium-dip (Li-dip), namely, the drop in A(Li) observed in the main sequence stars in temperature range of 6\,400-6\,800\,K, has been confirmed in both cluster and field stars (e.g., \citealt{Boesgaard_1986, Deliyannis_2019}). The origin of the Li-dip at this narrow $\mathrm{\teff}$ range has been attributed to an interplay of mass-temperature dependent processes, most importantly, shallow surface convective zone and higher atmospheric mixing due to significant spin-down of initial PMS rotational velocity. \citet{Charbonnel_2021} recently showed that hot metal-rich field stars do not exhibit any lithium decrease using GALAH and AMBRE data. This finding is in agreement with the result in \citet{Gao_GALAH_2020} using warm field stars from GALAH, and \citet{GES_Randich_2020} using OC stars, and Romano et al. (2021) using both.

In Fig. \ref{fig:ISM_li_dip}, we further investigate the Li ISM, with a sample of stars on the warm side of the Li-dip (warm group). To select these stars we adopted the following criteria: S/N\,$>$\,75/pix, $\mathrm{\teff}$\,$>$\,6\,800\,K, 3.8$<$\,$\logg$\,$<$4.25\,dex, ~-1\,<\,[Fe/H]\,<\,0\,dex, A(Li)\,$>$\,1.0\,dex, eVRAD$<$1.0\,km\,s$^{-1}$, ~e$\mathrm{\teff}$\,<\,200\,K, e$\logg$\,<\,0.1\,dex, e[Fe/H]\,<\,0.2\,dex, and eA(Li)\,<\,0.2\,dex and also avoid peculiar stars and stars with emissions. We find stars with Li around 3.4\,dex at [Fe/H]$\sim0.2\,$dex, consistently with the peak at A(Li)$\sim$3.4\,dex reported by \citet{GES_Randich_2020}. We note the presence of super-solar [Fe/H] stars with lithium between 2.2 and 3.0\,dex. These stars could be old (>6-7 Gyr) and have depleted their lithium. To be able to confirm these stars have indeed migrated from inner regions, an estimate of their birth-radii would be needed (e.g., \citealt{minchev_2018MNRAS}).

We further investigate  the ISM evolution in the metallicity regime -1<[Fe/H]<0\,dex. All of these stars have Li abundance above the Spite plateau value and there is a clear increase of lithium with metallicity from 2.2 to 3.2\,dex. Given the small sample size, we cannot reliably confirm the presence/absence of a warm plateau at ~A(Li)\,=\,2.69\,dex (see GALAH survey, \citealt{Gao_GALAH_2020}), in the region of ~-1.0$<$[Fe/H]$<$-0.5\,dex. However, the mean A(Li) for the 33 stars present in that metallicity range is lower at ~A(Li)\,=\,2.44$\pm$0.12\,dex and show a gradient with metallicity. If we trust that the hot stars on the hot side of the dip are accurate tracers of the lithium ISM, we do not measure the usually reported steep rise of the ISM in the domain -1.0$<$[Fe/H]$<$-0.5\,dex (based on cool dwarfs), but, instead, a shallow increase. 

The consequence of such finding for the modeling of the lithium ISM on the domain -1$<$[Fe/H]$<$-0.5\,dex would be to take into account earlier Li production by more massive sources and a longer delay in the production of lithium by the long-lived sources (as suggested by the chemical evolution model of \citealt{Cescutti_2019}). \citet{Romano_2021A&A} arrived to the same conclusion based on GES-iDR6 data, suggesting a shorter delay in the production of lithium, claiming that nova white-dwarf progenitors must be in the range 3-8 \(\textup{M}_\odot\) rather than 1-8 \(\textup{M}_\odot\), as usually assumed (see Fig. 8 of \citealt{Romano_2021A&A}).

\subsection{Search for lithium-rich giants} \label{sec:li_g}

Standard stellar evolution models predict that the surface Li abundances of low-mass red giants after the first dredge-up decreases by $\sim$60 times to below ~A(Li)$\sim$1.50\,dex (e.g., \citealt{Lagarde_2012A&A}) when starting from an initial ~A(Li)\,=\,3.3\,dex (solar meteoritic value). Lithium-rich giants are rare objects and confirm that lithium can be produced in stellar interiors (see e.g., \citealt{magrini_2021A&A}, and references therein); this results from the \citet{Cameron1971} mechanism. These authors proposed that the reaction $^{3}$H+$\alpha\rightarrow$ $^{7}$Be+$\gamma$ produces $^{7}$Be, which is then rapidly transported outwards by convection and non-standard mixing processes to lower temperatures, where it decays into $^{7}$Li. The Li-rich giants are believed to play a role in the enrichment of the ISM \citep{Romano_2001}. Stellar Li enrichment is also possible due to external sources such as the measured over-abundance of Li as a result of a mass transfer process in a binary system, where the companion produces Li through the Cameron-Fowler mechanism. Planet engulfment was also proposed to explain such high lithium abundance in giants, although it seems this mechanism can increase the abundance only up to ~A(Li)\,$\sim$\,2.2\,dex \citep{AguileraGomez2016ApJ}. We refer  to \citet{Casey_2016MNRAS} for a review on the enrichment processes in Li-rich giants.

\begin{figure*}[ht!]
        \centering
        \includegraphics[width=\linewidth]{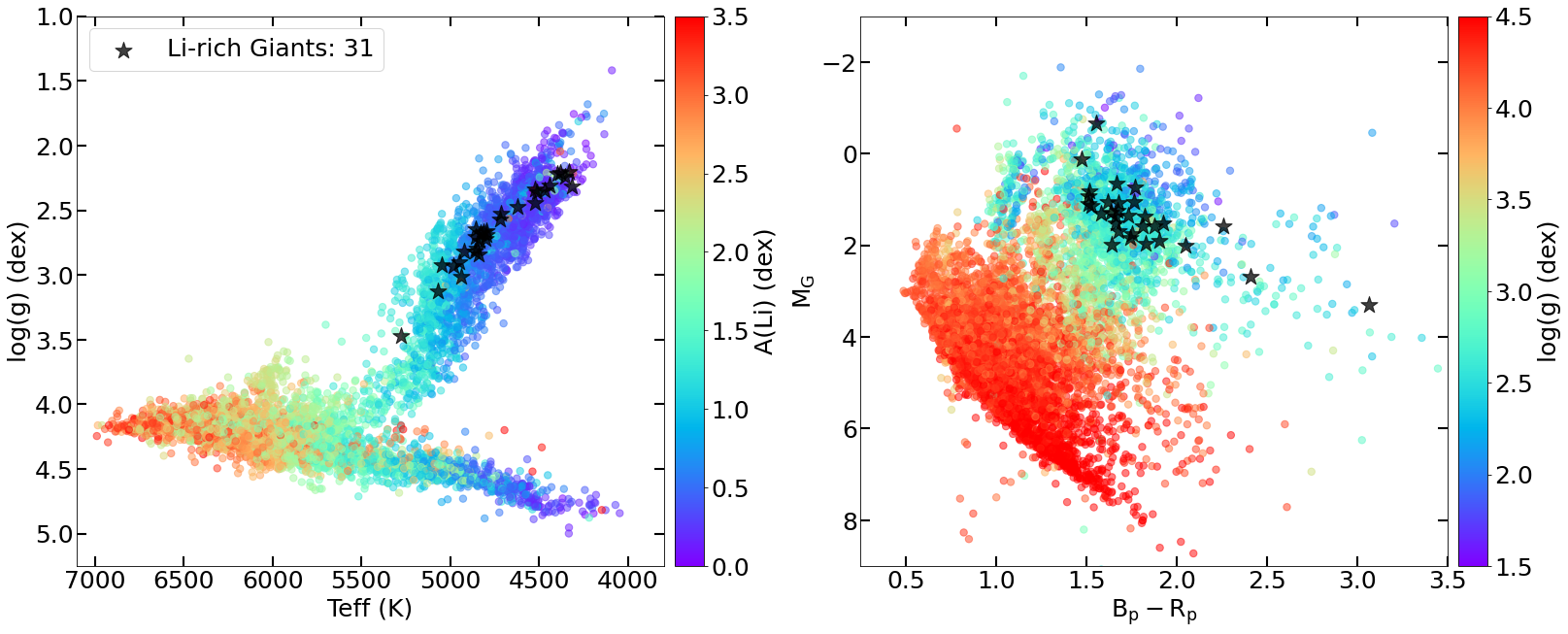}
        \caption{Kiel-diagram showing the newly-discovered Li-rich giants (black stars) along with the training sample color-coded according to their Li abundance (left). Gaia Color-Magnitude diagram for the same stars (right). The training sample stars are colored by their surface gravities.}
        \label{fig:Li_giants_CMD}
\end{figure*}

Our training sample contains just 38 lithium rich giants, considering a strict condition of ~$\logg$\,<\,3.2\,de and A(Li)\,>\,2.0\,dex. It is important that the CNN is able to identify these rare objects, as they are of a great scientific interest. The Li-rich giants have previously been reported in earlier Gaia-ESO papers \citep{Casey_2016MNRAS, Smiljanic_2018A&A, Sanna_2020A&A} and some of them are present in our training sample. In addition, we report the discovery of 31 new lithium rich giants by CNN in the observed sample (see Fig. \ref{fig:Li_giants_CMD}). These stars were not reported in previous Gaia-ESO papers. We also checked the GALAH survey catalog in the southern sky of Li-rich giants by \cite{Martell_2021MNRAS} and found no match.

To identify the Li-rich giants, we selected stars with ~$\mathrm{\teff}$\,<\,5\,500\,K, ~$\logg$\,<\,3.5\,dex and ~A(Li)\,>\,2.0\,dex, for which GES-iDR6 has not provided either one or any of the labels. To assure a reliable parameter estimation, we further selected spectra with low CNN uncertainties of ~e$\mathrm{\teff}$\,<\,50\,K, e$\logg$\,<\,0.1\,dex, e[Fe/H]\,<\,0.1\,dex and eA(Li)\,<\,0.1\,dex, and S/N\,>\,25/pix and ~E\_VRAD\,<\,0.5\,km\,s$^{-1}$. We also checked for good photometry in Gaia EDR3 \citep{gaia2020} by selecting ~RUWE\,$\leq$\,1.4. The CNAME and atmospheric parameters for the 33 stars are listed in Table \ref{table: li_rich_giants}. Out of the 31 Li-rich giants, half of the stars have A(Li) between 2.0-3.0\,dex with half have~A(Li)$>$3.0\,dex with a maximum lithium abundance of 3.88\,dex. One of the Li-rich giants is a fast-rotator with \textit{v}sin\textit{i}$=$12.1\,km\,s$^{-1}$; giants with high \textit{v}sin\textit{i} and A(Li) can indicate planetary engulfment and needs further study. We additionally confirmed that our Li-rich giants are not misclassified objects (e.g., PMS stars) using the $\gamma-$index of \citet{Damiani2014}.

As seen in Fig. \ref{fig:Li_giants_CMD}, our new Li-rich giants seem to be distributed along the whole giant branch, although a clear concentration is seen at the position of the red clump. However, in recent years, a view 
has emerged staring that Li-rich giants can be found only in the He-core burning red clump phase \citep{Deepak_2019MNRAS, Deepak_2021MNRAS, Martell_2021MNRAS}. Further analyses of our new sample is essential for investigating their properties and evaluating the possible mechanisms for their Li enrichment. Further investigations on these 31 Li-rich giants could be complemented by very precise asteroseismic log(\textit{g}) (see for instance \citealt{zhou2022} with LAMOST data), if available with surveys such as TESS and PLATO \citep{Singh_2021ApJ}.

\begin{table}[ht]
\begin{center}
\caption{31 newly discovered GES Li-rich giants and their CNN associated atmospheric parameters: $\mathrm{\teff}$ (K),  log(\textit{g}) (dex), \text{[Fe/H]} (dex); and lithium abundances, A(Li) (dex). Table is ordered by A(Li).}
\begin{tabular}{c c c c c} 
 \hline\hline
 CNAME & $\mathrm{\teff}$ & log(\textit{g}) & \text{[Fe/H]} & A(Li) \\ [0.5ex] 
 \hline 
 07434938-3841399 &  4841.0 &    2.84 & -0.31 &   3.88 \\
 10495937-6345553 &  4805.0 &    2.72 & -0.16 &   3.83 \\
 07464933-3750081 &  4948.0 &    2.90 & -0.20 &   3.62 \\
 08064077-4736441 &  4797.0 &    2.66 & -0.10 &   3.56 \\
 16271097-2455213 &  4920.0 &    2.82 & -0.45 &   3.55 \\
 06410348+0905141 &  5071.0 &    3.13 & -0.18 &   3.50 \\
 07493206-3759457 &  4799.0 &    2.69 & -0.24 &   3.48 \\
 10430727-6456318 &  4619.0 &    2.47 & -0.16 &   3.43 \\
 08110435-4853491 &  4831.0 &    2.68 & -0.30 &   3.42 \\
 10400095-6419586 &  4525.0 &    2.34 &  0.01 &   3.29 \\
 08084532-4701292 &  4836.0 &    2.74 & -0.19 &   3.26 \\
 07462219-3712141 &  4862.0 &    2.82 & -0.20 &   3.22 \\
 06273069-0440141 &  4714.0 &    2.53 & -0.68 &   3.21 \\
 08512566-4135067 &  4331.0 &    2.20 &  0.23 &   3.15 \\
 08102172-4845417 &  4514.0 &    2.36 & -0.06 &   3.14 \\
 06255393-0457404 &  4981.0 &    2.93 & -0.29 &   3.03 \\
 08083354-4711111 &  4441.0 &    2.31 &  0.10 &   3.00 \\
 07442999-3812166 &  4857.0 &    2.65 & -0.24 &   2.98 \\
 10350175-6405092 &  4469.0 &    2.35 &  0.11 &   2.88 \\
 07475310-3733040 &  4853.0 &    2.80 & -0.21 &   2.86 \\
 10483936-6327542 &  4383.0 &    2.21 &  0.08 &   2.74 \\
 10420066-6421333 &  4397.0 &    2.22 &  0.08 &   2.73 \\
 11130526-7617396 &  4815.0 &    2.67 & -0.32 &   2.73 \\
 11123294-7727006 &  4315.0 &    2.31 &  0.16 &   2.61 \\
 10575316-7634459 &  4858.0 &    2.70 & -0.21 &   2.42 \\
 07472841-3850499 &  5276.0 &    3.47 & -0.12 &   2.35 \\
 10513847-6335341 &  4352.0 &    2.24 &  0.31 &   2.29 \\
 07472390-3856376 &  5049.0 &    2.93 & -0.24 &   2.27 \\
 06272996-0518528 &  4522.0 &    2.44 & -0.02 &   2.20 \\
 08075108-4744027 &  4719.0 &    2.57 & -0.13 &   2.19 \\
 07483625-3724338 &  4939.0 &    3.01 & -0.15 &   2.03 \\
 \hline\hline
\end{tabular}
\label{table: li_rich_giants}
\end{center}
\end{table}

\section{Summary and future prospects}\label{sec:conclusion}

To prepare the ground for the future 4MOST and WEAVE spectroscopic surveys, we developed a convolutional neural network approach for determining atmospheric parameters ($\mathrm{\teff}$, log(\textit{g}), [Fe/H]) and lithium abundances from GES stellar spectra. We built a training set of 7\,031 stars, based on high-quality stellar labels from GES iDR6. The main results are summarized as follows:

- Our CNN shows very good performance, even though we masked H$_{\alpha}$ and despite the fact that the wavelength range in GIRAFFE HR15N setup is not considered optimal for determinations of atmospheric parameters \citep{Lanzafame2015}. These results indicate that our trained CNN models are competent and have learned the available spectral features. The CNN is able to provide results with typical uncertainties of $\sim$35 \,K for $\mathrm{\teff}$, 0.05\,dex for log(\textit{g}), 0.03\,dex for [Fe/H], and 0.06\,dex for A(Li).

- Overall, the CNN predictions show a very good agreement in comparison with the GES-iDR6 input labels. The CNN achieves a good performance for all S/N values, including the low S/N ($\approx$ 20/pix) spectra. Thanks to the large variety of rotational velocities in the training sample, the CNN is able to accurately predict atmospheric parameters, even for the fast rotators for which the spectral features are broadened and can be blended with neighbouring lines. As CNN is sensitive to even small systematics in the input data, we found that large uncertainties in $\mathrm{V_{rad}}$ (>0.5\,km\,s$^{-1}$) can degrade the CNN performances.

- Gaia benchmark stars within the training label range are accurately predicted within 1-sigma by CNN while those outside show some systematics. The origin of such a discrepancy could be a lack of metal-poor stars (both dwarfs and giants) in the training set. It  could also come from the fact that metal-poor stars are more difficult to parametrize due to weaker lines and possible NLTE effect. 

- The catalog of atmospheric parameters and Li abundances for $\sim\,40\,000$ stars is publicly available\footnote{doi:// \emph{to be added upon paper acceptance}}. In addition, we have made the CNN code, spectra and labels available to the community\footnote{\href{https://github.com/SamirNepal/Li_CNN_2022}{https://github.com/SamirNepal/Li\_CNN\_2022}}.

- The CNN atmospheric parameters and lithium predictions agree very well with GALAH DR3, within 250\,K for $\mathrm{\teff}$, 0.3\,dex for log(\textit{g}), 0.2\,dex for [Fe/H], 0.3\,dex for A(Li). Systematic offsets are present between the GALAH DR3 and CNN (also with respect to input GES-iDR6 labels) due to the different instrument setup, spectroscopic pipelines, and calibration strategies. We show that the CNN atmospheric parameters match up nicely with asteroseismic results from CoRoT. We also demonstrate that CNN can correct wrongly assigned labels.

- We have verified that the CNN is learning from relevant spectral features for the atmospheric parameters (e.g., the Quintet is sensitive to log(\textit{g})) and found that CNN is able to single out the lithium line among hundreds of other lines, for precisely determining lithium. Using correlations for inferring elemental abundances without spectral features should be avoided.

- We investigated the ISM chemical evolution of lithium, with the stars on the hot side of the lithium dip (more representative of the ISM). Our findings suggest that the usually reported steep rise of the upper boundary of lithium is not visible on the domain -1\,<\,[Fe/H]\,<\,0\,dex, exhibiting a shallower rise of the ISM. This suggests that earlier Li production by more massive sources and a longer delay in the production of Li by the long-lived sources for enriching the ISM should be taken in account, as claimed by recent chemical evolution modeling \citep{Cescutti_2019, Romano_2021A&A}. In addition, there is no decrease in the lithium boundary with [Fe/H]\,>\,0\,dex, but we report the presence of stars with lithium between 2.2 and 3.0\,dex which are likely to have depleted their lithium content.

- We report the discovery of 33 new Li-rich giants. A follow-up study using asteroseimic data for these stars could provide an insight on stellar Li production and mixing mechanisms. 4MOST is expected to discover thousands of these objects, making it possible to study these peculiar stars over a large Galactic volume, for instance, in the bulge, and metallicity range.\\

Our work confirms that CNNs are efficient for deriving lithium abundances based on HR15N spectra, namely, very similar data as 4MOST and WEAVE. It gives excellent perspectives for data analysis with CNN in the context of these two surveys.
However, several improvements could be made in order to refine CNN performance.
For instance, in order to increase the diversity in the training sample, adding the spectra of binary stars and properly dealing with emission features could be helpful.

For the future use of CNNs, it will be crucial to build the training sets proactively, namely, not only relying on sets we build for a given survey, but carefully filling in regions of the HR diagram with proper targets. In particular, attention should be paid to populating the metal-poor tail of the training set in order to avoid biases. In such a way, the training set limits would be extended and a larger label space could be probed -- as the current application is clearly limited within the available training set limits.
In a future work, it would be interesting to explore Bayesian NNs and different types of loss functions such as the negative log likelihood to provide better uncertainty estimates.

One important aspect of spectroscopy that was not taken into account in this project are the NLTE effects coupled with a 3D structure of the atmosphere that can affect lithium abundance measurements. Several studies have published grids of NLTE corrections for lithium abundances, such as \citet{lind2009}, and more recently \citet{wang2021}. This NLTE-3D corrections affect mainly the cool-giants (up to +0.3\,dex) in the high-lithium regime. For metal-rich dwarfs, the typical correction is on the order of -0.1\,dex, for 5\,000<$\mathrm{\teff}$<6\,500\,K (see also Figures 1 \& 2 of \citealt{Magrini2021a}). A potential future task could be to include these NLTE corrections to the training-set lithium label, but we expect no major change in the results presented in this work. In the context of future surveys, 3D NLTE measurements should be performed homogeneously for as many elements as possible. For instance, $\alpha-$elements such as O, Mg, and Ti will be measurable by 4MIDABLE-HR and are affected by 3D NLTE in a non-negligible way \citep{Bergemann2021, Bergemann2017, Sitnova2018, Bergemann2012}.

Concerning the optimization of the training set, properly including M stars with strong TiO bands in the training set will allow us to accurately parametrize this type of object. It will be a necessity for 4MOST, which is planned to observe (among other targets) open-clusters.
Regarding the sensitivity of CNN to $\mathrm{\vrad}$, future surveys observing with multiple spectrographs should take care to provide accurate radial velocities in order to minimize the possible systematics during the training phase.

 On this study, we show that lithium abundances in solar-type stars with lithium lower than 1.25\,dex can not be measured precisely at the GIRAFFE HR15 resolution ($\sim20\,000$). For the future use of CNN (or ML in general) for stellar abundance measurements, it will be necessary to develop an objective criterion for deciding whether an abundance is a real detection or an upper limit.

\begin{acknowledgements}
We thank the anonymous referee for the very constructive comments and suggestions. This work was partly supported by the European Union FP7 programme through ERC grant number 320360 and by the Leverhulme Trust through grant RPG-2012-541. We acknowledge the support from INAF and Ministero dell' Istruzione, dell' Universit\`a' e della Ricerca (MIUR) in the form of the grant "Premiale VLT 2012". The results presented here benefit from discussions held during the Gaia-ESO workshops and conferences, supported by the ESF (European Science Foundation) through the GREAT Research Network Programme. This work has made use of data from the European Space Agency (ESA) mission {\it Gaia} (\url{https://www.cosmos.esa.int/gaia}), processed by the {\it Gaia} Data Processing and Analysis Consortium (DPAC, \url{https://www.cosmos.esa.int/web/gaia/dpac/consortium}). Funding for the DPAC has been provided by national institutions, in particular the institutions participating in the {\it Gaia} Multilateral Agreement. S.N. is grateful for the unwavering support of his family. M.~L.~L.~Dantas and R. Smiljanic acknowledge support by the National Science Centre, Poland, project 2019/34/E/ST9/00133. T.B. was supported by grant No. 2018-04857 from the Swedish Research Council. M.B. is supported through the Lise Meitner grant from the Max Planck Society. We acknowledge support by the Collaborative Research centre SFB 881 (projects A5, A10), Heidelberg University, of the Deutsche Forschungsgemeinschaft (DFG, German Research Foundation). This project has received funding from the European Research Council (ERC) under the European Union's Horizon 2020 research and innovation programme (Grant agreement No. 949173). This work made use of \textsc{overleaf}\footnote{\url{https://www.overleaf.com/}} for preparing this document, and of the following \textsc{python} packages (not previously mentioned): \textsc{matplotlib} \citep{Hunter2007}, \textsc{numpy} \citep{Harris2020}, \textsc{pandas} \citep{mckinney-proc-scipy-2010}, \textsc{seaborn} \citep{Waskom2021}. This work also benefited from \textsc{topcat} \citep{Taylor2005}.

\end{acknowledgements}

\bibliographystyle{aa}
\bibliography{nepal_lithium_ges}

\begin{thebibliography}{126}
\expandafter\ifx\csname natexlab\endcsname\relax\def\natexlab#1{#1}\fi

\bibitem[{Abadi {et~al.}(2015)Abadi, Agarwal, Barham, Brevdo, Chen, Citro,
  Corrado, Davis, Dean, Devin, Ghemawat, Goodfellow, Harp, Irving, Isard, Jia,
  Jozefowicz, Kaiser, Kudlur, Levenberg, Man\'{e}, Monga, Moore, Murray, Olah,
  Schuster, Shlens, Steiner, Sutskever, Talwar, Tucker, Vanhoucke, Vasudevan,
  Vi\'{e}gas, Vinyals, Warden, Wattenberg, Wicke, Yu, \&
  Zheng}]{tensorflow2015}
Abadi, M., Agarwal, A., Barham, P., {et~al.} 2015, {TensorFlow}: Large-Scale
  Machine Learning on Heterogeneous Systems, software available from
  tensorflow.org

\bibitem[{{Aguilera-G{\'o}mez} {et~al.}(2016){Aguilera-G{\'o}mez},
  {Chanam{\'e}}, {Pinsonneault}, \& {Carlberg}}]{AguileraGomez2016ApJ}
{Aguilera-G{\'o}mez}, C., {Chanam{\'e}}, J., {Pinsonneault}, M.~H., \&
  {Carlberg}, J.~K. 2016, \apj, 829, 127

\bibitem[{{Ambrosch} {et~al.}(2022){Ambrosch}, {Guiglion}, {Mikolaitis},
  {Chiappini}, {Tautvai{\v{s}}ien{\.{e}}}, {Nepal}, {Gilmore}, {Randich},
  {Bensby}, {Bergemann}, {Morbidelli}, {Pancino}, {Sacco}, {Smiljanic},
  {Zaggia}, {Jofr{\'e}}, \& {Jim{\'e}nez-Esteban}}]{Ambrosch_2022arXiv}
{Ambrosch}, M., {Guiglion}, G., {Mikolaitis}, {\v{S}}., {et~al.} 2022, arXiv
  e-prints, arXiv:2208.08872

\bibitem[{{Anders} {et~al.}(2018){Anders}, {Chiappini}, {Santiago},
  {Matijevi{\v{c}}}, {Queiroz}, {Steinmetz}, \& {Guiglion}}]{tSNE_Anders}
{Anders}, F., {Chiappini}, C., {Santiago}, B.~X., {et~al.} 2018, \aap, 619,
  A125

\bibitem[{{Bailer-Jones} {et~al.}(1997){Bailer-Jones}, {Irwin}, {Gilmore}, \&
  {von Hippel}}]{Bailer-Jones_1997MNRAS}
{Bailer-Jones}, C. A.~L., {Irwin}, M., {Gilmore}, G., \& {von Hippel}, T. 1997,
  \mnras, 292, 157

\bibitem[{{Bailer-Jones} {et~al.}(1998){Bailer-Jones}, {Irwin}, \& {von
  Hippel}}]{Bailer-Jones_1998MNRAS}
{Bailer-Jones}, C. A.~L., {Irwin}, M., \& {von Hippel}, T. 1998, \mnras, 298,
  361

\bibitem[{{Bensby} \& {Lind}(2018)}]{Bensby_2018}
{Bensby}, T. \& {Lind}, K. 2018, \aap, 615, A151

\bibitem[{{Bergemann} {et~al.}(2017){Bergemann}, {Collet}, {Amarsi}, {Kovalev},
  {Ruchti}, \& {Magic}}]{Bergemann2017}
{Bergemann}, M., {Collet}, R., {Amarsi}, A.~M., {et~al.} 2017, \apj, 847, 15

\bibitem[{{Bergemann} {et~al.}(2021){Bergemann}, {Hoppe}, {Semenova},
  {Carlsson}, {Yakovleva}, {Voronov}, {Bautista}, {Nemer}, {Belyaev},
  {Leenaarts}, {Mashonkina}, {Reiners}, \& {Ellwarth}}]{Bergemann2021}
{Bergemann}, M., {Hoppe}, R., {Semenova}, E., {et~al.} 2021, \mnras, 508, 2236

\bibitem[{{Bergemann} {et~al.}(2012){Bergemann}, {Lind}, {Collet}, {Magic}, \&
  {Asplund}}]{Bergemann2012}
{Bergemann}, M., {Lind}, K., {Collet}, R., {Magic}, Z., \& {Asplund}, M. 2012,
  \mnras, 427, 27

\bibitem[{{Bialek} {et~al.}(2020){Bialek}, {Fabbro}, {Venn}, {Kumar},
  {O'Briain}, \& {Yi}}]{Bialek_2020MNRAS}
{Bialek}, S., {Fabbro}, S., {Venn}, K.~A., {et~al.} 2020, \mnras, 498, 3817

\bibitem[{Bishop(1995)}]{Bishop1995}
Bishop, C.~M. 1995, Neural Networks for Pattern Recognition (USA: Oxford
  University Press, Inc.)

\bibitem[{{Blanco-Cuaresma} {et~al.}(2014){Blanco-Cuaresma}, {Soubiran},
  {Jofr{\'e}}, \& {Heiter}}]{gbs_2014_BC}
{Blanco-Cuaresma}, S., {Soubiran}, C., {Jofr{\'e}}, P., \& {Heiter}, U. 2014,
  \aap, 566, A98

\bibitem[{{Boesgaard} \& {Tripicco}(1986)}]{Boesgaard_1986}
{Boesgaard}, A.~M. \& {Tripicco}, M.~J. 1986, Astrophysical Journal Letters,
  302, L49

\bibitem[{{Bonifacio} \& {Molaro}(1997)}]{Bonifacio_1997MNRAS}
{Bonifacio}, P. \& {Molaro}, P. 1997, \mnras, 285, 847

\bibitem[{{Bragaglia} {et~al.}(2022){Bragaglia}, {Alfaro}, {Flaccomio},
  {Blomme}, {Donati}, {Costado}, {Damiani}, {Franciosini}, {Prisinzano},
  {Randich}, {Friel}, {Hatztidimitriou}, {Vallenari}, {Spagna},
  {Balaguer-Nunez}, {Bonito}, {Cantat Gaudin}, {Casamiquela}, {Jeffries},
  {Jordi}, {Magrini}, {Drew}, {Jackson}, {Abbas}, {Caramazza}, {Hayes},
  {Jim{\'e}nez-Esteban}, {Re Fiorentin}, {Wright}, {Bayo}, {Bensby},
  {Bergemann}, {Gilmore}, {Gonneau}, {Heiter}, {Hourihane}, {Pancino}, {Sacco},
  {Smiljanic}, {Zaggia}, \& {Vink}}]{Bragaglia_2022A&A}
{Bragaglia}, A., {Alfaro}, E.~J., {Flaccomio}, E., {et~al.} 2022, \aap, 659,
  A200

\bibitem[{{Brown} {et~al.}(1989){Brown}, {Sneden}, {Lambert}, \&
  {Dutchover}}]{Brown1989ApJS}
{Brown}, J.~A., {Sneden}, C., {Lambert}, D.~L., \& {Dutchover}, Edward, J.
  1989, \apjs, 71, 293

\bibitem[{{Buder} {et~al.}(2021){Buder}, {Sharma}, {Kos}, {Amarsi},
  {Nordlander}, {Lind}, {Martell}, {Asplund}, {Bland-Hawthorn}, {Casey}, {de
  Silva}, {D'Orazi}, {Freeman}, {Hayden}, {Lewis}, {Lin}, {Schlesinger},
  {Simpson}, {Stello}, {Zucker}, {Zwitter}, {Beeson}, {Buck}, {Casagrande},
  {Clark}, {{\v{C}}otar}, {da Costa}, {de Grijs}, {Feuillet}, {Horner},
  {Kafle}, {Khanna}, {Kobayashi}, {Liu}, {Montet}, {Nandakumar}, {Nataf},
  {Ness}, {Spina}, {Tepper-Garc{\'\i}a}, {Ting}, {Traven},
  {Vogrin{\v{c}}i{\v{c}}}, {Wittenmyer}, {Wyse}, {{\v{Z}}erjal}, \& {Galah
  Collaboration}}]{galah_dr3_2021}
{Buder}, S., {Sharma}, S., {Kos}, J., {et~al.} 2021, \mnras

\bibitem[{{Cameron} \& {Fowler}(1971)}]{Cameron1971}
{Cameron}, A.~G.~W. \& {Fowler}, W.~A. 1971, The Astrophysical Journal, 164,
  111

\bibitem[{{Casey} {et~al.}(2016){Casey}, {Ruchti}, {Masseron}, {Randich},
  {Gilmore}, {Lind}, {Kennedy}, {Koposov}, {Hourihane}, {Franciosini}, {Lewis},
  {Magrini}, {Morbidelli}, {Sacco}, {Worley}, {Feltzing}, {Jeffries},
  {Vallenari}, {Bensby}, {Bragaglia}, {Flaccomio}, {Francois}, {Korn},
  {Lanzafame}, {Pancino}, {Recio-Blanco}, {Smiljanic}, {Carraro}, {Costado},
  {Damiani}, {Donati}, {Frasca}, {Jofr{\'e}}, {Lardo}, {de Laverny}, {Monaco},
  {Prisinzano}, {Sbordone}, {Sousa}, {Tautvai{\v{s}}ien{\.{e}}}, {Zaggia},
  {Zwitter}, {Delgado Mena}, {Chorniy}, {Martell}, {Silva Aguirre}, {Miglio},
  {Chiappini}, {Montalban}, {Morel}, \& {Valentini}}]{Casey_2016MNRAS}
{Casey}, A.~R., {Ruchti}, G., {Masseron}, T., {et~al.} 2016, \mnras, 461, 3336

\bibitem[{{Castro-Ginard} {et~al.}(2020){Castro-Ginard}, {Jordi}, {Luri},
  {{\'A}lvarez Cid-Fuentes}, {Casamiquela}, {Anders}, {Cantat-Gaudin},
  {Mongui{\'o}}, {Balaguer-N{\'u}{\~n}ez}, {Sol{\`a}}, \&
  {Badia}}]{Castro-Ginard_2020A&A}
{Castro-Ginard}, A., {Jordi}, C., {Luri}, X., {et~al.} 2020, \aap, 635, A45

\bibitem[{{Cescutti} \& {Molaro}(2019)}]{Cescutti_2019}
{Cescutti}, G. \& {Molaro}, P. 2019, \mnras, 482, 4372

\bibitem[{{Charbonnel} \& {Balachandran}(2000)}]{Charbonnel_2000A&A}
{Charbonnel}, C. \& {Balachandran}, S.~C. 2000, \aap, 359, 563

\bibitem[{{Charbonnel} {et~al.}(2021){Charbonnel}, {Borisov}, {de Laverny}, \&
  {Prantzos}}]{Charbonnel_2021}
{Charbonnel}, C., {Borisov}, S., {de Laverny}, P., \& {Prantzos}, N. 2021,
  \aap, 649, L10

\bibitem[{Chollet {et~al.}(2015)}]{chollet2015keras}
Chollet, F. {et~al.} 2015, Keras, \url{https://keras.io}

\bibitem[{{Dalton}(2016)}]{WEAVE_2016}
{Dalton}, G. 2016, in Astronomical Society of the Pacific Conference Series,
  Vol. 507, Multi-Object Spectroscopy in the Next Decade: Big Questions, Large
  Surveys, and Wide Fields, ed. I.~{Skillen}, M.~{Balcells}, \& S.~{Trager}, 97

\bibitem[{{Damiani} {et~al.}(2014){Damiani}, {Prisinzano}, {Micela}, {Randich},
  {Gilmore}, {Drew}, {Jeffries}, {Fr{\'e}mat}, {Alfaro}, {Bensby}, {Bragaglia},
  {Flaccomio}, {Lanzafame}, {Pancino}, {Recio-Blanco}, {Sacco}, {Smiljanic},
  {Jackson}, {de Laverny}, {Morbidelli}, {Worley}, {Hourihane}, {Costado},
  {Jofr{\'e}}, {Lind}, \& {Maiorca}}]{Damiani2014}
{Damiani}, F., {Prisinzano}, L., {Micela}, G., {et~al.} 2014, \aap, 566, A50

\bibitem[{{Dantas} {et~al.}(2022){Dantas}, {Guiglion}, {Smiljanic}, {Romano},
  {Magrini}, {Bensby}, {Chiappini}, {Franciosini}, {Nepal},
  {Tautvai{\v{s}}ien{\.{e}}}, {Gilmore}, {Randich}, {Lanzafame}, {Heiter},
  {Morbidelli}, {Prisinzano}, \& {Zaggia}}]{Dantas_2022arXiv}
{Dantas}, M.~L.~L., {Guiglion}, G., {Smiljanic}, R., {et~al.} 2022, arXiv
  e-prints, arXiv:2211.14132

\bibitem[{{D'Antona} \& {Matteucci}(1991)}]{DAntona_1991A&A}
{D'Antona}, F. \& {Matteucci}, F. 1991, \aap, 248, 62

\bibitem[{{de Jong} {et~al.}(2019){de Jong}, {Agertz}, {Berbel}, {Aird},
  {Alexander}, {Amarsi}, {Anders}, {Andrae}, {Ansarinejad}, {Ansorge},
  {Antilogus}, {Anwand-Heerwart}, {Arentsen}, {Arnadottir}, {Asplund}, {Auger},
  {Azais}, {Baade}, {Baker}, {Baker}, {Balbinot}, {Baldry}, {Banerji},
  {Barden}, {Barklem}, {Barth{\'e}l{\'e}my-Mazot}, {Battistini}, {Bauer},
  {Bell}, {Bellido-Tirado}, {Bellstedt}, {Belokurov}, {Bensby}, {Bergemann},
  {Bestenlehner}, {Bielby}, {Bilicki}, {Blake}, {Bland-Hawthorn}, {Boeche},
  {Boland}, {Boller}, {Bongard}, {Bongiorno}, {Bonifacio}, {Boudon}, {Brooks},
  {Brown}, {Brown}, {Br{\"u}ggen}, {Brynnel}, {Brzeski}, {Buchert},
  {Buschkamp}, {Caffau}, {Caillier}, {Carrick}, {Casagrande}, {Case}, {Casey},
  {Cesarini}, {Cescutti}, {Chapuis}, {Chiappini}, {Childress}, {Christlieb},
  {Church}, {Cioni}, {Cluver}, {Colless}, {Collett}, {Comparat}, {Cooper},
  {Couch}, {Courbin}, {Croom}, {Croton}, {Daguis{\'e}}, {Dalton}, {Davies},
  {Davis}, {de Laverny}, {Deason}, {Dionies}, {Disseau}, {Doel}, {D{\"o}scher},
  {Driver}, {Dwelly}, {Eckert}, {Edge}, {Edvardsson}, {Youssoufi}, {Elhaddad},
  {Enke}, {Erfanianfar}, {Farrell}, {Fechner}, {Feiz}, {Feltzing}, {Ferreras},
  {Feuerstein}, {Feuillet}, {Finoguenov}, {Ford}, {Fotopoulou}, {Fouesneau},
  {Frenk}, {Frey}, {Gaessler}, {Geier}, {Gentile Fusillo}, {Gerhard},
  {Giannantonio}, {Giannone}, {Gibson}, {Gillingham},
  {Gonz{\'a}lez-Fern{\'a}ndez}, {Gonzalez-Solares}, {Gottloeber}, {Gould},
  {Grebel}, {Gueguen}, {Guiglion}, {Haehnelt}, {Hahn}, {Hansen}, {Hartman},
  {Hauptner}, {Hawkins}, {Haynes}, {Haynes}, {Heiter}, {Helmi}, {Aguayo},
  {Hewett}, {Hinton}, {Hobbs}, {Hoenig}, {Hofman}, {Hook}, {Hopgood},
  {Hopkins}, {Hourihane}, {Howes}, {Howlett}, {Huet}, {Irwin}, {Iwert},
  {Jablonka}, {Jahn}, {Jahnke}, {Jarno}, {Jin}, {Jofre}, {Johl}, {Jones},
  {J{\"o}nsson}, {Jordan}, {Karovicova}, {Khalatyan}, {Kelz}, {Kennicutt},
  {King}, {Kitaura}, {Klar}, {Klauser}, {Kneib}, {Koch}, {Koposov},
  {Kordopatis}, {Korn}, {Kosmalski}, {Kotak}, {Kovalev}, {Kreckel}, {Kripak},
  {Krumpe}, {Kuijken}, {Kunder}, {Kushniruk}, {Lam}, {Lamer}, {Laurent},
  {Lawrence}, {Lehmitz}, {Lemasle}, {Lewis}, {Li}, {Lidman}, {Lind}, {Liske},
  {Lizon}, {Loveday}, {Ludwig}, {McDermid}, {Maguire}, {Mainieri}, {Mali},
  {Mandel}, {Mandel}, {Mannering}, {Martell}, {Martinez Delgado}, {Matijevic},
  {McGregor}, {McMahon}, {McMillan}, {Mena}, {Merloni}, {Meyer}, {Michel},
  {Micheva}, {Migniau}, {Minchev}, {Monari}, {Muller}, {Murphy},
  {Muthukrishna}, {Nandra}, {Navarro}, {Ness}, {Nichani}, {Nichol}, {Nicklas},
  {Niederhofer}, {Norberg}, {Obreschkow}, {Oliver}, {Owers}, {Pai},
  {Pankratow}, {Parkinson}, {Paschke}, {Paterson}, {Pecontal}, {Parry},
  {Phillips}, {Pillepich}, {Pinard}, {Pirard}, {Piskunov}, {Plank},
  {Pl{\"u}schke}, {Pons}, {Popesso}, {Power}, {Pragt}, {Pramskiy}, {Pryer},
  {Quattri}, {Queiroz}, {Quirrenbach}, {Rahurkar}, {Raichoor}, {Ramstedt},
  {Rau}, {Recio-Blanco}, {Reiss}, {Renaud}, {Revaz}, {Rhode}, {Richard},
  {Richter}, {Rix}, {Robotham}, {Roelfsema}, {Romaniello}, {Rosario},
  {Rothmaier}, {Roukema}, {Ruchti}, {Rupprecht}, {Rybizki}, {Ryde}, {Saar},
  {Sadler}, {Sahl{\'e}n}, {Salvato}, {Sassolas}, {Saunders}, {Saviauk},
  {Sbordone}, {Schmidt}, {Schnurr}, {Scholz}, {Schwope}, {Seifert}, {Shanks},
  {Sheinis}, {Sivov}, {Sk{\'u}lad{\'o}ttir}, {Smartt}, {Smedley}, {Smith},
  {Smith}, {Sorce}, {Spitler}, {Starkenburg}, {Steinmetz}, {Stilz}, {Storm},
  {Sullivan}, {Sutherland}, {Swann}, {Tamone}, {Taylor}, {Teillon}, {Tempel},
  {ter Horst}, {Thi}, {Tolstoy}, {Trager}, {Traven}, {Tremblay}, {Tresse},
  {Valentini}, {van de Weygaert}, {van den Ancker}, {Veljanoski}, {Venkatesan},
  {Wagner}, {Wagner}, {Walcher}, {Waller}, {Walton}, {Wang}, {Winkler},
  {Wisotzki}, {Worley}, {Worseck}, {Xiang}, {Xu}, {Yong}, {Zhao}, {Zheng},
  {Zscheyge}, \& {Zucker}}]{deJong2019Msngr}
{de Jong}, R.~S., {Agertz}, O., {Berbel}, A.~A., {et~al.} 2019, The Messenger,
  175, 3

\bibitem[{{Deepak} \& {Lambert}(2021)}]{Deepak_2021MNRAS}
{Deepak} \& {Lambert}, D.~L. 2021, \mnras, 507, 205

\bibitem[{{Deepak} \& {Reddy}(2019)}]{Deepak_2019MNRAS}
{Deepak} \& {Reddy}, B.~E. 2019, \mnras, 484, 2000

\bibitem[{{Delgado Mena} {et~al.}(2015){Delgado Mena}, {Bertr{\'a}n de Lis},
  {Adibekyan}, {Sousa}, {Figueira}, {Mortier}, {Gonz{\'a}lez Hern{\'a}ndez},
  {Tsantaki}, {Israelian}, \& {Santos}}]{Delgado_2015}
{Delgado Mena}, E., {Bertr{\'a}n de Lis}, S., {Adibekyan}, V.~Z., {et~al.}
  2015, \aap, 576, A69

\bibitem[{{Deliyannis} {et~al.}(2019){Deliyannis}, {Anthony-Twarog},
  {Lee-Brown}, \& {Twarog}}]{Deliyannis_2019}
{Deliyannis}, C.~P., {Anthony-Twarog}, B.~J., {Lee-Brown}, D.~B., \& {Twarog},
  B.~A. 2019, \aj, 158, 163

\bibitem[{{Fabbro} {et~al.}(2018){Fabbro}, {Venn}, {O'Briain}, {Bialek},
  {Kielty}, {Jahandar}, \& {Monty}}]{Fabbro_2018MNRAS}
{Fabbro}, S., {Venn}, K.~A., {O'Briain}, T., {et~al.} 2018, \mnras, 475, 2978

\bibitem[{{Fields}(2011)}]{fields2011}
{Fields}, B.~D. 2011, Annual Review of Nuclear and Particle Science, 61, 47

\bibitem[{{Filipi Gon{\c{c}}alves dos Santos} \&
  {Papa}(2022)}]{dosSantos_2022arXiv}
{Filipi Gon{\c{c}}alves dos Santos}, C. \& {Papa}, J.~P. 2022, arXiv e-prints,
  arXiv:2201.03299

\bibitem[{{Franciosini} {et~al.}(2022){Franciosini}, {Randich}, {de Laverny},
  {Biazzo}, {Feuillet}, {Frasca}, {Lind}, {Prisinzano}, {Tautvai{\v{s}}iene},
  {Lanzafame}, {Smiljanic}, {Gonneau}, {Magrini}, {Pancino}, {Guiglion},
  {Sacco}, {Sanna}, {Gilmore}, {Bonifacio}, {Jeffries}, {Micela}, {Prusti},
  {Alfaro}, {Bensby}, {Bragaglia}, {Fran{\c{c}}ois}, {Korn}, {Van Eck}, {Bayo},
  {Bergemann}, {Carraro}, {Heiter}, {Hourihane}, {Jofr{\'e}}, {Lewis},
  {Martayan}, {Monaco}, {Morbidelli}, {Worley}, \& {Zaggia}}]{Franciosini_2022}
{Franciosini}, E., {Randich}, S., {de Laverny}, P., {et~al.} 2022, \aap, 668,
  A49

\bibitem[{{Fu} {et~al.}(2018){Fu}, {Romano}, {Bragaglia}, {Mucciarelli},
  {Lind}, {Delgado Mena}, {Sousa}, {Randich}, {Bressan}, {Sbordone}, {Martell},
  {Korn}, {Abia}, {Smiljanic}, {Jofr{\'e}}, {Pancino},
  {Tautvai{\v{s}}ien{\.{e}}}, {Tang}, {Magrini}, {Lanzafame}, {Carraro},
  {Bensby}, {Damiani}, {Alfaro}, {Flaccomio}, {Morbidelli}, {Zaggia}, {Lardo},
  {Monaco}, {Frasca}, {Donati}, {Drazdauskas}, {Chorniy}, {Bayo}, \&
  {Kordopatis}}]{Fu_2018}
{Fu}, X., {Romano}, D., {Bragaglia}, A., {et~al.} 2018, \aap, 610, A38

\bibitem[{Fukushima \& Miyake(1982)}]{fukushima1982neocognitron}
Fukushima, K. \& Miyake, S. 1982, in Competition and cooperation in neural nets
  (Springer), 267--285

\bibitem[{{Gaia Collaboration} {et~al.}(2021){Gaia Collaboration}, {Brown},
  {Vallenari}, {Prusti}, {de Bruijne}, {Babusiaux}, {Biermann}, {Creevey},
  {Evans}, {Eyer}, {Hutton}, {Jansen}, {Jordi}, {Klioner}, {Lammers},
  {Lindegren}, {Luri}, {Mignard}, {Panem}, {Pourbaix}, {Randich}, {Sartoretti},
  {Soubiran}, {Walton}, {Arenou}, {Bailer-Jones}, {Bastian}, {Cropper},
  {Drimmel}, {Katz}, {Lattanzi}, {van Leeuwen}, {Bakker}, {Cacciari},
  {Casta{\~n}eda}, {De Angeli}, {Ducourant}, {Fabricius}, {Fouesneau},
  {Fr{\'e}mat}, {Guerra}, {Guerrier}, {Guiraud}, {Jean-Antoine Piccolo},
  {Masana}, {Messineo}, {Mowlavi}, {Nicolas}, {Nienartowicz}, {Pailler},
  {Panuzzo}, {Riclet}, {Roux}, {Seabroke}, {Sordo}, {Tanga}, {Th{\'e}venin},
  {Gracia-Abril}, {Portell}, {Teyssier}, {Altmann}, {Andrae}, {Bellas-Velidis},
  {Benson}, {Berthier}, {Blomme}, {Brugaletta}, {Burgess}, {Busso}, {Carry},
  {Cellino}, {Cheek}, {Clementini}, {Damerdji}, {Davidson}, {Delchambre},
  {Dell'Oro}, {Fern{\'a}ndez-Hern{\'a}ndez}, {Galluccio}, {Garc{\'\i}a-Lario},
  {Garcia-Reinaldos}, {Gonz{\'a}lez-N{\'u}{\~n}ez}, {Gosset}, {Haigron},
  {Halbwachs}, {Hambly}, {Harrison}, {Hatzidimitriou}, {Heiter},
  {Hern{\'a}ndez}, {Hestroffer}, {Hodgkin}, {Holl}, {Jan{\ss}en}, {Jevardat de
  Fombelle}, {Jordan}, {Krone-Martins}, {Lanzafame}, {L{\"o}ffler}, {Lorca},
  {Manteiga}, {Marchal}, {Marrese}, {Moitinho}, {Mora}, {Muinonen}, {Osborne},
  {Pancino}, {Pauwels}, {Petit}, {Recio-Blanco}, {Richards}, {Riello},
  {Rimoldini}, {Robin}, {Roegiers}, {Rybizki}, {Sarro}, {Siopis}, {Smith},
  {Sozzetti}, {Ulla}, {Utrilla}, {van Leeuwen}, {van Reeven}, {Abbas}, {Abreu
  Aramburu}, {Accart}, {Aerts}, {Aguado}, {Ajaj}, {Altavilla}, {{\'A}lvarez},
  {{\'A}lvarez Cid-Fuentes}, {Alves}, {Anderson}, {Anglada Varela}, {Antoja},
  {Audard}, {Baines}, {Baker}, {Balaguer-N{\'u}{\~n}ez}, {Balbinot}, {Balog},
  {Barache}, {Barbato}, {Barros}, {Barstow}, {Bartolom{\'e}}, {Bassilana},
  {Bauchet}, {Baudesson-Stella}, {Becciani}, {Bellazzini}, {Bernet}, {Bertone},
  {Bianchi}, {Blanco-Cuaresma}, {Boch}, {Bombrun}, {Bossini}, {Bouquillon},
  {Bragaglia}, {Bramante}, {Breedt}, {Bressan}, {Brouillet}, {Bucciarelli},
  {Burlacu}, {Busonero}, {Butkevich}, {Buzzi}, {Caffau}, {Cancelliere},
  {C{\'a}novas}, {Cantat-Gaudin}, {Carballo}, {Carlucci}, {Carnerero},
  {Carrasco}, {Casamiquela}, {Castellani}, {Castro-Ginard}, {Castro Sampol},
  {Chaoul}, {Charlot}, {Chemin}, {Chiavassa}, {Cioni}, {Comoretto}, {Cooper},
  {Cornez}, {Cowell}, {Crifo}, {Crosta}, {Crowley}, {Dafonte}, {Dapergolas},
  {David}, {David}, {de Laverny}, {De Luise}, {De March}, {De Ridder}, {de
  Souza}, {de Teodoro}, {de Torres}, {del Peloso}, {del Pozo}, {Delbo},
  {Delgado}, {Delgado}, {Delisle}, {Di Matteo}, {Diakite}, {Diener},
  {Distefano}, {Dolding}, {Eappachen}, {Edvardsson}, {Enke}, {Esquej}, {Fabre},
  {Fabrizio}, {Faigler}, {Fedorets}, {Fernique}, {Fienga}, {Figueras},
  {Fouron}, {Fragkoudi}, {Fraile}, {Franke}, {Gai}, {Garabato},
  {Garcia-Gutierrez}, {Garc{\'\i}a-Torres}, {Garofalo}, {Gavras}, {Gerlach},
  {Geyer}, {Giacobbe}, {Gilmore}, {Girona}, {Giuffrida}, {Gomel}, {Gomez},
  {Gonzalez-Santamaria}, {Gonz{\'a}lez-Vidal}, {Granvik},
  {Guti{\'e}rrez-S{\'a}nchez}, {Guy}, {Hauser}, {Haywood}, {Helmi}, {Hidalgo},
  {Hilger}, {H{\l}adczuk}, {Hobbs}, {Holland}, {Huckle}, {Jasniewicz},
  {Jonker}, {Juaristi Campillo}, {Julbe}, {Karbevska}, {Kervella}, {Khanna},
  {Kochoska}, {Kontizas}, {Kordopatis}, {Korn}, {Kostrzewa-Rutkowska},
  {Kruszy{\'n}ska}, {Lambert}, {Lanza}, {Lasne}, {Le Campion}, {Le Fustec},
  {Lebreton}, {Lebzelter}, {Leccia}, {Leclerc}, {Lecoeur-Taibi}, {Liao},
  {Licata}, {Lindstr{\o}m}, {Lister}, {Livanou}, {Lobel}, {Madrero Pardo},
  {Managau}, {Mann}, {Marchant}, {Marconi}, {Marcos Santos}, {Marinoni},
  {Marocco}, {Marshall}, {Martin Polo}, {Mart{\'\i}n-Fleitas}, {Masip},
  {Massari}, {Mastrobuono-Battisti}, {Mazeh}, {McMillan}, {Messina},
  {Michalik}, {Millar}, {Mints}, {Molina}, {Molinaro}, {Moln{\'a}r},
  {Montegriffo}, {Mor}, {Morbidelli}, {Morel}, {Morris}, {Mulone}, {Munoz},
  {Muraveva}, {Murphy}, {Musella}, {Noval}, {Ord{\'e}novic}, {Orr{\`u}},
  {Osinde}, {Pagani}, {Pagano}, {Palaversa}, {Palicio}, {Panahi}, {Pawlak},
  {Pe{\~n}alosa Esteller}, {Penttil{\"a}}, {Piersimoni}, {Pineau}, {Plachy},
  {Plum}, {Poggio}, {Poretti}, {Poujoulet}, {Pr{\v{s}}a}, {Pulone}, {Racero},
  {Ragaini}, {Rainer}, {Raiteri}, {Rambaux}, {Ramos}, {Ramos-Lerate}, {Re
  Fiorentin}, {Regibo}, {Reyl{\'e}}, {Ripepi}, {Riva}, {Rixon}, {Robichon},
  {Robin}, {Roelens}, {Rohrbasser}, {Romero-G{\'o}mez}, {Rowell}, {Royer},
  {Rybicki}, {Sadowski}, {Sagrist{\`a} Sell{\'e}s}, {Sahlmann}, {Salgado},
  {Salguero}, {Samaras}, {Sanchez Gimenez}, {Sanna}, {Santove{\~n}a},
  {Sarasso}, {Schultheis}, {Sciacca}, {Segol}, {Segovia}, {S{\'e}gransan},
  {Semeux}, {Shahaf}, {Siddiqui}, {Siebert}, {Siltala}, {Slezak}, {Smart},
  {Solano}, {Solitro}, {Souami}, {Souchay}, {Spagna}, {Spoto}, {Steele},
  {Steidelm{\"u}ller}, {Stephenson}, {S{\"u}veges}, {Szabados}, {Szegedi-Elek},
  {Taris}, {Tauran}, {Taylor}, {Teixeira}, {Thuillot}, {Tonello}, {Torra},
  {Torra}, {Turon}, {Unger}, {Vaillant}, {van Dillen}, {Vanel}, {Vecchiato},
  {Viala}, {Vicente}, {Voutsinas}, {Weiler}, {Wevers}, {Wyrzykowski}, {Yoldas},
  {Yvard}, {Zhao}, {Zorec}, {Zucker}, {Zurbach}, \& {Zwitter}}]{gaia2020}
{Gaia Collaboration}, {Brown}, A.~G.~A., {Vallenari}, A., {et~al.} 2021, \aap,
  649, A1

\bibitem[{{Gaia Collaboration} {et~al.}(2016){Gaia Collaboration}, {Prusti},
  {de Bruijne}, {Brown}, {Vallenari}, {Babusiaux}, {Bailer-Jones}, {Bastian},
  {Biermann}, {Evans}, {Eyer}, {Jansen}, {Jordi}, {Klioner}, {Lammers},
  {Lindegren}, {Luri}, {Mignard}, {Milligan}, {Panem}, {Poinsignon},
  {Pourbaix}, {Randich}, {Sarri}, {Sartoretti}, {Siddiqui}, {Soubiran},
  {Valette}, {van Leeuwen}, {Walton}, {Aerts}, {Arenou}, {Cropper}, {Drimmel},
  {H{\o}g}, {Katz}, {Lattanzi}, {O'Mullane}, {Grebel}, {Holland}, {Huc},
  {Passot}, {Bramante}, {Cacciari}, {Casta{\~n}eda}, {Chaoul}, {Cheek}, {De
  Angeli}, {Fabricius}, {Guerra}, {Hern{\'a}ndez}, {Jean-Antoine-Piccolo},
  {Masana}, {Messineo}, {Mowlavi}, {Nienartowicz}, {Ord{\'o}{\~n}ez-Blanco},
  {Panuzzo}, {Portell}, {Richards}, {Riello}, {Seabroke}, {Tanga},
  {Th{\'e}venin}, {Torra}, {Els}, {Gracia-Abril}, {Comoretto},
  {Garcia-Reinaldos}, {Lock}, {Mercier}, {Altmann}, {Andrae}, {Astraatmadja},
  {Bellas-Velidis}, {Benson}, {Berthier}, {Blomme}, {Busso}, {Carry},
  {Cellino}, {Clementini}, {Cowell}, {Creevey}, {Cuypers}, {Davidson}, {De
  Ridder}, {de Torres}, {Delchambre}, {Dell'Oro}, {Ducourant}, {Fr{\'e}mat},
  {Garc{\'\i}a-Torres}, {Gosset}, {Halbwachs}, {Hambly}, {Harrison}, {Hauser},
  {Hestroffer}, {Hodgkin}, {Huckle}, {Hutton}, {Jasniewicz}, {Jordan},
  {Kontizas}, {Korn}, {Lanzafame}, {Manteiga}, {Moitinho}, {Muinonen},
  {Osinde}, {Pancino}, {Pauwels}, {Petit}, {Recio-Blanco}, {Robin}, {Sarro},
  {Siopis}, {Smith}, {Smith}, {Sozzetti}, {Thuillot}, {van Reeven}, {Viala},
  {Abbas}, {Abreu Aramburu}, {Accart}, {Aguado}, {Allan}, {Allasia},
  {Altavilla}, {{\'A}lvarez}, {Alves}, {Anderson}, {Andrei}, {Anglada Varela},
  {Antiche}, {Antoja}, {Ant{\'o}n}, {Arcay}, {Atzei}, {Ayache}, {Bach},
  {Baker}, {Balaguer-N{\'u}{\~n}ez}, {Barache}, {Barata}, {Barbier}, {Barblan},
  {Baroni}, {Barrado y Navascu{\'e}s}, {Barros}, {Barstow}, {Becciani},
  {Bellazzini}, {Bellei}, {Bello Garc{\'\i}a}, {Belokurov}, {Bendjoya},
  {Berihuete}, {Bianchi}, {Bienaym{\'e}}, {Billebaud}, {Blagorodnova},
  {Blanco-Cuaresma}, {Boch}, {Bombrun}, {Borrachero}, {Bouquillon}, {Bourda},
  {Bouy}, {Bragaglia}, {Breddels}, {Brouillet}, {Br{\"u}semeister},
  {Bucciarelli}, {Budnik}, {Burgess}, {Burgon}, {Burlacu}, {Busonero}, {Buzzi},
  {Caffau}, {Cambras}, {Campbell}, {Cancelliere}, {Cantat-Gaudin}, {Carlucci},
  {Carrasco}, {Castellani}, {Charlot}, {Charnas}, {Charvet}, {Chassat},
  {Chiavassa}, {Clotet}, {Cocozza}, {Collins}, {Collins}, {Costigan}, {Crifo},
  {Cross}, {Crosta}, {Crowley}, {Dafonte}, {Damerdji}, {Dapergolas}, {David},
  {David}, {De Cat}, {de Felice}, {de Laverny}, {De Luise}, {De March}, {de
  Martino}, {de Souza}, {Debosscher}, {del Pozo}, {Delbo}, {Delgado},
  {Delgado}, {di Marco}, {Di Matteo}, {Diakite}, {Distefano}, {Dolding}, {Dos
  Anjos}, {Drazinos}, {Dur{\'a}n}, {Dzigan}, {Ecale}, {Edvardsson}, {Enke},
  {Erdmann}, {Escolar}, {Espina}, {Evans}, {Eynard Bontemps}, {Fabre},
  {Fabrizio}, {Faigler}, {Falc{\~a}o}, {Farr{\`a}s Casas}, {Faye}, {Federici},
  {Fedorets}, {Fern{\'a}ndez-Hern{\'a}ndez}, {Fernique}, {Fienga}, {Figueras},
  {Filippi}, {Findeisen}, {Fonti}, {Fouesneau}, {Fraile}, {Fraser}, {Fuchs},
  {Furnell}, {Gai}, {Galleti}, {Galluccio}, {Garabato}, {Garc{\'\i}a-Sedano},
  {Gar{\'e}}, {Garofalo}, {Garralda}, {Gavras}, {Gerssen}, {Geyer}, {Gilmore},
  {Girona}, {Giuffrida}, {Gomes}, {Gonz{\'a}lez-Marcos},
  {Gonz{\'a}lez-N{\'u}{\~n}ez}, {Gonz{\'a}lez-Vidal}, {Granvik}, {Guerrier},
  {Guillout}, {Guiraud}, {G{\'u}rpide}, {Guti{\'e}rrez-S{\'a}nchez}, {Guy},
  {Haigron}, {Hatzidimitriou}, {Haywood}, {Heiter}, {Helmi}, {Hobbs},
  {Hofmann}, {Holl}, {Holland}, {Hunt}, {Hypki}, {Icardi}, {Irwin}, {Jevardat
  de Fombelle}, {Jofr{\'e}}, {Jonker}, {Jorissen}, {Julbe}, {Karampelas},
  {Kochoska}, {Kohley}, {Kolenberg}, {Kontizas}, {Koposov}, {Kordopatis},
  {Koubsky}, {Kowalczyk}, {Krone-Martins}, {Kudryashova}, {Kull}, {Bachchan},
  {Lacoste-Seris}, {Lanza}, {Lavigne}, {Le Poncin-Lafitte}, {Lebreton},
  {Lebzelter}, {Leccia}, {Leclerc}, {Lecoeur-Taibi}, {Lemaitre}, {Lenhardt},
  {Leroux}, {Liao}, {Licata}, {Lindstr{\o}m}, {Lister}, {Livanou}, {Lobel},
  {L{\"o}ffler}, {L{\'o}pez}, {Lopez-Lozano}, {Lorenz}, {Loureiro},
  {MacDonald}, {Magalh{\~a}es Fernandes}, {Managau}, {Mann}, {Mantelet},
  {Marchal}, {Marchant}, {Marconi}, {Marie}, {Marinoni}, {Marrese},
  {Marschalk{\'o}}, {Marshall}, {Mart{\'\i}n-Fleitas}, {Martino}, {Mary},
  {Matijevi{\v{c}}}, {Mazeh}, {McMillan}, {Messina}, {Mestre}, {Michalik},
  {Millar}, {Miranda}, {Molina}, {Molinaro}, {Molinaro}, {Moln{\'a}r},
  {Moniez}, {Montegriffo}, {Monteiro}, {Mor}, {Mora}, {Morbidelli}, {Morel},
  {Morgenthaler}, {Morley}, {Morris}, {Mulone}, {Muraveva}, {Musella},
  {Narbonne}, {Nelemans}, {Nicastro}, {Noval}, {Ord{\'e}novic},
  {Ordieres-Mer{\'e}}, {Osborne}, {Pagani}, {Pagano}, {Pailler}, {Palacin},
  {Palaversa}, {Parsons}, {Paulsen}, {Pecoraro}, {Pedrosa}, {Pentik{\"a}inen},
  {Pereira}, {Pichon}, {Piersimoni}, {Pineau}, {Plachy}, {Plum}, {Poujoulet},
  {Pr{\v{s}}a}, {Pulone}, {Ragaini}, {Rago}, {Rambaux}, {Ramos-Lerate},
  {Ranalli}, {Rauw}, {Read}, {Regibo}, {Renk}, {Reyl{\'e}}, {Ribeiro},
  {Rimoldini}, {Ripepi}, {Riva}, {Rixon}, {Roelens}, {Romero-G{\'o}mez},
  {Rowell}, {Royer}, {Rudolph}, {Ruiz-Dern}, {Sadowski}, {Sagrist{\`a}
  Sell{\'e}s}, {Sahlmann}, {Salgado}, {Salguero}, {Sarasso}, {Savietto},
  {Schnorhk}, {Schultheis}, {Sciacca}, {Segol}, {Segovia}, {Segransan},
  {Serpell}, {Shih}, {Smareglia}, {Smart}, {Smith}, {Solano}, {Solitro},
  {Sordo}, {Soria Nieto}, {Souchay}, {Spagna}, {Spoto}, {Stampa}, {Steele},
  {Steidelm{\"u}ller}, {Stephenson}, {Stoev}, {Suess}, {S{\"u}veges}, {Surdej},
  {Szabados}, {Szegedi-Elek}, {Tapiador}, {Taris}, {Tauran}, {Taylor},
  {Teixeira}, {Terrett}, {Tingley}, {Trager}, {Turon}, {Ulla}, {Utrilla},
  {Valentini}, {van Elteren}, {Van Hemelryck}, {van Leeuwen}, {Varadi},
  {Vecchiato}, {Veljanoski}, {Via}, {Vicente}, {Vogt}, {Voss}, {Votruba},
  {Voutsinas}, {Walmsley}, {Weiler}, {Weingrill}, {Werner}, {Wevers},
  {Whitehead}, {Wyrzykowski}, {Yoldas}, {{\v{Z}}erjal}, {Zucker}, {Zurbach},
  {Zwitter}, {Alecu}, {Allen}, {Allende Prieto}, {Amorim},
  {Anglada-Escud{\'e}}, {Arsenijevic}, {Azaz}, {Balm}, {Beck}, {Bernstein},
  {Bigot}, {Bijaoui}, {Blasco}, {Bonfigli}, {Bono}, {Boudreault}, {Bressan},
  {Brown}, {Brunet}, {Bunclark}, {Buonanno}, {Butkevich}, {Carret}, {Carrion},
  {Chemin}, {Ch{\'e}reau}, {Corcione}, {Darmigny}, {de Boer}, {de Teodoro}, {de
  Zeeuw}, {Delle Luche}, {Domingues}, {Dubath}, {Fodor}, {Fr{\'e}zouls},
  {Fries}, {Fustes}, {Fyfe}, {Gallardo}, {Gallegos}, {Gardiol}, {Gebran},
  {Gomboc}, {G{\'o}mez}, {Grux}, {Gueguen}, {Heyrovsky}, {Hoar}, {Iannicola},
  {Isasi Parache}, {Janotto}, {Joliet}, {Jonckheere}, {Keil}, {Kim},
  {Klagyivik}, {Klar}, {Knude}, {Kochukhov}, {Kolka}, {Kos}, {Kutka}, {Lainey},
  {LeBouquin}, {Liu}, {Loreggia}, {Makarov}, {Marseille}, {Martayan},
  {Martinez-Rubi}, {Massart}, {Meynadier}, {Mignot}, {Munari}, {Nguyen},
  {Nordlander}, {Ocvirk}, {O'Flaherty}, {Olias Sanz}, {Ortiz}, {Osorio},
  {Oszkiewicz}, {Ouzounis}, {Palmer}, {Park}, {Pasquato}, {Peltzer}, {Peralta},
  {P{\'e}turaud}, {Pieniluoma}, {Pigozzi}, {Poels}, {Prat}, {Prod'homme},
  {Raison}, {Rebordao}, {Risquez}, {Rocca-Volmerange}, {Rosen}, {Ruiz-Fuertes},
  {Russo}, {Sembay}, {Serraller Vizcaino}, {Short}, {Siebert}, {Silva},
  {Sinachopoulos}, {Slezak}, {Soffel}, {Sosnowska}, {Strai{\v{z}}ys}, {ter
  Linden}, {Terrell}, {Theil}, {Tiede}, {Troisi}, {Tsalmantza}, {Tur},
  {Vaccari}, {Vachier}, {Valles}, {Van Hamme}, {Veltz}, {Virtanen}, {Wallut},
  {Wichmann}, {Wilkinson}, {Ziaeepour}, \& {Zschocke}}]{gaia2016}
{Gaia Collaboration}, {Prusti}, T., {de Bruijne}, J.~H.~J., {et~al.} 2016,
  \aap, 595, A1

\bibitem[{{Gao} {et~al.}(2019){Gao}, {Shi}, {Yan}, {Yan}, {Xiang}, {Zhou},
  {Li}, \& {Zhao}}]{Gao_LAMOST_2019}
{Gao}, Q., {Shi}, J.-R., {Yan}, H.-L., {et~al.} 2019, The Astrophysical Journal
  Supplement Series, 245, 33

\bibitem[{{Gao} {et~al.}(2020){Gao}, {Lind}, {Amarsi}, {Buder},
  {Bland-Hawthorn}, {Campbell}, {Asplund}, {Casey}, {de Silva}, {Freeman},
  {Hayden}, {Lewis}, {Martell}, {Simpson}, {Sharma}, {Zucker}, {Zwitter},
  {Horner}, {Munari}, {Nordlander}, {Stello}, {Ting}, {Traven}, {Wittenmyer},
  \& {GALAH Collaboration}}]{Gao_GALAH_2020}
{Gao}, X., {Lind}, K., {Amarsi}, A.~M., {et~al.} 2020, Monthly Notices of the
  Royal Astronomical Society, 497, L30

\bibitem[{{Gilmore} {et~al.}(2012){Gilmore}, {Randich}, {Asplund}, {Binney},
  {Bonifacio}, {Drew}, {Feltzing}, {Ferguson}, {Jeffries}, {Micela},
  {Negueruela}, {Prusti}, {Rix}, {Vallenari}, {Alfaro}, {Allende-Prieto},
  {Babusiaux}, {Bensby}, {Blomme}, {Bragaglia}, {Flaccomio}, {Fran{\c{c}}ois},
  {Irwin}, {Koposov}, {Korn}, {Lanzafame}, {Pancino}, {Paunzen},
  {Recio-Blanco}, {Sacco}, {Smiljanic}, {Van Eck}, {Walton}, {Aden}, {Aerts},
  {Affer}, {Alcala}, {Altavilla}, {Alves}, {Antoja}, {Arenou}, {Argiroffi},
  {Asensio Ramos}, {Bailer-Jones}, {Balaguer-Nunez}, {Bayo}, {Barbuy},
  {Barisevicius}, {Barrado y Navascues}, {Battistini}, {Bellas Velidis},
  {Bellazzini}, {Belokurov}, {Bergemann}, {Bertelli}, {Biazzo}, {Bienayme},
  {Bland-Hawthorn}, {Boeche}, {Bonito}, {Boudreault}, {Bouvier}, {Brandao},
  {Brown}, {de Bruijne}, {Burleigh}, {Caballero}, {Caffau}, {Calura},
  {Capuzzo-Dolcetta}, {Caramazza}, {Carraro}, {Casagrande}, {Casewell},
  {Chapman}, {Chiappini}, {Chorniy}, {Christlieb}, {Cignoni}, {Cocozza},
  {Colless}, {Collet}, {Collins}, {Correnti}, {Covino}, {Crnojevic}, {Cropper},
  {Cunha}, {Damiani}, {David}, {Delgado}, {Duffau}, {Edvardsson}, {Eldridge},
  {Enke}, {Eriksson}, {Evans}, {Eyer}, {Famaey}, {Fellhauer}, {Ferreras},
  {Figueras}, {Fiorentino}, {Flynn}, {Folha}, {Franciosini}, {Frasca},
  {Freeman}, {Fremat}, {Friel}, {Gaensicke}, {Gameiro}, {Garzon}, {Geier},
  {Geisler}, {Gerhard}, {Gibson}, {Gomboc}, {Gomez}, {Gonzalez-Fernandez},
  {Gonzalez Hernandez}, {Gosset}, {Grebel}, {Greimel}, {Groenewegen},
  {Grundahl}, {Guarcello}, {Gustafsson}, {Hadrava}, {Hatzidimitriou}, {Hambly},
  {Hammersley}, {Hansen}, {Haywood}, {Heber}, {Heiter}, {Held}, {Helmi},
  {Hensler}, {Herrero}, {Hill}, {Hodgkin}, {Huelamo}, {Huxor}, {Ibata},
  {Jackson}, {de Jong}, {Jonker}, {Jordan}, {Jordi}, {Jorissen}, {Katz},
  {Kawata}, {Keller}, {Kharchenko}, {Klement}, {Klutsch}, {Knude}, {Koch},
  {Kochukhov}, {Kontizas}, {Koubsky}, {Lallement}, {de Laverny}, {van Leeuwen},
  {Lemasle}, {Lewis}, {Lind}, {Lindstrom}, {Lobel}, {Lopez Santiago}, {Lucas},
  {Ludwig}, {Lueftinger}, {Magrini}, {Maiz Apellaniz}, {Maldonado}, {Marconi},
  {Marino}, {Martayan}, {Martinez-Valpuesta}, {Matijevic}, {McMahon},
  {Messina}, {Meyer}, {Miglio}, {Mikolaitis}, {Minchev}, {Minniti}, {Moitinho},
  {Momany}, {Monaco}, {Montalto}, {Monteiro}, {Monier}, {Montes}, {Mora},
  {Moraux}, {Morel}, {Mowlavi}, {Mucciarelli}, {Munari}, {Napiwotzki},
  {Nardetto}, {Naylor}, {Naze}, {Nelemans}, {Okamoto}, {Ortolani}, {Pace},
  {Palla}, {Palous}, {Parker}, {Penarrubia}, {Pillitteri}, {Piotto}, {Posbic},
  {Prisinzano}, {Puzeras}, {Quirrenbach}, {Ragaini}, {Read}, {Read}, {Reyle},
  {De Ridder}, {Robichon}, {Robin}, {Roeser}, {Romano}, {Royer}, {Ruchti},
  {Ruzicka}, {Ryan}, {Ryde}, {Santos}, {Sanz Forcada}, {Sarro Baro},
  {Sbordone}, {Schilbach}, {Schmeja}, {Schnurr}, {Schoenrich}, {Scholz},
  {Seabroke}, {Sharma}, {De Silva}, {Smith}, {Solano}, {Sordo}, {Soubiran},
  {Sousa}, {Spagna}, {Steffen}, {Steinmetz}, {Stelzer}, {Stempels},
  {Tabernero}, {Tautvaisiene}, {Thevenin}, {Torra}, {Tosi}, {Tolstoy}, {Turon},
  {Walker}, {Wambsganss}, {Worley}, {Venn}, {Vink}, {Wyse}, {Zaggia},
  {Zeilinger}, {Zoccali}, {Zorec}, {Zucker}, {Zwitter}, \& {Gaia-ESO Survey
  Team}}]{GES_2012Msngr}
{Gilmore}, G., {Randich}, S., {Asplund}, M., {et~al.} 2012, The Messenger, 147,
  25

\bibitem[{{Gilmore} {et~al.}(2022){Gilmore}, {Randich}, {Worley}, {Hourihane},
  {Gonneau}, {Sacco}, {Lewis}, {Magrini}, {Fran{\c{c}}ois}, {Jeffries},
  {Koposov}, {Bragaglia}, {Alfaro}, {Allende Prieto}, {Blomme}, {Korn},
  {Lanzafame}, {Pancino}, {Recio-Blanco}, {Smiljanic}, {Van Eck}, {Zwitter},
  {Bensby}, {Flaccomio}, {Irwin}, {Franciosini}, {Morbidelli}, {Damiani},
  {Bonito}, {Friel}, {Vink}, {Prisinzano}, {Abbas}, {Hatzidimitriou}, {Held},
  {Jordi}, {Paunzen}, {Spagna}, {Jackson}, {Ma{\'\i}z Apell{\'a}niz},
  {Asplund}, {Bonifacio}, {Feltzing}, {Binney}, {Drew}, {Ferguson}, {Micela},
  {Negueruela}, {Prusti}, {Rix}, {Vallenari}, {Bergemann}, {Casey}, {de
  Laverny}, {Frasca}, {Hill}, {Lind}, {Sbordone}, {Sousa}, {Adibekyan},
  {Caffau}, {Daflon}, {Feuillet}, {Gebran}, {Gonzalez Hernandez}, {Guiglion},
  {Herrero}, {Lobel}, {Merle}, {Mikolaitis}, {Montes}, {Morel}, {Ruchti},
  {Soubiran}, {Tabernero}, {Tautvai{\v{s}}ien{\.{e}}}, {Traven}, {Valentini},
  {Van der Swaelmen}, {Villanova}, {Viscasillas V{\'a}zquez}, {Bayo}, {Biazzo},
  {Carraro}, {Edvardsson}, {Heiter}, {Jofr{\'e}}, {Marconi}, {Martayan},
  {Masseron}, {Monaco}, {Walton}, {Zaggia}, {Aguirre B{\o}rsen-Koch}, {Alves},
  {Balaguer-Nunez}, {Barklem}, {Barrado}, {Bellazzini}, {Berlanas}, {Binks},
  {Bressan}, {Capuzzo-Dolcetta}, {Casagrande}, {Casamiquela}, {Collins},
  {D'Orazi}, {Dantas}, {Debattista}, {Delgado-Mena}, {Di Marcantonio},
  {Drazdauskas}, {Evans}, {Famaey}, {Franchini}, {Fr{\'e}mat}, {Fu}, {Geisler},
  {Gerhard}, {Gonz{\'a}lez Solares}, {Grebel}, {Guti{\'e}rrez Albarr{\'a}n},
  {Jim{\'e}nez-Esteban}, {J{\"o}nsson}, {Khachaturyants}, {Kordopatis}, {Kos},
  {Lagarde}, {Ludwig}, {Mahy}, {Mapelli}, {Marfil}, {Martell}, {Messina},
  {Miglio}, {Minchev}, {Moitinho}, {Montalban}, {Monteiro}, {Morossi},
  {Mowlavi}, {Mucciarelli}, {Murphy}, {Nardetto}, {Ortolani}, {Paletou},
  {Palou{\v{s}}}, {Pickering}, {Quirrenbach}, {Re Fiorentin}, {Read}, {Romano},
  {Ryde}, {Sanna}, {Santos}, {Seabroke}, {Spina}, {Steinmetz}, {Stonkut{\'e}},
  {Sutorius}, {Th{\'e}venin}, {Tosi}, {Tsantaki}, {Wright}, {Wyse}, {Zoccali},
  {Zorec}, \& {Zucker}}]{Gilmore2022}
{Gilmore}, G., {Randich}, S., {Worley}, C.~C., {et~al.} 2022, \aap, 666, A120

\bibitem[{{Gratton} \& {D'Antona}(1989)}]{Gratton_1989A&A}
{Gratton}, R.~G. \& {D'Antona}, F. 1989, \aap, 215, 66

\bibitem[{{Grevesse} {et~al.}(2007){Grevesse}, {Asplund}, \&
  {Sauval}}]{Grevesse_2007SSRv}
{Grevesse}, N., {Asplund}, M., \& {Sauval}, A.~J. 2007, \ssr, 130, 105

\bibitem[{{Guiglion} {et~al.}(2019){Guiglion}, {Chiappini}, {Romano},
  {Matteucci}, {Anders}, {Steinmetz}, {Minchev}, {de Laverny}, \&
  {Recio-Blanco}}]{guiglion2019}
{Guiglion}, G., {Chiappini}, C., {Romano}, D., {et~al.} 2019, \aap, 623, A99

\bibitem[{{Guiglion} {et~al.}(2016){Guiglion}, {de Laverny}, {Recio-Blanco},
  {Worley}, {De Pascale}, {Masseron}, {Prantzos}, \&
  {Mikolaitis}}]{guiglion2016}
{Guiglion}, G., {de Laverny}, P., {Recio-Blanco}, A., {et~al.} 2016, \aap, 595,
  A18

\bibitem[{{Guiglion} {et~al.}(2020){Guiglion}, {Matijevi{\v{c}}}, {Queiroz},
  {Valentini}, {Steinmetz}, {Chiappini}, {Grebel}, {McMillan}, {Kordopatis},
  {Kunder}, {Zwitter}, {Khalatyan}, {Anders}, {Enke}, {Minchev}, {Monari},
  {Wyse}, {Bienaym{\'e}}, {Bland-Hawthorn}, {Gibson}, {Navarro}, {Parker},
  {Reid}, {Seabroke}, \& {Siebert}}]{guiglion2020}
{Guiglion}, G., {Matijevi{\v{c}}}, G., {Queiroz}, A.~B.~A., {et~al.} 2020,
  \aap, 644, A168

\bibitem[{Harris {et~al.}(2020)Harris, Millman, van~der Walt, Gommers,
  Virtanen, Cournapeau, Wieser, Taylor, Berg, Smith, Kern, Picus, Hoyer, van
  Kerkwijk, Brett, Haldane, del R{\'{i}}o, Wiebe, Peterson,
  G{\'{e}}rard-Marchant, Sheppard, Reddy, Weckesser, Abbasi, Gohlke, \&
  Oliphant}]{Harris2020}
Harris, C.~R., Millman, K.~J., van~der Walt, S.~J., {et~al.} 2020, Nature, 585,
  357

\bibitem[{{Heiter} {et~al.}(2015){Heiter}, {Jofr{\'e}}, {Gustafsson}, {Korn},
  {Soubiran}, \& {Th{\'e}venin}}]{gbs_2015_heiter}
{Heiter}, U., {Jofr{\'e}}, P., {Gustafsson}, B., {et~al.} 2015, \aap, 582, A49

\bibitem[{{Heiter} {et~al.}(2021){Heiter}, {Lind}, {Bergemann}, {Asplund},
  {Mikolaitis}, {Barklem}, {Masseron}, {de Laverny}, {Magrini}, {Edvardsson},
  {J{\"o}nsson}, {Pickering}, {Ryde}, {Bayo Ar{\'a}n}, {Bensby}, {Casey},
  {Feltzing}, {Jofr{\'e}}, {Korn}, {Pancino}, {Damiani}, {Lanzafame}, {Lardo},
  {Monaco}, {Morbidelli}, {Smiljanic}, {Worley}, {Zaggia}, {Randich}, \&
  {Gilmore}}]{heiter_2021}
{Heiter}, U., {Lind}, K., {Bergemann}, M., {et~al.} 2021, \aap, 645, A106

\bibitem[{{Hong-liang} \& {Jian-rong}(2022)}]{Hong-liang2022ChA&A}
{Hong-liang}, Y. \& {Jian-rong}, S. 2022, \caa, 46, 1

\bibitem[{{Hunter}(2007)}]{Hunter2007}
{Hunter}, J.~D. 2007, Computing in Science and Engineering, 9, 90

\bibitem[{{Izzo} {et~al.}(2015){Izzo}, {Della Valle}, {Mason}, {Matteucci},
  {Romano}, {Pasquini}, {Vanzi}, {Jordan}, {Fernandez}, {Bluhm}, {Brahm},
  {Espinoza}, \& {Williams}}]{Izzo_2015ApJ}
{Izzo}, L., {Della Valle}, M., {Mason}, E., {et~al.} 2015, \apjl, 808, L14

\bibitem[{{Jackson} {et~al.}(2015){Jackson}, {Jeffries}, {Lewis}, {Koposov},
  {Sacco}, {Randich}, {Gilmore}, {Asplund}, {Binney}, {Bonifacio}, {Drew},
  {Feltzing}, {Ferguson}, {Micela}, {Neguerela}, {Prusti}, {Rix}, {Vallenari},
  {Alfaro}, {Allende Prieto}, {Babusiaux}, {Bensby}, {Blomme}, {Bragaglia},
  {Flaccomio}, {Francois}, {Hambly}, {Irwin}, {Korn}, {Lanzafame}, {Pancino},
  {Recio-Blanco}, {Smiljanic}, {Van Eck}, {Walton}, {Bayo}, {Bergemann},
  {Carraro}, {Costado}, {Damiani}, {Edvardsson}, {Franciosini}, {Frasca},
  {Heiter}, {Hill}, {Hourihane}, {Jofr{\'e}}, {Lardo}, {de Laverny}, {Lind},
  {Magrini}, {Marconi}, {Martayan}, {Masseron}, {Monaco}, {Morbidelli},
  {Prisinzano}, {Sbordone}, {Sousa}, {Worley}, \& {Zaggia}}]{Jackson2015}
{Jackson}, R.~J., {Jeffries}, R.~D., {Lewis}, J., {et~al.} 2015, \aap, 580, A75

\bibitem[{{Jofr{\'e}} {et~al.}(2015){Jofr{\'e}}, {Heiter}, {Soubiran},
  {Blanco-Cuaresma}, {Masseron}, {Nordlander}, {Chemin}, {Worley}, {Van Eck},
  {Hourihane}, {Gilmore}, {Adibekyan}, {Bergemann}, {Cantat-Gaudin},
  {Delgado-Mena}, {Gonz{\'a}lez Hern{\'a}ndez}, {Guiglion}, {Lardo}, {de
  Laverny}, {Lind}, {Magrini}, {Mikolaitis}, {Montes}, {Pancino},
  {Recio-Blanco}, {Sordo}, {Sousa}, {Tabernero}, \&
  {Vallenari}}]{Jofre2015_alpha}
{Jofr{\'e}}, P., {Heiter}, U., {Soubiran}, C., {et~al.} 2015, \aap, 582, A81

\bibitem[{{Jofr{\'e}} {et~al.}(2014){Jofr{\'e}}, {Heiter}, {Soubiran},
  {Blanco-Cuaresma}, {Worley}, {Pancino}, {Cantat-Gaudin}, {Magrini},
  {Bergemann}, {Gonz{\'a}lez Hern{\'a}ndez}, {Hill}, {Lardo}, {de Laverny},
  {Lind}, {Masseron}, {Montes}, {Mucciarelli}, {Nordlander}, {Recio Blanco},
  {Sobeck}, {Sordo}, {Sousa}, {Tabernero}, {Vallenari}, \& {Van
  Eck}}]{gbs_2014_jofre}
{Jofr{\'e}}, P., {Heiter}, U., {Soubiran}, C., {et~al.} 2014, \aap, 564, A133

\bibitem[{{Jofr{\'e}} {et~al.}(2018){Jofr{\'e}}, {Heiter}, {Tucci Maia},
  {Soubiran}, {Worley}, {Hawkins}, {Blanco-Cuaresma}, \&
  {Rodrigo}}]{gbs_v2_2018}
{Jofr{\'e}}, P., {Heiter}, U., {Tucci Maia}, M., {et~al.} 2018, Research Notes
  of the American Astronomical Society, 2, 152

\bibitem[{{Kingma} \& {Ba}(2014)}]{Kingma_ADAM_2014arXiv}
{Kingma}, D.~P. \& {Ba}, J. 2014, arXiv e-prints, arXiv:1412.6980

\bibitem[{{Kusakabe} {et~al.}(2019){Kusakabe}, {Cheoun}, {Kim}, {Hashimoto},
  {Ono}, {Nomoto}, {Suzuki}, {Kajino}, \& {Mathews}}]{Kusakabe2019}
{Kusakabe}, M., {Cheoun}, M.-K., {Kim}, K.~S., {et~al.} 2019, \apj, 872, 164

\bibitem[{{Lagarde} {et~al.}(2012){Lagarde}, {Decressin}, {Charbonnel},
  {Eggenberger}, {Ekstr{\"o}m}, \& {Palacios}}]{Lagarde_2012A&A}
{Lagarde}, N., {Decressin}, T., {Charbonnel}, C., {et~al.} 2012, \aap, 543,
  A108

\bibitem[{{Lambert} \& {Reddy}(2004)}]{Lambert_2004}
{Lambert}, D.~L. \& {Reddy}, B.~E. 2004, \mnras, 349, 757

\bibitem[{{Lanzafame} {et~al.}(2015){Lanzafame}, {Frasca}, {Damiani},
  {Franciosini}, {Cottaar}, {Sousa}, {Tabernero}, {Klutsch}, {Spina}, {Biazzo},
  {Prisinzano}, {Sacco}, {Randich}, {Brugaletta}, {Delgado Mena}, {Adibekyan},
  {Montes}, {Bonito}, {Gameiro}, {Alcal{\'a}}, {Gonz{\'a}lez Hern{\'a}ndez},
  {Jeffries}, {Messina}, {Meyer}, {Gilmore}, {Asplund}, {Binney}, {Bonifacio},
  {Drew}, {Feltzing}, {Ferguson}, {Micela}, {Negueruela}, {Prusti}, {Rix},
  {Vallenari}, {Alfaro}, {Allende Prieto}, {Babusiaux}, {Bensby}, {Blomme},
  {Bragaglia}, {Flaccomio}, {Francois}, {Hambly}, {Irwin}, {Koposov}, {Korn},
  {Smiljanic}, {Van Eck}, {Walton}, {Bayo}, {Bergemann}, {Carraro}, {Costado},
  {Edvardsson}, {Heiter}, {Hill}, {Hourihane}, {Jackson}, {Jofr{\'e}}, {Lardo},
  {Lewis}, {Lind}, {Magrini}, {Marconi}, {Martayan}, {Masseron}, {Monaco},
  {Morbidelli}, {Sbordone}, {Worley}, \& {Zaggia}}]{Lanzafame2015}
{Lanzafame}, A.~C., {Frasca}, A., {Damiani}, F., {et~al.} 2015, \aap, 576, A80

\bibitem[{LeCun \& Bengio(1995)}]{lecun-bengio-95a}
LeCun, Y. \& Bengio, Y. 1995, in The Handbook of Brain Theory and Neural
  Networks, ed. M.~A. Arbib (MIT Press)

\bibitem[{{Lecun} {et~al.}(2015){Lecun}, {Bengio}, \&
  {Hinton}}]{Lecun_2015Natur}
{Lecun}, Y., {Bengio}, Y., \& {Hinton}, G. 2015, \nat, 521, 436

\bibitem[{LeCun {et~al.}(1989)LeCun, Boser, Denker, Henderson, Howard, Hubbard,
  \& Jackel}]{LeCun_backpropagation_1989}
LeCun, Y., Boser, B., Denker, J.~S., {et~al.} 1989, Neural Computation, 1, 541

\bibitem[{{Leung} \& {Bovy}(2019)}]{Leung_2019MNRAS}
{Leung}, H.~W. \& {Bovy}, J. 2019, \mnras, 483, 3255

\bibitem[{{Lima} {et~al.}(2022){Lima}, {Sodr{\'e}}, {Bom}, {Teixeira},
  {Nakazono}, {Buzzo}, {Queiroz}, {Herpich}, {Castellon}, {Dantas}, {Dors},
  {Souza}, {Akras}, {Jim{\'e}nez-Teja}, {Kanaan}, {Ribeiro}, \&
  {Schoennell}}]{Lima2022A&C}
{Lima}, E.~V.~R., {Sodr{\'e}}, L., {Bom}, C.~R., {et~al.} 2022, Astronomy and
  Computing, 38, 100510

\bibitem[{{Lin} \& {Wu}(2021)}]{Lin_2021PhRvD}
{Lin}, Y.-C. \& {Wu}, J.-H.~P. 2021, \prd, 103, 063034

\bibitem[{{Lind} {et~al.}(2009){Lind}, {Asplund}, \& {Barklem}}]{lind2009}
{Lind}, K., {Asplund}, M., \& {Barklem}, P.~S. 2009, \aap, 503, 541

\bibitem[{{Lindegren} {et~al.}(2021){Lindegren}, {Klioner}, {Hern{\'a}ndez},
  {Bombrun}, {Ramos-Lerate}, {Steidelm{\"u}ller}, {Bastian}, {Biermann}, {de
  Torres}, {Gerlach}, {Geyer}, {Hilger}, {Hobbs}, {Lammers}, {McMillan},
  {Stephenson}, {Casta{\~n}eda}, {Davidson}, {Fabricius}, {Gracia-Abril},
  {Portell}, {Rowell}, {Teyssier}, {Torra}, {Bartolom{\'e}}, {Clotet},
  {Garralda}, {Gonz{\'a}lez-Vidal}, {Torra}, {Abbas}, {Altmann}, {Anglada
  Varela}, {Balaguer-N{\'u}{\~n}ez}, {Balog}, {Barache}, {Becciani}, {Bernet},
  {Bertone}, {Bianchi}, {Bouquillon}, {Brown}, {Bucciarelli}, {Busonero},
  {Butkevich}, {Buzzi}, {Cancelliere}, {Carlucci}, {Charlot}, {Cioni},
  {Crosta}, {Crowley}, {del Peloso}, {del Pozo}, {Drimmel}, {Esquej}, {Fienga},
  {Fraile}, {Gai}, {Garcia-Reinaldos}, {Guerra}, {Hambly}, {Hauser},
  {Jan{\ss}en}, {Jordan}, {Kostrzewa-Rutkowska}, {Lattanzi}, {Liao}, {Licata},
  {Lister}, {L{\"o}ffler}, {Marchant}, {Masip}, {Mignard}, {Mints}, {Molina},
  {Mora}, {Morbidelli}, {Murphy}, {Pagani}, {Panuzzo}, {Pe{\~n}alosa Esteller},
  {Poggio}, {Re Fiorentin}, {Riva}, {Sagrist{\`a} Sell{\'e}s}, {Sanchez
  Gimenez}, {Sarasso}, {Sciacca}, {Siddiqui}, {Smart}, {Souami}, {Spagna},
  {Steele}, {Taris}, {Utrilla}, {van Reeven}, \& {Vecchiato}}]{Lindegren2021}
{Lindegren}, L., {Klioner}, S.~A., {Hern{\'a}ndez}, J., {et~al.} 2021, \aap,
  649, A2

\bibitem[{{Lodders} \& {Palme}(2009)}]{Lodders_2009M&PSA}
{Lodders}, K. \& {Palme}, H. 2009, Meteoritics and Planetary Science
  Supplement, 72, 5154

\bibitem[{{Magrini} {et~al.}(2021{\natexlab{a}}){Magrini}, {Lagarde},
  {Charbonnel}, {Franciosini}, {Randich}, {Smiljanic}, {Casali}, {Viscasillas
  V{\'a}zquez}, {Spina}, {Biazzo}, {Pasquini}, {Bragaglia}, {Van der Swaelmen},
  {Tautvai{\v{s}}ien{\.{e}}}, {Inno}, {Sanna}, {Prisinzano}, {Degl'Innocenti},
  {Prada Moroni}, {Roccatagliata}, {Tognelli}, {Monaco}, {de Laverny},
  {Delgado-Mena}, {Baratella}, {D'Orazi}, {Vallenari}, {Gonneau}, {Worley},
  {Jim{\'e}nez-Esteban}, {Jofre}, {Bensby}, {Fran{\c{c}}ois}, {Guiglion},
  {Bayo}, {Jeffries}, {Binks}, {Gilmore}, {Damiani}, {Korn}, {Pancino},
  {Sacco}, {Hourihane}, {Morbidelli}, \& {Zaggia}}]{Magrini2021a}
{Magrini}, L., {Lagarde}, N., {Charbonnel}, C., {et~al.} 2021{\natexlab{a}},
  \aap, 651, A84

\bibitem[{{Magrini} {et~al.}(2021{\natexlab{b}}){Magrini}, {Smiljanic},
  {Franciosini}, {Pasquini}, {Randich}, {Casali}, {Viscasillas V{\'a}zquez},
  {Bragaglia}, {Spina}, {Biazzo}, {Tautvai{\v{s}}ien{\.{e}}}, {Masseron}, {Van
  der Swaelmen}, {Pancino}, {Jim{\'e}nez-Esteban}, {Guiglion}, {Martell},
  {Bensby}, {D'Orazi}, {Baratella}, {Korn}, {Jofre}, {Gilmore}, {Worley},
  {Hourihane}, {Gonneau}, {Sacco}, \& {Morbidelli}}]{magrini_2021A&A}
{Magrini}, L., {Smiljanic}, R., {Franciosini}, E., {et~al.} 2021{\natexlab{b}},
  \aap, 655, A23

\bibitem[{{Margalef-Bentabol} {et~al.}(2020){Margalef-Bentabol},
  {Huertas-Company}, {Charnock}, {Margalef-Bentabol}, {Bernardi}, {Dubois},
  {Storey-Fisher}, \& {Zanisi}}]{Margalef-Bentabol_2020MNRAS}
{Margalef-Bentabol}, B., {Huertas-Company}, M., {Charnock}, T., {et~al.} 2020,
  \mnras, 496, 2346

\bibitem[{{Martell} {et~al.}(2021){Martell}, {Simpson}, {Balasubramaniam},
  {Buder}, {Sharma}, {Hon}, {Stello}, {Ting}, {Asplund}, {Bland-Hawthorn}, {De
  Silva}, {Freeman}, {Hayden}, {Kos}, {Lewis}, {Lind}, {Zucker}, {Zwitter},
  {Campbell}, {{\v{C}}otar}, {Horner}, {Montet}, \&
  {Wittenmyer}}]{Martell_2021MNRAS}
{Martell}, S.~L., {Simpson}, J.~D., {Balasubramaniam}, A.~G., {et~al.} 2021,
  \mnras, 505, 5340

\bibitem[{{Matijevi{\v{c}}} {et~al.}(2017){Matijevi{\v{c}}}, {Chiappini},
  {Grebel}, {Wyse}, {Zwitter}, {Bienaym{\'e}}, {Bland-Hawthorn}, {Freeman},
  {Gibson}, {Gilmore}, {Helmi}, {Kordopatis}, {Kunder}, {Munari}, {Navarro},
  {Parker}, {Reid}, {Seabroke}, {Siviero}, {Steinmetz}, \&
  {Watson}}]{Matijevic_2017A&A}
{Matijevi{\v{c}}}, G., {Chiappini}, C., {Grebel}, E.~K., {et~al.} 2017, \aap,
  603, A19

\bibitem[{{Matteucci} {et~al.}(1995){Matteucci}, {D'Antona}, \&
  {Timmes}}]{matteucci1995}
{Matteucci}, F., {D'Antona}, F., \& {Timmes}, F.~X. 1995, \aap, 303, 460

\bibitem[{{McKellar}(1940)}]{McKellar_1940PASP}
{McKellar}, A. 1940, \pasp, 52, 407

\bibitem[{{M}c{K}inney(2010)}]{mckinney-proc-scipy-2010}
{M}c{K}inney, W. 2010, in {P}roceedings of the 9th {P}ython in {S}cience
  {C}onference, ed. {S}t\'efan van~der {W}alt \& {J}arrod {M}illman, 56 -- 61

\bibitem[{{Miglio} {et~al.}(2021){Miglio}, {Chiappini}, {Mackereth}, {Davies},
  {Brogaard}, {Casagrande}, {Chaplin}, {Girardi}, {Kawata}, {Khan}, {Izzard},
  {Montalb{\'a}n}, {Mosser}, {Vincenzo}, {Bossini}, {Noels}, {Rodrigues},
  {Valentini}, \& {Mandel}}]{Miglio_2021}
{Miglio}, A., {Chiappini}, C., {Mackereth}, J.~T., {et~al.} 2021, \aap, 645,
  A85

\bibitem[{{Minchev} {et~al.}(2018){Minchev}, {Anders}, {Recio-Blanco},
  {Chiappini}, {de Laverny}, {Queiroz}, {Steinmetz}, {Adibekyan}, {Carrillo},
  {Cescutti}, {Guiglion}, {Hayden}, {de Jong}, {Kordopatis}, {Majewski},
  {Martig}, \& {Santiago}}]{minchev_2018MNRAS}
{Minchev}, I., {Anders}, F., {Recio-Blanco}, A., {et~al.} 2018, \mnras, 481,
  1645

\bibitem[{{Ness} {et~al.}(2015){Ness}, {Hogg}, {Rix}, {Ho}, \&
  {Zasowski}}]{Ness_Cannon_2015ApJ}
{Ness}, M., {Hogg}, D.~W., {Rix}, H.~W., {Ho}, A. Y.~Q., \& {Zasowski}, G.
  2015, \apj, 808, 16

\bibitem[{{O'Briain} {et~al.}(2021){O'Briain}, {Ting}, {Fabbro}, {Yi}, {Venn},
  \& {Bialek}}]{Teaghan_2021ApJ}
{O'Briain}, T., {Ting}, Y.-S., {Fabbro}, S., {et~al.} 2021, \apj, 906, 130

\bibitem[{{Pancino} {et~al.}(2017){Pancino}, {Lardo}, {Altavilla}, {Marinoni},
  {Ragaini}, {Cocozza}, {Bellazzini}, {Sabbi}, {Zoccali}, {Donati}, {Heiter},
  {Koposov}, {Blomme}, {Morel}, {S{\'\i}mon-D{\'\i}az}, {Lobel}, {Soubiran},
  {Montalban}, {Valentini}, {Casey}, {Blanco-Cuaresma}, {Jofr{\'e}}, {Worley},
  {Magrini}, {Hourihane}, {Fran{\c{c}}ois}, {Feltzing}, {Gilmore}, {Randich},
  {Asplund}, {Bonifacio}, {Drew}, {Jeffries}, {Micela}, {Vallenari}, {Alfaro},
  {Allende Prieto}, {Babusiaux}, {Bensby}, {Bragaglia}, {Flaccomio}, {Hambly},
  {Korn}, {Lanzafame}, {Smiljanic}, {Van Eck}, {Walton}, {Bayo}, {Carraro},
  {Costado}, {Damiani}, {Edvardsson}, {Franciosini}, {Frasca}, {Lewis},
  {Monaco}, {Morbidelli}, {Prisinzano}, {Sacco}, {Sbordone}, {Sousa}, {Zaggia},
  \& {Koch}}]{Pancino_2017A&A}
{Pancino}, E., {Lardo}, C., {Altavilla}, G., {et~al.} 2017, \aap, 598, A5

\bibitem[{{Pasquini} {et~al.}(2002){Pasquini}, {Avila}, {Blecha}, {Cacciari},
  {Cayatte}, {Colless}, {Damiani}, {de Propris}, {Dekker}, {di Marcantonio},
  {Farrell}, {Gillingham}, {Guinouard}, {Hammer}, {Kaufer}, {Hill}, {Marteaud},
  {Modigliani}, {Mulas}, {North}, {Popovic}, {Rossetti}, {Royer}, {Santin},
  {Schmutzer}, {Simond}, {Vola}, {Waller}, \& {Zoccali}}]{Pasquini_2002Msngr}
{Pasquini}, L., {Avila}, G., {Blecha}, A., {et~al.} 2002, The Messenger, 110, 1

\bibitem[{Pedregosa {et~al.}(2011)Pedregosa, Varoquaux, Gramfort, Michel,
  Thirion, Grisel, Blondel, Prettenhofer, Weiss, Dubourg, Vanderplas, Passos,
  Cournapeau, Brucher, Perrot, \& Duchesnay}]{scikitlearn}
Pedregosa, F., Varoquaux, G., Gramfort, A., {et~al.} 2011, Journal of Machine
  Learning Research, 12, 2825

\bibitem[{{Petrillo} {et~al.}(2017){Petrillo}, {Tortora}, {Chatterjee},
  {Vernardos}, {Koopmans}, {Verdoes Kleijn}, {Napolitano}, {Covone},
  {Schneider}, {Grado}, \& {McFarland}}]{Petrillo_2017MNRAS}
{Petrillo}, C.~E., {Tortora}, C., {Chatterjee}, S., {et~al.} 2017, \mnras, 472,
  1129

\bibitem[{{Pinsonneault}(1997)}]{Pinsonneault_1997ARA&A}
{Pinsonneault}, M. 1997, \araa, 35, 557

\bibitem[{{Pitrou} {et~al.}(2018){Pitrou}, {Coc}, {Uzan}, \&
  {Vangioni}}]{pitrou2018precision}
{Pitrou}, C., {Coc}, A., {Uzan}, J.-P., \& {Vangioni}, E. 2018, \physrep, 754,
  1

\bibitem[{{Prantzos} {et~al.}(2017){Prantzos}, {de Laverny}, {Guiglion},
  {Recio-Blanco}, \& {Worley}}]{prantzos2017}
{Prantzos}, N., {de Laverny}, P., {Guiglion}, G., {Recio-Blanco}, A., \&
  {Worley}, C.~C. 2017, \aap, 606, A132

\bibitem[{{Ram{\'\i}rez} {et~al.}(2012){Ram{\'\i}rez}, {Fish}, {Lambert}, \&
  {Allende Prieto}}]{Ramirez_2012}
{Ram{\'\i}rez}, I., {Fish}, J.~R., {Lambert}, D.~L., \& {Allende Prieto}, C.
  2012, \apj, 756, 46

\bibitem[{{Randich} {et~al.}(2013){Randich}, {Gilmore}, \& {Gaia-ESO
  Consortium}}]{Randich_2013Msngr}
{Randich}, S., {Gilmore}, G., \& {Gaia-ESO Consortium}. 2013, The Messenger,
  154, 47

\bibitem[{{Randich} {et~al.}(2022){Randich}, {Gilmore}, {Magrini}, {Sacco},
  {Jackson}, {Jeffries}, {Worley}, {Hourihane}, {Gonneau}, {Viscasillas
  Vazquez}, {Franciosini}, {Lewis}, {Alfaro}, {Allende Prieto}, {Bensby},
  {Blomme}, {Bragaglia}, {Flaccomio}, {Fran{\c{c}}ois}, {Irwin}, {Koposov},
  {Korn}, {Lanzafame}, {Pancino}, {Recio-Blanco}, {Smiljanic}, {Van Eck},
  {Zwitter}, {Asplund}, {Bonifacio}, {Feltzing}, {Binney}, {Drew}, {Ferguson},
  {Micela}, {Negueruela}, {Prusti}, {Rix}, {Vallenari}, {Bayo}, {Bergemann},
  {Biazzo}, {Carraro}, {Casey}, {Damiani}, {Frasca}, {Heiter}, {Hill},
  {Jofr{\'e}}, {de Laverny}, {Lind}, {Marconi}, {Martayan}, {Masseron},
  {Monaco}, {Morbidelli}, {Prisinzano}, {Sbordone}, {Sousa}, {Zaggia},
  {Adibekyan}, {Bonito}, {Caffau}, {Daflon}, {Feuillet}, {Gebran}, {Gonzalez
  Hernandez}, {Guiglion}, {Herrero}, {Lobel}, {Maiz Apellaniz}, {Merle},
  {Mikolaitis}, {Montes}, {Morel}, {Soubiran}, {Spina}, {Tabernero},
  {Tautvai{\v{s}}iene}, {Traven}, {Valentini}, {Van der Swaelmen}, {Villanova},
  {Wright}, {Abbas}, {Aguirre B{\o}rsen-Koch}, {Alves}, {Balaguer-Nunez},
  {Barklem}, {Barrado}, {Berlanas}, {Binks}, {Bressan}, {Capuzzo-Dolcetta},
  {Casagrande}, {Casamiquela}, {Collins}, {D'Orazi}, {Dantas}, {Debattista},
  {Delgado-Mena}, {Di Marcantonio}, {Drazdauskas}, {Evans}, {Famaey},
  {Franchini}, {Fr{\'e}mat}, {Friel}, {Fu}, {Geisler}, {Gerhard}, {Gonzalez
  Solares}, {Grebel}, {Gutierrez Albarran}, {Hatzidimitriou}, {Held},
  {Jim{\'e}nez-Esteban}, {J{\"o}nsson}, {Jordi}, {Khachaturyants},
  {Kordopatis}, {Kos}, {Lagarde}, {Mahy}, {Mapelli}, {Marfil}, {Martell},
  {Messina}, {Miglio}, {Minchev}, {Moitinho}, {Montalban}, {Monteiro},
  {Morossi}, {Mowlavi}, {Mucciarelli}, {Murphy}, {Nardetto}, {Ortolani},
  {Paletou}, {Palou{\v{s}}}, {Paunzen}, {Pickering}, {Quirrenbach}, {Re
  Fiorentin}, {Read}, {Romano}, {Ryde}, {Sanna}, {Santos}, {Seabroke},
  {Spagna}, {Steinmetz}, {Stonkut{\'e}}, {Sutorius}, {Th{\'e}venin}, {Tosi},
  {Tsantaki}, {Vink}, {Wright}, {Wyse}, {Zoccali}, {Zorec}, {Zucker}, \&
  {Walton}}]{randich2022}
{Randich}, S., {Gilmore}, G., {Magrini}, L., {et~al.} 2022, \aap, 666, A121

\bibitem[{{Randich} \& {Magrini}(2021)}]{Randich_2021}
{Randich}, S. \& {Magrini}, L. 2021, Frontiers in Astronomy and Space Sciences,
  8, 6

\bibitem[{{Randich} {et~al.}(2020){Randich}, {Pasquini}, {Franciosini},
  {Magrini}, {Jackson}, {Jeffries}, {d'Orazi}, {Romano}, {Sanna},
  {Tautvai{\v{s}}ien{\.{e}}}, {Tsantaki}, {Wright}, {Gilmore}, {Bensby},
  {Bragaglia}, {Pancino}, {Smiljanic}, {Bayo}, {Carraro}, {Gonneau},
  {Hourihane}, {Morbidelli}, \& {Worley}}]{GES_Randich_2020}
{Randich}, S., {Pasquini}, L., {Franciosini}, E., {et~al.} 2020, \aap, 640, L1

\bibitem[{{Reeves} {et~al.}(1970){Reeves}, {Fowler}, \&
  {Hoyle}}]{Reeves_1970Natur}
{Reeves}, H., {Fowler}, W.~A., \& {Hoyle}, F. 1970, \nat, 226, 727

\bibitem[{{Romano} {et~al.}(2021){Romano}, {Magrini}, {Randich}, {Casali},
  {Bonifacio}, {Jeffries}, {Matteucci}, {Franciosini}, {Spina}, {Guiglion},
  {Chiappini}, {Mucciarelli}, {Ventura}, {Grisoni}, {Bellazzini}, {Bensby},
  {Bragaglia}, {de Laverny}, {Korn}, {Martell}, {Tautvai{\v{s}}ien{\.{e}}},
  {Carraro}, {Gonneau}, {Jofr{\'e}}, {Pancino}, {Smiljanic}, {Vallenari}, {Fu},
  {Guti{\'e}rrez Albarr{\'a}n}, {Jim{\'e}nez-Esteban}, {Montes}, {Damiani},
  {Bergemann}, \& {Worley}}]{Romano_2021A&A}
{Romano}, D., {Magrini}, L., {Randich}, S., {et~al.} 2021, \aap, 653, A72

\bibitem[{{Romano} {et~al.}(1999){Romano}, {Matteucci}, {Molaro}, \&
  {Bonifacio}}]{Romano_1999A&A}
{Romano}, D., {Matteucci}, F., {Molaro}, P., \& {Bonifacio}, P. 1999, \aap,
  352, 117

\bibitem[{{Romano} {et~al.}(2001){Romano}, {Matteucci}, {Ventura}, \&
  {D'Antona}}]{Romano_2001}
{Romano}, D., {Matteucci}, F., {Ventura}, P., \& {D'Antona}, F. 2001, \aap,
  374, 646

\bibitem[{{Sackmann} \& {Boothroyd}(1999)}]{Sackmann_1999ApJ}
{Sackmann}, I.~J. \& {Boothroyd}, A.~I. 1999, \apj, 510, 217

\bibitem[{{Sanna} {et~al.}(2020){Sanna}, {Franciosini}, {Pancino},
  {Mucciarelli}, {Tsantaki}, {Charbonnel}, {Smiljanic}, {Fu}, {Bragaglia},
  {Lagarde}, {Tautvai{\v{s}}iene}, {Magrini}, {Randich}, {Bensby}, {Korn},
  {Bayo}, {Bergemann}, {Carraro}, \& {Morbidelli}}]{Sanna_2020A&A}
{Sanna}, N., {Franciosini}, E., {Pancino}, E., {et~al.} 2020, \aap, 639, L2

\bibitem[{{Singh} {et~al.}(2021){Singh}, {Reddy}, {Campbell}, {Kumar}, \&
  {Vrard}}]{Singh_2021ApJ}
{Singh}, R., {Reddy}, B.~E., {Campbell}, S.~W., {Kumar}, Y.~B., \& {Vrard}, M.
  2021, \apjl, 913, L4

\bibitem[{{Sitnova} {et~al.}(2018){Sitnova}, {Mashonkina}, \&
  {Ryabchikova}}]{Sitnova2018}
{Sitnova}, T.~M., {Mashonkina}, L.~I., \& {Ryabchikova}, T.~A. 2018, \mnras,
  477, 3343

\bibitem[{{Smiljanic} {et~al.}(2018){Smiljanic}, {Franciosini}, {Bragaglia},
  {Tautvai{\v{s}}ien{\.{e}}}, {Fu}, {Pancino}, {Adibekyan}, {Sousa}, {Randich},
  {Montalb{\'a}n}, {Pasquini}, {Magrini}, {Drazdauskas}, {Garc{\'\i}a},
  {Mathur}, {Mosser}, {R{\'e}gulo}, {de Assis Peralta}, {Hekker}, {Feuillet},
  {Valentini}, {Morel}, {Martell}, {Gilmore}, {Feltzing}, {Vallenari},
  {Bensby}, {Korn}, {Lanzafame}, {Recio-Blanco}, {Bayo}, {Carraro}, {Costado},
  {Frasca}, {Jofr{\'e}}, {Lardo}, {de Laverny}, {Lind}, {Masseron}, {Monaco},
  {Morbidelli}, {Prisinzano}, {Sbordone}, \& {Zaggia}}]{Smiljanic_2018A&A}
{Smiljanic}, R., {Franciosini}, E., {Bragaglia}, A., {et~al.} 2018, \aap, 617,
  A4

\bibitem[{{Smiljanic} {et~al.}(2014){Smiljanic}, {Korn}, {Bergemann}, {Frasca},
  {Magrini}, {Masseron}, {Pancino}, {Ruchti}, {San Roman}, {Sbordone}, {Sousa},
  {Tabernero}, {Tautvai{\v{s}}ien{\.{e}}}, {Valentini}, {Weber}, {Worley},
  {Adibekyan}, {Allende Prieto}, {Barisevi{\v{c}}ius}, {Biazzo},
  {Blanco-Cuaresma}, {Bonifacio}, {Bragaglia}, {Caffau}, {Cantat-Gaudin},
  {Chorniy}, {de Laverny}, {Delgado-Mena}, {Donati}, {Duffau}, {Franciosini},
  {Friel}, {Geisler}, {Gonz{\'a}lez Hern{\'a}ndez}, {Gruyters}, {Guiglion},
  {Hansen}, {Heiter}, {Hill}, {Jacobson}, {Jofre}, {J{\"o}nsson}, {Lanzafame},
  {Lardo}, {Ludwig}, {Maiorca}, {Mikolaitis}, {Montes}, {Morel}, {Mucciarelli},
  {Mu{\~n}oz}, {Nordlander}, {Pasquini}, {Puzeras}, {Recio-Blanco}, {Ryde},
  {Sacco}, {Santos}, {Serenelli}, {Sordo}, {Soubiran}, {Spina}, {Steffen},
  {Vallenari}, {Van Eck}, {Villanova}, {Gilmore}, {Randich}, {Asplund},
  {Binney}, {Drew}, {Feltzing}, {Ferguson}, {Jeffries}, {Micela}, {Negueruela},
  {Prusti}, {Rix}, {Alfaro}, {Babusiaux}, {Bensby}, {Blomme}, {Flaccomio},
  {Fran{\c{c}}ois}, {Irwin}, {Koposov}, {Walton}, {Bayo}, {Carraro}, {Costado},
  {Damiani}, {Edvardsson}, {Hourihane}, {Jackson}, {Lewis}, {Lind}, {Marconi},
  {Martayan}, {Monaco}, {Morbidelli}, {Prisinzano}, \&
  {Zaggia}}]{Smiljanic_2014A&A}
{Smiljanic}, R., {Korn}, A.~J., {Bergemann}, M., {et~al.} 2014, \aap, 570, A122

\bibitem[{{Sneden} {et~al.}(2012){Sneden}, {Bean}, {Ivans}, {Lucatello}, \&
  {Sobeck}}]{MOOG_sneden_2012}
{Sneden}, C., {Bean}, J., {Ivans}, I., {Lucatello}, S., \& {Sobeck}, J. 2012,
  {MOOG: LTE line analysis and spectrum synthesis}

\bibitem[{{Spite} \& {Spite}(1982)}]{spite1982abundance}
{Spite}, F. \& {Spite}, M. 1982, \aap, 115, 357

\bibitem[{{Stonkut{\.{e}}} {et~al.}(2016){Stonkut{\.{e}}}, {Koposov}, {Howes},
  {Feltzing}, {Worley}, {Gilmore}, {Ruchti}, {Kordopatis}, {Randich},
  {Zwitter}, {Bensby}, {Bragaglia}, {Smiljanic}, {Costado},
  {Tautvai{\v{s}}ien{\.{e}}}, {Casey}, {Korn}, {Lanzafame}, {Pancino},
  {Franciosini}, {Hourihane}, {Jofr{\'e}}, {Lardo}, {Lewis}, {Magrini},
  {Monaco}, {Morbidelli}, {Sacco}, \& {Sbordone}}]{Stonkute_2016MNRAS}
{Stonkut{\.{e}}}, E., {Koposov}, S.~E., {Howes}, L.~M., {et~al.} 2016, \mnras,
  460, 1131

\bibitem[{{Taylor}(2005)}]{Taylor2005}
{Taylor}, M.~B. 2005, in Astronomical Society of the Pacific Conference Series,
  Vol. 347, Astronomical Data Analysis Software and Systems XIV, ed.
  P.~{Shopbell}, M.~{Britton}, \& R.~{Ebert}, 29

\bibitem[{{Ting} {et~al.}(2018){Ting}, {Conroy}, {Rix}, \&
  {Asplund}}]{Ting_2018ApJ_no_Oxygen}
{Ting}, Y.-S., {Conroy}, C., {Rix}, H.-W., \& {Asplund}, M. 2018, \apj, 860,
  159

\bibitem[{{Ting} {et~al.}(2017){Ting}, {Conroy}, {Rix}, \&
  {Cargile}}]{Ting_2017ApJ_no_Oxygen}
{Ting}, Y.-S., {Conroy}, C., {Rix}, H.-W., \& {Cargile}, P. 2017, \apj, 843, 32

\bibitem[{{Ting} {et~al.}(2019){Ting}, {Conroy}, {Rix}, \&
  {Cargile}}]{Ting_Payne_2019ApJ}
{Ting}, Y.-S., {Conroy}, C., {Rix}, H.-W., \& {Cargile}, P. 2019, \apj, 879, 69

\bibitem[{{Valenti} \& {Piskunov}(1996)}]{SME_1996A&AS}
{Valenti}, J.~A. \& {Piskunov}, N. 1996, \aaps, 118, 595

\bibitem[{{Valentini} {et~al.}(2016){Valentini}, {Chiappini}, {Miglio},
  {Montalb{\'a}n}, {Rodrigues}, {Mosser}, {Anders}, {the CoRoT RG Group}, \&
  {GES Consortium}}]{valentini_corot_2016}
{Valentini}, M., {Chiappini}, C., {Miglio}, A., {et~al.} 2016, Astronomische
  Nachrichten, 337, 970

\bibitem[{Van~der Maaten \& Hinton(2008)}]{tSNE2008}
Van~der Maaten, L. \& Hinton, G. 2008, Journal of machine learning research, 9

\bibitem[{{{\v{C}}otar} {et~al.}(2021){{\v{C}}otar}, {Zwitter}, {Traven},
  {Bland-Hawthorn}, {Buder}, {Hayden}, {Kos}, {Lewis}, {Martell}, {Nordlander},
  {Stello}, {Horner}, {Ting}, {{\v{Z}}erjal}, \& {Galah
  Collaboration}}]{Klemen_2021MNRAS}
{{\v{C}}otar}, K., {Zwitter}, T., {Traven}, G., {et~al.} 2021, \mnras, 500,
  4849

\bibitem[{{Wang} {et~al.}(2021){Wang}, {Nordlander}, {Asplund}, {Amarsi},
  {Lind}, \& {Zhou}}]{wang2021}
{Wang}, E.~X., {Nordlander}, T., {Asplund}, M., {et~al.} 2021, \mnras, 500,
  2159

\bibitem[{Waskom(2021)}]{Waskom2021}
Waskom, M.~L. 2021, Journal of Open Source Software, 6, 3021

\bibitem[{{Woosley} \& {Weaver}(1995)}]{WoosleyWeaver1995}
{Woosley}, S.~E. \& {Weaver}, T.~A. 1995, \apjs, 101, 181

\bibitem[{{Xiang} {et~al.}(2019){Xiang}, {Ting}, {Rix}, {Sandford}, {Buder},
  {Lind}, {Liu}, {Shi}, \& {Zhang}}]{Xiang_2019ApJS}
{Xiang}, M., {Ting}, Y.-S., {Rix}, H.-W., {et~al.} 2019, \apjs, 245, 34

\bibitem[{{Zhang} {et~al.}(2019){Zhang}, {Zhao}, {Yang}, {Wang}, \&
  {Zuo}}]{Zhang_2019PASP}
{Zhang}, X., {Zhao}, G., {Yang}, C.~Q., {Wang}, Q.~X., \& {Zuo}, W.~B. 2019,
  \pasp, 131, 094202

\bibitem[{{Zhou} {et~al.}(2022){Zhou}, {Wang}, {Yan}, {Huang}, {Zhang}, {Ting},
  {Zhang}, \& {Shi}}]{zhou2022}
{Zhou}, Y., {Wang}, C., {Yan}, H., {et~al.} 2022, \apj, 931, 136

\end{thebibliography}

\appendix

\section{CNN model technical details} \label{sec:cnn_technical}

In the following sections, we detail the technical aspect of the CNN, in particular, the architecture, the choice of hyperparameters, and model generalization.

\subsection{Convolution and the fully connected layers}

Convolution layers are the central part of the CNN class of neural networks, as they are key to identifying patterns and features in input data \citep{fukushima1982neocognitron, LeCun_backpropagation_1989}. The 1D stellar spectra we use are characterized by absorption features governed by the physical properties of the stellar atmosphere. The CNN's goal is then to learn how these spectral features correlate with the stellar labels. The convolution layer, consisting of a collection of filters, when convolved with the 1D input from the previous layer, is able to extract the features. During the learning process, these filter parameters are optimized. After extensive tests, we adopted the model with 3 Conv1D layers with eight, six, and four filters, respectively. Using multiple filters in each convolution layer is similar to looking at the same object with different perspectives. 

After the first and second convolution layers, we apply a Maxpooling process, which reduces the feature map size by half. This is very useful to reduce the overall training parameters, which also reduces training time, while the network focuses on important features. Maxpooling isn't applied after the third convolution layer to avoid losing too much information.

At the heart of every neural network are the fully connected layers (or dense layers) \citep{Lecun_2015Natur}. They make up the central component that adds complexity and meaning to the functional approximation of the relationship between (in our case) the input spectrum and the output labels. As shown in Fig. \ref{fig:cnn_model}, the features learned from the input spectrum by the convolution layers are passed to the dense layers. This combination of convolution and dense layers ensures that the model learns from the whole spectral range instead of just the individual spectral features.

Our architecture contains three dense layers and one output layer (also a dense layer). The four feature maps from the last Conv1D layers are flattened before being fed to the first dense layer. The first dense layer has 64 neurons and receives input from the 6\,788 neurons of the flattened layer. The second and third dense layers have 128 and 32 neurons, respectively. The output dense layer is naturally composed of four neurons corresponding to the four training labels. Our choice of the number of layers and neurons is based on a good deal of experimentation, with the goal of having a CNN that is complex enough, without mitigating the training performance. 

\subsection{Choice of hyperparameters} \label{sec:hyper}

Hyperparameters are set at the beginning of the training and remain the same throughout the training, as opposed to the learnable model parameters such as the weights and biases. Here we discuss some important hyperparameters:

\begin{enumerate}
  \item \textbf{Weight initialization:} The weights of all parameters in the model have to be initialized before the training and neural networks are very sensitive to the initial weight values, as poor initialization can lead to a non-convergence. We adopted the intensively used ``golrot uniform" that initializes weights from a uniform distribution within a certain range.
  
  \item \textbf{Activation functions: } Activation functions are the mathematical functions that decide whether a neuron is activated or not. It adds non-linearity to the network and decides the output of any node or layer depending on the input. Each layer is activated using the ``Leaky-ReLu" activation function and for the output layer, we use the ``linear" activation. 
    
  \item \textbf{Epochs:} One complete pass of the training data through the network is called an epoch. Multiple epochs are needed for a good training. We allow for large number of training epochs until the training and test loss curves flatten out and stopped by using the EarlyStopping process (see Fig. \ref{fig:loss_acc}).
  
  \item \textbf{Batch size:} This refers to the number of data items used for one update of the model parameters during a single training epoch. The ``mini batch stochastic gradient descent" learning algorithm updates the model weights multiple times depending on the batch size in a single training epoch. It is an excellent way to lower the training time. A good choice for the batch size also provides regularization and stability during the training. We adopted a batch size of 64 as a balance between good approximation of the training set and faster training time.
  
  \item \textbf{Learning rate:} The learning rate ($\eta$) is the amount by which the weights are updated during the training and affect both the smooth convergence and training time. We tested several values of $\eta$ and found that the best performances, for our model, are achieved for $\eta = 0.0001$.
\end{enumerate}

\subsection{Model generalization: Avoiding over- or under-fitting} \label{sec:generalize}

The generalization and proper convergence of the model during the training is important to avoid over or underfitting and to ensure that the training progresses smoothly \citep{dosSantos_2022arXiv}. Our choice of convolution and dense layers ensures that the model does not under-fit the training data, hence, attention must be paid\ to avoiding overfitting the model. For this purpose, we employ the following regularization, dropouts, and early-stopping procedures detailed below.

In each of the three convolution layers, the L2 Regularization function is applied, allowing for a penalization of the loss function (see Sect. \ref{sec:train}) by adding to it a squared magnitude of model weights as a penalty term. The penalty term minimizes the model weights and ascertains that less significant features in the spectrum do not significantly affect the label prediction.

We applied a dropout layer on the inputs of the three inner dense layers. At each training epoch (explained below in Sect. \ref{sec:hyper}), a certain number of neurons are randomly selected and their contribution to the activation of neurons in subsequent layers is temporally removed. This forces the network to learn from the whole wavelength range of the spectrum as the model weights do not rely only on a very few spectral features and do not neglect less significant features. In Fig. \ref{fig:cnn_model}, we can see that 20\% of the neurons are dropped prior to the dense layers.

While training the CNN model, it is recommended that the training stop once the validation performance starts to degrade. For this task, we employed a callback called EarlyStopping in the model. This callback monitors the validation and test loss at the end of each training epoch and once the loss degrades or stagnates, over the last 25 epochs, the training is stopped and the model weights of the best training epoch are saved. 

Besides these techniques, the noise in the real observational data also plays an important role: noise in the training data acts as a regularizer and reduces over-fitting \cite{Bishop1995}, allowing for a faster training. Model based networks that do not use real observations but synthetic data instead, such as The Payne \citep{Ting_Payne_2019ApJ} using noise free spectra and StarNet \citep{Fabbro_2018MNRAS} with added Gaussian noise, are usually not representative of the inherent correlated noise of real spectra. Interstellar extinction, atmospheric extinction, and instrumental signatures are not simulated in the synthetic spectra and can lead to a significant synthetic gap. Furthermore, these synthetic data are also normally homogeneous in terms of labels, which is also not a true representation of the observations. The data-driven CNN employed in our study is  able to deal with the real noise efficiently. The noise in the data lead to a more efficient regularization and reduced generalization errors.

\section{Performance at different S/N} \label{sec: snr_trend}

\begin{figure*}[h!]
        \centering
        \includegraphics[width=\linewidth]{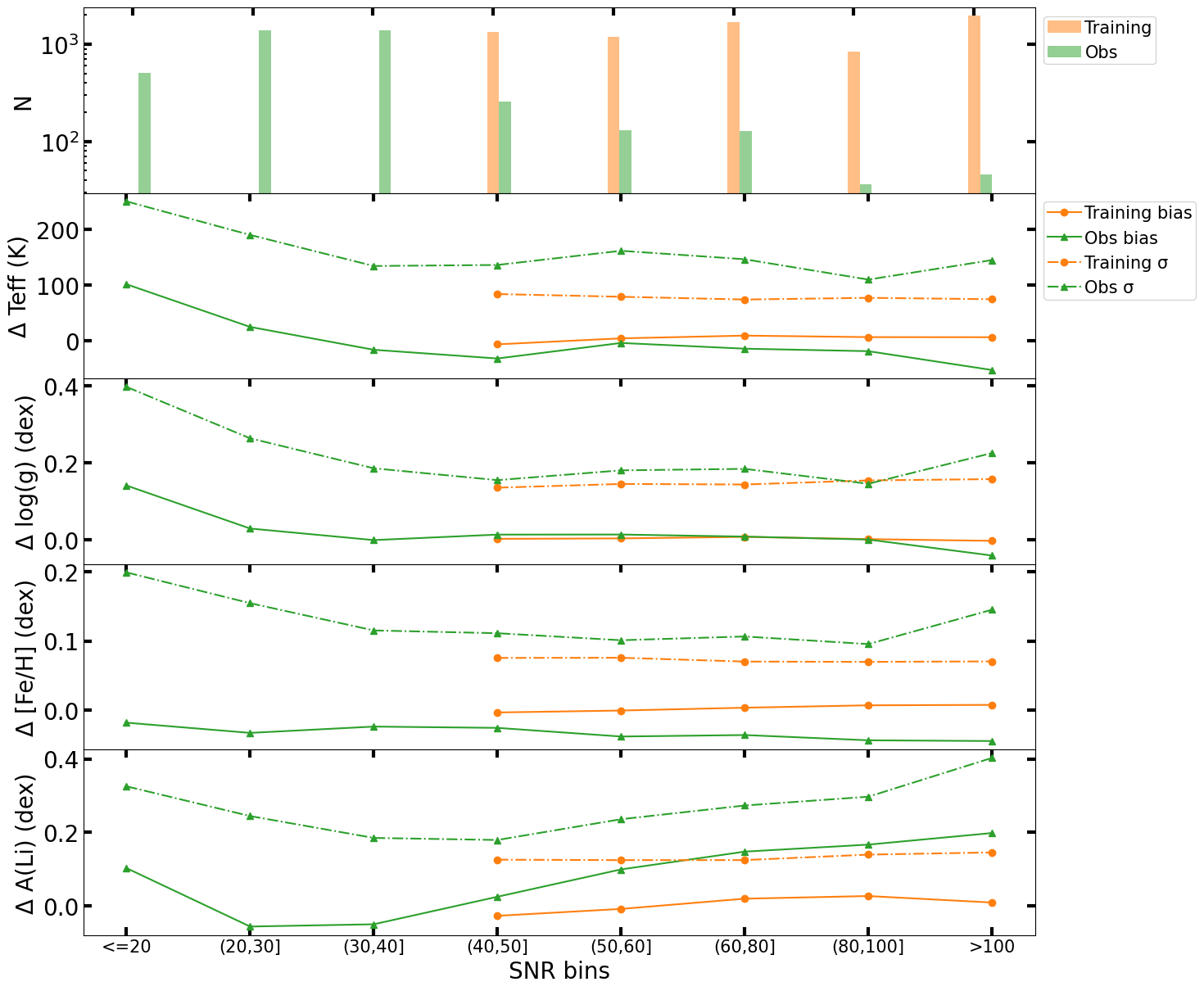}
        \caption{Histogram showing the number of spectra in each S/N bin for the training (yellow, 7031 stars) and observed (green, 5096 stars) samples (top panel). Bias (Bias = mean (CNN-iDR6), solid) and dispersion ($\sigma$ = std(CNN-iDR6), dash-dot) as a function of S/N (bottom 4 panels). The observed sample is selected within training label limits, eVRAD $<$ 1.0 km/s, with no GES flags and with $\mathrm{UPPER\_COMBINED\_LI1 = 0.0}$. For the observed sample, we have very few stars in the two highest S/N bins in comparison to the lower S/N bins.}
        \label{fig:bias_disp_snr}
\end{figure*}

Here, we investigate the robustness of CNN predictions for different S/N regimes. As illustrated in \figurename~\ref{fig:bias_disp_snr}, the mean bias and mean scatter between CNN and GES predictions remains constant across the different S/N bins for our four labels, in the training sample. For the observed sample, even though there are fewer stars in the higher S/N bins, CNN performances are similar compared to the training sample for the atmospheric parameters. In the bins S/N $\leq$ 30, the values are slightly higher for the observed sample, but are expected considering the level of noise and spectral resolution. The bias and $\sigma$ for Lithium show an increasing trend for S/N $>$ 50 in the observed sample due to low statistics for these bins. Also, the iDR6 A(Li) values for these stars mostly lie at the edges of our training A(Li) range, namely, above 3.0 dex or below 1.0 dex. We conclude that is robust to the noise in both the training and observed samples.

\end{document}